\documentclass[reqno,a4paper,final]{amsart}
\usepackage{amsmath,amsfonts,amssymb,amsthm,amstext,amscd,eucal}
\usepackage{bbold}
\usepackage{array}
\usepackage[latin1]{inputenc}
\usepackage[all]{xy}
\usepackage{hyperref}
\newcommand{\half}{\tfrac{1}{2}}
\renewcommand{\d}{\partial}
\newcommand{\fg}{\mathfrak{g}}
\newcommand{\fso}{\mathfrak{so}}
\newcommand{\SO}{\mathrm{SO}}
\renewcommand{\O}{\mathrm{O}}
\newcommand{\Cl}{\mathrm{C}\ell}
\newcommand{\Spin}{\mathrm{Spin}}
\newcommand{\Sp}{\mathrm{Sp}}
\newcommand{\SL}{\mathrm{SL}}
\newcommand{\SU}{\mathrm{SU}}
\newcommand{\RR}{\mathbb{R}}
\newcommand{\CC}{\mathbb{C}}
\newcommand{\CP}{\mathbb{CP}}
\newcommand{\KK}{\mathbb{K}}
\newcommand{\VV}{\mathbb{V}}
\newcommand{\ZZ}{\mathbb{Z}}
\newcommand{\eL}{\mathcal{L}}
\DeclareMathOperator{\AdS}{AdS}
\DeclareMathOperator{\dvol}{dvol}
\newcommand{\be}{\boldsymbol{e}}
\newcommand{\bv}{\boldsymbol{v}}
\newcommand{\bx}{\boldsymbol{x}}
\newcommand{\id}{\mathbb{1}}
\allowdisplaybreaks[1]
\numberwithin{equation}{section}
\hypersetup{%
  pdftitle = {Supersymmetric Kaluza--Klein reductions of AdS
    backgrounds},
  pdfkeywords = {anti de Sitter, quotients, supergravity,
    classifications, Kaluza-Klein, spin structures, supersymmetry},
  pdfauthor  = {José Figueroa-O'Farrill, Joan Simón}
}
\begin{document}
\title[Supersymmetric reductions of AdS backgrounds]{Supersymmetric
  Kaluza--Klein reductions of AdS backgrounds}
\author[Figueroa-O'Farrill]{José Figueroa-O'Farrill}
\address{School of Mathematics, The University of Edinburgh, United
  Kingdom}
\email{j.m.figueroa@ed.ac.uk}
\author[Simón]{Joan Simón}
\address{The Weizmann Institute of Physical Sciences, Rehovot, Israel}
\address{Department of Physics and Astronomy, University of
  Pennsylvania, United States}
\address{Kavli Institute of Theoretical Physics, University of
California, Santa Barbara, United States}
\email{jsimon@bokchoy.hep.upenn.edu}
\thanks{EMPG-03-24, WIS/01/04-JAN-DPP,UPR-1061-T, NSF-KITP-03-120}
\begin{abstract}
  This paper contains a classification of smooth Kaluza--Klein
  reductions (by one-parameter subgroups) of the maximally
  supersymmetric anti de~Sitter backgrounds of supergravity theories.
  We present a classification of one-parameter subgroups of isometries
  of anti de~Sitter spaces, discuss the causal properties of their
  orbits on these manifolds, and discuss their action on the space of
  Killing spinors.  We analyse the problem of which quotients admit a
  spin structure.  We then apply these results to write down the list
  of smooth everywhere spacelike supersymmetric quotients of $\AdS_3
  \times S^3 (\times \RR^4)$, $\AdS_4 \times S^7$, $\AdS_5 \times S^5$
  and $\AdS_7 \times S^4$, and the fraction of supersymmetry preserved
  by each quotient.  The results are summarised in tables which should
  be useful on their own.  The paper also includes a discussion of
  supersymmetry of singular quotients.
\end{abstract}

\maketitle

\tableofcontents

\section{Introduction and motivation}
\label{sec:intro}

In \cite{FigSimFlat} we started a programme of classification of
quotients of supersymmetric backgrounds of M-theory, of which this
paper is part.  Our initial motivation was the exploration of the then
novel sectors of type II string theory which emerged by the embedding
of the Melvin universe \cite{Melvin} into string theory
\cite{GibbonsWiltshire, GibbonsMaeda, DGGH1, DGGH2, DGGH3, CGS, CG,
  GSflux, Saffin, CHC, EmparanFlux, BrecherSaffin, Uranga,
  EmpGut, RTFlux}.  These are the so-called fluxbranes.  Their
supergravity description is in terms of the Kaluza--Klein reduction of
eleven-dimensional Minkowski spacetime along the orbits of suitable
one-parameter subgroups of the Poincaré group, and in
\cite{FigSimFlat} we classified all such reductions leading to a
smooth quotient and, in particular, we identified those for which the
quotient was supersymmetric.  These we christened generalised
supersymmetric fluxbranes.  One such fluxbrane, the nullbrane, is
obtained by quotienting by a combination of a translation and a null
rotation and such a quotient, when discretised, gives rise to an exact
time-dependent background of string theory
\cite{JoanNull,LMS1,LMS2,FabMcG,HorPol} describing a regular stable
supersymmetric cosmology interpolating smoothly between a big crunch
phase and a big bang phase.  It can be understood as a
desingularisation of the parabolic orbifold of \cite{HorowitzSteif}.

We continued our programme by classifying the supersymmetric
Kaluza--Klein reductions of the elementary half-BPS backgrounds of
M-theory: the M2- and M5-branes \cite{FigSimBranes} and the
gravitational wave and the Kaluza--Klein monopole \cite{FigSimGrav}.
This yielded a number of novel string backgrounds of type II string
theory which can be interpreted as intersections of type II branes
with generalised fluxbranes.  It also yielded two other classes of
backgrounds without a clear physical interpretation.  The first such
class, which were termed `exotic' in \cite{FigSimBranes,FigSimGrav},
can be shown to suffer from causal pathologies such as closed
lightlike and timelike curves \cite{Liatjoan}, despite being obtained
as quotients by a freely acting, everywhere spacelike Killing vector.
The second class involves the action of a `null translation' along the
brane worldvolume plus rotations in transverse planes.  Even though
these are free from closed causal curves, their interpretation remains
an open question.  Crucial to this analysis is the asymptotic flatness
of the spacetimes, which also holds in the case of intersecting
branes, to which the analysis can be extended \cite{FigRajSim}.

In this work, we shall extend this programme to the Kaluza--Klein
reductions of Freund--Rubin vacua of the form $\AdS_{p+1} \times S^q$.
These vacua can be understood as near-horizon geometries of the
asymptotically flat brane backgrounds, and in this limit there is a
symmetry enhancement giving rise to an increased number of allowed
quotients.  Some of these quotients are also allowed in the brane
backgrounds \cite{FigSimBranes,FigSimGrav} themselves and can be thus
interpreted as near-horizon geometries of the intersection of branes
with fluxbranes and nullbranes.

Using the AdS/CFT correspondence \cite{Malda}, D-branes in fluxbrane
backgrounds provide a connection with non-commutative (and non-local)
gauge theories, the so-called dipole theories \cite{BerGanDipoles,
  DGR, BDGKR, DSJDipole, AliYav}, whereas D-branes in the nullbrane
vacuum (or their dual descriptions) provide explicit time-dependent
backgrounds in string theory.  See \cite{AkiSav,JoanNullAdS} for field
theory dual proposals.  Other interesting quotients are concerned with
the possible existence of higher dimensional analogues of the family
of BTZ black holes in asymptotically $\AdS_3$ spacetimes \cite{BTZ,
  BHTZ}.  It is interesting to understand the geometry of these new
quotients as a first step to extending the AdS/CFT correspondence to
lorentzian orbifolds of $\AdS$, in particular, in the context of time
dependent backgrounds \cite{JoanNullAdS}, as was done for the selfdual
orbifolds \cite{CH} in \cite{BalNaqSim}.

The main result of this paper is then a classification of the smooth
supersymmetric quotients of Freund--Rubin backgrounds of AdS type by
one-parameter subgroups of isometries.  This work, together with
\cite{FOMRS}, completes some of the partial results in the literature
concerning the geometry of quotients of AdS
\cite{JoanNullAdS,HorMar,BehLus,GhoMuk,Cai,BDHRS,BRSbtz,mcinnes2003,
  mcinnes2004,FHLT}.  Using the techniques developed in \cite{FOMRS},
it is straightforward to derive the explicit geometries of the
corresponding Kaluza--Klein backgrounds associated with the
one-parameter subgroups studied in the present paper.  We choose not
to do this in the present paper, but may return to it elsewhere.

At a purely technical level, the classification problem in this paper
is solved by reducing to that of flat background, as was done in
\cite{FigSimBranes,FigSimGrav,FigRajSim}, with one crucial difference.
In the latter case the flat space in question is the asymptotic
spacetime of the brane background and hence we could bring to bear the
results of \cite{FigSimFlat}.  Here the flat spaces are $\RR^{2,p}$
and $\RR^{q+1}$ into which we can embed $\AdS_{p+1}$ (locally) and
$S^q$ as quadrics, and this makes the classification problem more
complicated as it is necessary to classify adjoint orbits of
$\fso(2,p)$.  This has been done by Boubel \cite{Boubel} as a special
case of a more general problem.  Boubel's method is thus perhaps
unnecessarily complex for the case at hand and partial results are
obtained in \cite{BHTZ} for $p=2$ and \cite{HolstPeldan} for $p=3$.
An elementary derivation for general $p$ appears in a recent paper of
Madden and Ross \cite{MaddenRoss}.  We will base our classification on
a refinement of these results.

Having obtained the possible quotients, the issue of supersymmetry is
easily dealt with by exploiting the one--to--one equivariant
correspondence between Killing spinors on $\AdS_{p+1} \times S^q$ and
parallel spinors in a flat space, here $\RR^{2,p} \times \RR^{q+1}$.
This follows from Bär's cone construction \cite{Baer} and its
lorentzian extension \cite{KathHabil} as will be explained in a more
general context in a forthcoming paper \cite{FigLeiSim} devoted to a
classification of supersymmetric Freund--Rubin backgrounds.

This paper is organised as follows.  In Section~\ref{sec:so2n} we set
up the problem of classifying one-parameter groups of isometries of
anti-de~Sitter spaces.  This is equivalent to the determination of the
orbits in (the projectivisation of) the Lie algebra $\fso(2,p)$ under
the action of the adjoint group $\SO(2,p)$.  This is in turn a special
case of the problem of determining the normal forms of skew-symmetric
linear transformations in a pseudo-euclidean space, which has been
solved recently by Boubel \cite{Boubel}, whose results we use.  Each
normal form is a direct sum of a finite number (with parameters) of
elementary blocks, whose enumeration is the purpose of
Section~\ref{sec:blocks}.  Great care has been exercised in
distinguishing between blocks which are related by an
orientation-reversing transformation.  Our results therefore
constitute a (necessary) refinement of the results of \cite{Boubel}
and of \cite{MaddenRoss}, and on the results of
\cite{BHTZ,HolstPeldan} for low-dimensional anti-de~Sitter spaces.
The elementary blocks are displayed matricially in
Tables~\ref{tab:blocks123}, \ref{tab:blocks4} and \ref{tab:blocks56},
and as elements of $\fso(2,p)$ in Table~\ref{tab:blocksSO}.  In
Section~\ref{sec:so2p} we classify the orbits of $\fso(2,p)$ under
$\SO(2,p)$ for the cases of interest $p=2,3,4,6$.  For each such $p$,
we determine the corresponding Killing vectors acting on $\AdS_{p+1}$
and determine their causal character on $\AdS_{p+1}$, paying close
attention to those Killing vectors whose norm in $\AdS_{p+1}$ is
bounded below, as only such vectors can give rise to quotients of the
Freund--Rubin backgrounds of the type $\AdS_{p+1} \times S^q$ which
are free from causal pathologies.  At the end of this section we will
have an enumeration of possible Killing vectors.  We have been careful
in enumerating them consistently with the embedding $\fso(2,p) \subset
\fso(2,p+1)$ or equivalently with the foliation of $\AdS_{p+2}$ by
$\AdS_{p+1}$ leaves.  In Section~\ref{sec:susy} we analyse the issue
of what happens to supersymmetry under quotients for Freund--Rubin
backgrounds of the form $\AdS_{p+1} \times S^q$.  This is done from
the point of view of supergravity; although we do comment on the
phenomenon of `supersymmetry without supersymmetry' which illustrates
the difference between supersymmetry in supergravity and in M-theory.
We first derive a criterion for the existence of a spin structure in
the quotient which reduces to a simple calculation in a Clifford
algebra.  We also summarise the representation-theoretical approach of
\cite{JMFKilling} to determining the Killing spinors of an AdS
background (summarised in Table~\ref{tab:Killing} for the backgrounds
of interest) in order to set up the problem of determining the
supersymmetry preserved under a quotient.  In Section~\ref{sec:squots}
we apply the preceding technology to classify the smooth spacelike
supersymmetric quotients of the Freund--Rubin backgrounds of several
supergravity theories.  In Section~\ref{sec:ads3s3} we tackle the
Freund--Rubin $\AdS_3 \times S^3$ vacuum of six-dimensional $(1,0)$
and $(2,0)$ supergravities and (primarily) its half-BPS lift $\AdS_3
\times S^3 \times \RR^4$ to IIB supergravity.  In
Sections~\ref{sec:ads4s7}, \ref{sec:ads5s5} and \ref{sec:ads7s4} we do
the same for the Freund--Rubin vacua of eleven-dimensional and IIB
supergravities.  These results are summarised in
Tables~\ref{tab:AdS3S3}, \ref{tab:AdS4S7}, \ref{tab:AdS5S5} and
\ref{tab:AdS7S4}.  For the benefit of the impatient reader, we
summarise the notation in those tables as follows:
\begin{itemize}
\item $\be_{ij} = \be_i \wedge \be_j \in \Lambda^2 \RR^{2,p}\cong
  \fso(2,p)$ with $\be_1$ and $\be_2$ timelike and the rest spacelike;
  and
\item $R_{ij} \in \fso(q+1)$ is the infinitesimal generator of rotations
  in the $(ij)$ plane in $\RR^{q+1}$.
\end{itemize}
In this way it should be possible to use our results without the
time-consuming---albeit ultimately rewarding---task of reading the
rest of the paper.

The paper concludes in Section~\ref{sec:singular} with a discussion of
supersymmetry for singular quotients.  This section is not as detailed
as the rest of the paper, and in it we simply state generic conditions
which allow the preservation of supersymmetry in a quotient that is
not necessarily smooth.  The results are again summarised in tabular
form.

\section{One-parameter subgroups of isometries of $\AdS_{p+1}$}
\label{sec:so2n}

In this section we set up the classification problem of one-parameter
subgroups of isometries of $\AdS_{p+1}$ and outline our strategy to
solve it.

For the purposes of this paper, $\AdS_{p+1}$ shall denote a
\emph{simply-connected} lorentzian spaceform with negative constant
sectional curvature.  Equivalently, it is the universal covering space
of the quadric
\begin{equation}
  \label{eq:adsquadric}
  -(t_1)^2 - (t_2)^2 + (x_1)^2 + \cdots + (x_p)^2 = - R^2
\end{equation}
in $\RR^{2,p} = \{(t_1,t_2,x_1,\dots,x_p)\}$.  Strictly speaking, this
is $\AdS_{p+1}$ with radius of curvature $R$, where $R$ can be any
positive number.  The quadric has closed timelike curves; for example,
\begin{equation*}
  t \mapsto (R \cos t, R\sin t, 0, \dots, 0)~,
\end{equation*}
and in the universal covering space these curves are unwrapped.  The
isometry group of the quadric is $\O(2,p)$, which acts linearly on
$\RR^{2,p}$ and preserves the quadric.  In the Freund--Rubin
supergravity backgrounds which are the focus of this work, the $\AdS$
factor contains more information than just the metric: it also has an
orientation, hence the true symmetry group of the quadric is
$\SO(2,p)$.  This, however, is not the isometry group of the
simply-connected $\AdS_{p+1}$.  The reason is simple.  The group
$\SO(2,p)$ has a maximal connected compact subgroup $\SO(2) \times
\SO(p)$.  The $\SO(2)$ factor is generated by the Killing vector $t_1
\frac{\d}{\d t_2} - t_2 \frac{\d}{\d t_1}$, of which the closed
timelike curves above are orbits.  However in $\AdS_{p+1}$ these
curves are not closed, hence this Killing vector does not generate a
circle subgroup but an $\RR$ subgroup.  Indeed, the symmetry group of
$\AdS_{p+1}$ is actually the infinite cover $\widetilde{\SO}(2,p)$ of
$\SO(2,p)$, in which the above $\SO(2)$ subgroup is unwrapped to an
$\RR$-subgroup. Of course, both $\SO(2,p)$ and $\widetilde{\SO}(2,p)$
share the same Lie algebra $\fso(2,p)$ and, furthermore, their adjoint
actions agree, since $\widetilde{\SO}(2,p)$ is a central extension of
$\SO(2,p)$ (see the discussion in Section~\ref{sec:sotilde}) and
the adjoint representation is trivial for the centre, whence the
action factors to $\SO(2,p)$.

Every one-parameter subgroup of isometries is generated by some
element in the Lie algebra of isometries.  Indeed, one-parameter
subgroups of a Lie group $G$ are in one-to-one correspondence with the
projectivised Lie algebra $\mathbb{P}\fg$.  In other words, two
elements $X$ and $Y$ of $\fg$ generate the same subgroup if they are
collinear.  Moreover, when $G$ acts by isometries on some space, we
will also identify one-parameter groups which are related by
conjugation in $G$  (or more generally by conjugation by the isometry
group of the space in question, which may be larger than $G$) since
the corresponding orbits will be isometric.  This means that we are
interested in classifying the equivalence classes of elements $X \in
\fg$ under
\begin{equation}
  \label{eq:equiv}
  X \sim t g X g^{-1}\qquad \text{where $t\in\RR^\times$ and $g\in
    G$.}
\end{equation}

In the case of interest, where $G=\widetilde{\SO}(2,p)$, it will
actually be enough to classify the equivalence classes of elements $X
\in \fso(2,p)$ under \eqref{eq:equiv} with $G=\SO(2,p)$.  There are
many partial results in the literature on this type of problem,
culminating more or less with the work of Boubel \cite{Boubel}.

Boubel has determined the normal forms for a (skew-)symmetric
endomorphism $B$ relative to a nondegenerate (skew-)symmetric bilinear
form $A$ in a $\KK$-vector space.  We are interested in the case where
$\KK = \RR$, $A$ is symmetric and $B$ is skew-symmetric.  Moreover we
are interested in the case where $A$ has signature $(2,p)$.  These
normal forms can be decomposed into elementary blocks to which $A$
and $B$ restrict, with $A$ still nondegenerate.  As a result the
blocks have signature $(m,n)$ where $m\leq 2$ and $n\leq p$.  The
tables of Boubel (as reproduced by Neukirchner \cite{Neukirchner} and
Leitner \cite{FelipeKilling}) classify these building blocks up to the
action of the orthogonal group of $A$; but in this analysis the
orientation plays no role and as a result, for the present purposes,
the classification contains discrete ambiguities which we will have to
resolve.

Boubel departs from the observation, as did others who have also
looked into this problem \cite{BHTZ,HolstPeldan}, that if
$\lambda\in\CC$ is an eigenvalue of $B$, then so are $-\lambda$,
$\bar\lambda$ and $-\bar\lambda$.  Several cases are therefore
possible, which can be labelled by the minimal polynomial $\mu(x)$ of
$B$:
\begin{enumerate}
\item $\lambda = 0$, with $\mu(x) = x^n$;
\item $\lambda = \beta \in \RR^\times$, with $\mu(x) = (x-\beta)^n
  (x+\beta)^n$;
\item $\lambda = i \varphi \in i\RR^\times$, with $\mu(x) =
  (x-i\varphi)^n (x+i\varphi)^n$; and
\item $\lambda = \beta + i \varphi \in \CC\setminus (\RR \cup i\RR)$,
  with $\mu(x) = (x- \lambda)^n (x+ \lambda)^n (x - \bar\lambda)^n (x+
  \bar\lambda)^n$.
\end{enumerate}
To each of these minimal polynomials there are associated (real)
Jordan normal forms.  The actual form depends on the dimension of the
block, which is a positive integer multiple $k$ of the index of
nilpotency $n$ in the above expressions.  Let $I_k$ denote the rank
$k$ identity matrix.

The normal form with minimal polynomial $\mu(x) = x^n$ and
characteristic polynomial $\chi(x) = x^{nk}$ is the $nk \times nk$
matrix:
\begin{equation*}
  \begin{pmatrix}
    0 & I_k & \\
      & \ddots & I_k \\
      &  & 0
  \end{pmatrix}~,
\end{equation*}
whereas the one with minimal polynomial $\mu(x) = (x-\beta)^n
(x+\beta)^n$ and characteristic polynomial $\chi(x) = \mu(x)^k$ is the
$2nk \times 2nk$ matrix:
\begin{equation*}
  \begin{pmatrix}
    M_{2k} & I_{2k} & \\
      & \ddots & I_{2k} \\
      &  & M_{2k}
  \end{pmatrix}~,
\end{equation*}
where $M_{2k}$ is the diagonal $2k \times 2k$ matrix with alternating
entries $+\beta$ and $-\beta$.

Similarly, the normal form with minimal polynomial $\mu(x) = (x-i\varphi)^n
(x+i\varphi)^n$ and characteristic polynomial $\chi(x) = \mu(x)^k$ is
the  $2kn \times 2kn$ matrix: 
\begin{equation*}
  \begin{pmatrix}
    J_{2k} & I_{2k} & \\
      & \ddots & I_{2k} \\
      &  & J_{2k}
  \end{pmatrix}~,
\end{equation*}
where $J_{2k}$ is the block-diagonal $2k \times 2k$ matrix consisting
of $k$ equal $2\times 2$ blocks consisting of the matrix
$\begin{pmatrix} 0 & -\varphi\\ \varphi & 0 \end{pmatrix}$.

Finally the normal form with minimal polynomial $\mu(x) = (x-
\lambda)^n (x+ \lambda)^n (x - \bar\lambda)^n (x+ \bar\lambda)^n$ and
characteristic polynomial $\chi(x) = \mu(x)^k$ is the $4nk \times 4nk$
matrix:
\begin{equation*}
  \begin{pmatrix}
     L_{4k} & I_{4k} & \\
      & \ddots & I_{4k} \\
      &  & L_{4k}
  \end{pmatrix}~,
\end{equation*}
where $L_{4k}(\beta,\varphi)$ is the block-diagonal $4k \times 4k$
matrix consisting of $k$ equal $4\times 4$ blocks consisting of the
matrix
\begin{equation*}
  \begin{pmatrix}
    \beta & - \varphi & & \\
    \varphi & \beta & & \\
    & & -\beta & \varphi \\
    & & -\varphi & -\beta
  \end{pmatrix}~.
\end{equation*}

The fact that $B$ is skew-symmetric implies conditions on the entries
of the (nondegenerate) inner-product of the elements of the Jordan
basis.  One can then argue that by changing basis in such a way that
the normal form remains unchanged, one can set to zero all inner
products which are not constrained by the Jordan form of $B$.  It is
then a simple matter, albeit a little tedious, to select the possible
normal forms with signature $(m,n)$ for $m\leq 2$.  These blocks are
tabulated in \cite{Boubel,Neukirchner,FelipeKilling} and an elementary
derivation can be found in \cite{MaddenRoss}.  Boubel's method is
equivalent (for the case under consideration) to the one employed in
\cite{BHTZ,HolstPeldan,MaddenRoss}.

From the supergravity point of view, they all share the drawback that
little attention is paid to whether one conjugates by $\SO(2,p)$ or
$\O(2,p)$.  We will therefore be forced to refine these results.  This
is done simply by taking each normal form in turn and investigating
whether the effect of an orientation-reversing orthogonal
transformation can be undone by an orientation-preserving orthogonal
transformation.  If this is not possible, then there are two normal
forms up to the action of the special orthogonal group, whereas if it
is possible then there is a single normal form.  In the following
section we summarise the results of these investigations.  The
calculations are routine.

\section{The elementary blocks}
\label{sec:blocks}

Let $\VV = \RR^{2,p}$ and $B$ be a skew-symmetric endomorphism of
$\VV$.  Associated with $B$ there will be an orthogonal decomposition
$\VV = \VV_1 \oplus \dots \oplus \VV_k$ into nondegenerate subspaces
stabilised by $B$.  Moreover we can assume that the $\VV_k$ are
indecomposable, so that the restriction of $B$ to any of the $\VV_k$
does not break down into further nondegenerate blocks.  In this way
every endomorphism $B$ will decompose into elementary blocks $B_k$,
namely their restrictions to each of the $\VV_k$, and conversely, from
a knowledge of all the possible blocks, we can assemble all the
possible endomorphisms $B$.  We are interested in the normal forms of
$B$ up to orientation-preserving isometries of $\RR^{2,p}$; that is,
up to the action of $\SO(2,p)$ and, as we will see, this requires
knowing the normal forms of the elementary blocks of signature
$(m,n)$ with $m\leq 2$ and $n\leq p$ up to the action of both
$\O(m,n)$ and $\SO(m,n)$.  We now describe the elementary blocks
which are summarised in Tables \ref{tab:blocks123}, \ref{tab:blocks4}
and \ref{tab:blocks56}.

There are two possible trivial blocks of size $1$, corresponding to
signatures $(0,1)$ and $(1,0)$.  Recall that the $\VV_k$ are
nondegenerate, hence if $B$ leaves invariant a null direction then this
signals the existence of a larger elementary block -- in fact, of
dimension at least $3$.  These two blocks are denoted $B^{(0,1)}$ and
$B^{(1,0)}$, respectively.

There are three possible blocks of size $2$, corresponding to
signatures $(0,2)$, $(1,1)$ and $(2,0)$.  The normal forms are
well-known, corresponding to a rotation in the case of definite
signature or a boost in the case of signature $(1,1)$.  We will follow
the mnemonic convention that boost parameters will be denoted $\beta$
and rotation parameters (i.e., angles) will be denoted $\varphi$.
These blocks are thus denoted $B^{(0,2)}(\varphi)$, $B^{(1,1)}(\beta)$
and $B^{(2,0)}(\varphi)$, respectively, with the understanding that
these parameters are never zero.  We may often belabour the point by
writing $B^{(0,2)}(\varphi\neq0)$, etc.  Let us now write these blocks
explicitly.  We will always write them relative to a
pseudo-orthonormal basis ordered in such a way that the timelike
directions appear first.  We will refer to such a basis as an
\emph{ordered frame}.  Hence relative to an ordered frame, the metric
of a $(1,1)$ block is diagonal with entries $(-1,1)$, for a $(2,0)$
block it is diagonal with entries $(-1,-1)$ and for a $(0,2)$ block
the diagonal entries are $(1,1)$.  Notice that ordered frames are in
general not compatible with the Jordan normal form of the
endomorphism.

Rather than writing the endomorphisms we prefer to write the
corresponding skew-symmetric bilinear forms; that is, the
corresponding elements of $\fso(m,n)$.  To be precise, let 
$\{\be_i\}$ be an ordered frame.  Let $B \be_j = \sum_k \be_k
B^k{}_j$.  Then the bilinear form associated to $B$ has entries
\begin{equation*}
  B_{ij} = \left< \be_i, B \be_j\right> = \sum_k \eta_{ik} B^k{}_j~,
\end{equation*}
where $\left<-,-\right>$ denotes the inner product and $\eta_{ik} =
\left<\be_i,\be_k\right>$ its components relative the frame.
Recapitulating, then, $B^{(m,n)}$, with or without parameters, will
label an element of $\fso(m,n)$ which, under the vector space
isomorphism $\fso(m,n) \cong \Lambda^2 \RR^{m,n}$, can be represented
as a skew-symmetric bilinear form.  With these preliminaries we can
now arrive at the first few entries in Table~\ref{tab:blocks123}.

There are three possible three-dimensional signatures $(0,3)$, $(1,2)$
and $(2,1)$ for a subspace of $\RR^{2,p}$.  Any skew-symmetric
endomorphism in a three-dimensional space will leave a direction
invariant; if this direction is timelike or spacelike then it induces
an orthogonal decomposition into smaller blocks; so we only have to
consider the case of the direction being null, which can only happen
in signatures $(1,2)$ or $(2,1)$.  In terms of Lorentz
transformations, such an endomorphism corresponds to a null rotation.
These are easier to write down in a light-cone basis, but for
uniformity we change basis to an ordered frame and arrive at the last
two entries in Table~\ref{tab:blocks123}.

\begin{table}[h!]
  \centering
  \setlength{\extrarowheight}{3pt}
  \renewcommand{\arraystretch}{1.3}
    \begin{tabular}{|>{$}l<{$}|>{$}l<{$}|>{$}l<{$}|}\hline
      \multicolumn{1}{|c|}{Name} &   \multicolumn{1}{c|}{Minimal polynomial} &
      \multicolumn{1}{c|}{Bilinear Form} \\
      \hline\hline
      B^{(0,1)} & x & [0]\\[3pt]
      B^{(1,0)} & x & [0]\\[3pt]
      B^{(0,2)}(\varphi\neq 0) & (x-i\varphi)(x+i\varphi) &
      \begin{bmatrix}0 & \varphi\\ -\varphi & 0\end{bmatrix}\\[3pt]
      B^{(1,1)}(\beta\neq 0) & (x-\beta)(x+\beta) &
      \begin{bmatrix}0 & -\beta \\ \beta & 0 \end{bmatrix}\\[3pt]
      B^{(2,0)}(\varphi\neq 0) & (x-i\varphi)(x+i\varphi) &
      \begin{bmatrix}0 & \varphi \\ - \varphi & 0 \end{bmatrix}\\[3pt]
      B^{(1,2)} & x^3 & \begin{bmatrix}0 & -1 & 0 \\ 1 & 0 & -1 \\
        0 & 1 & 0 \end{bmatrix}\\[3pt]
      B^{(2,1)} & x^3 & \begin{bmatrix}0 & 1 & 0 \\ -1 & 0 & 1 \\
        0 & -1 & 0 \end{bmatrix}\\[3pt]
      \hline
    \end{tabular}
  \vspace{8pt}
  \caption{Elementary blocks of size $\leq 3$ relative to an
    ordered frame.}
  \label{tab:blocks123}
\end{table}

There are three possible four-dimensional signatures for nondegenerate
subspaces of $\RR^{2,p}$, namely $(0,4)$, $(1,3)$ and $(2,2)$.
Clearly the first two break up into smaller blocks; so we only have to
consider signature $(2,2)$, where our lorentzian intuition begins to
be challenged.  One possibility is a combination of a simultaneous
boost in two orthogonal $(1,1)$-planes and a simultaneous rotation in
two orthogonal $(2,0)$- and $(0,2)$-planes.  This gives rise to the
blocks denoted $B^{(2,2)}_\pm(\beta,\varphi)$ in
Table~\ref{tab:blocks4}, where the parameters $\beta$ and $\varphi$
can both be chosen to be positive.  The blocks denoted $B^{(2,2)}_\pm$
are truly new to this signature and cannot be described in terms of
Lorentz transformations: they describe transformations which relate
two orthogonal null directions and hence they can first appear in
signature $(2,2)$.  The blocks $B^{(2,2)}_\pm(\beta)$ consist of a
deformation of $B^{(2,2)}_\pm$ by a simultaneous boost in two
orthogonal $(1,1)$ planes each containing one of the null directions
in $B^{(2,2)}_\pm$. Similarly the blocks $B^{(2,2)}_\pm(\varphi)$
consist of a deformation of $B^{(2,2)}_\pm$ by a simultaneous rotation
in a $(2,0)$-plane and in an orthogonal $(0,2)$-plane.

\begin{table}[h!]
  \centering
  \setlength{\extrarowheight}{3pt}
  \renewcommand{\arraystretch}{1.3}
  \begin{tiny}
    \begin{tabular}{|>{$}l<{$}|>{$}l<{$}|>{$}l<{$}|}\hline
      \multicolumn{1}{|c|}{Name} & \multicolumn{1}{c|}{Minimal Polynomial}  &
      \multicolumn{1}{c|}{Bilinear Form} \\
      \hline\hline
      B^{(2,2)}_\pm & x^2 &
      \begin{bmatrix}0 & \mp 1 & 1 & 0\\ \pm 1 & 0 & 0
        & \mp 1 \\ -1 & 0 & 0 & 1 \\ 0 & \pm 1 & -1 & 0
      \end{bmatrix}\\[5pt]
      B^{(2,2)}_\pm(\beta >0) & (x-\beta)^2(x+\beta)^2 &
      \begin{bmatrix}0 & \mp 1 & 1 & -\beta\\
        \pm 1 & 0 & \pm \beta & \mp 1 \\ - 1 & \mp \beta & 0 & 1 \\
        \beta & \pm 1 & -1 & 0 \end{bmatrix}\\[5pt]
      B^{(2,2)}_\pm(\varphi\neq 0) & (x-i\varphi)^2(x+i\varphi)^2 &
      \begin{bmatrix}0 & \mp 1\pm\varphi & 1 & 0\\ \pm 1\mp \varphi & 0 & 0
        & \mp 1 \\ -1 & 0 & 0 & 1 + \varphi \\ 0 & \pm 1 & -1 -
        \varphi & 0 \end{bmatrix}\\[5pt]
      B^{(2,2)}_\pm(\beta>0,\varphi>0) &
      (x-\lambda)(x+\lambda)(x-\bar\lambda)(x+\bar\lambda) & 
      \begin{bmatrix}0 & \pm \varphi & 0 & - \beta\\ \mp \varphi & 0 &
        \pm \beta & 0 \\ 0 & \mp \beta & 0 & -\varphi \\ \beta & 0 &
        \varphi & 0 \end{bmatrix}\\
      \hline
    \end{tabular}
  \end{tiny}
  \vspace{8pt}
  \caption{Elementary blocks of signature $(2,2)$ relative to an
    ordered frame, where $\lambda = \beta + i\varphi$.}
  \label{tab:blocks4}
\end{table}

There are three possible signatures for five-dimensional nondegenerate
subspaces of $\RR^{2,p}$: $(0,5)$, $(1,4)$ and $(2,3)$.  In the
euclidean and lorentzian cases $B$ always stabilises a nondegenerate
subspace, whence it decomposes into smaller blocks.  There is only one
possibility left, which is a $(2,3)$ block.  There is only such
possible indecomposable block up to $\SO(2,3)$ or indeed $\O(2,3)$,
which we denote $B^{(2,3)}$ and appears in Table~\ref{tab:blocks56}.
This corresponds to a transformation of the type $B^{(2,2)}_\pm$
together with a null rotation involving a conjugate null direction to
one of the ones in $B^{(2,2)}_\pm$ and a third spacelike direction.

Finally, there are three possible signatures for six-dimensional
nondegenerate subspaces of $\RR^{2,p}$: $(0,6)$, $(1,5)$ and $(2,4)$.
Again for the euclidean and lorentzian cases $B$ always stabilise a
nondegenerate subspace, whence it decomposes into smaller blocks.
Hence we must only consider the $(2,4)$ blocks.  It perhaps comes as a
surprise that there is an indecomposable endomorphism in this
signature: it is a combination of a simultaneous rotation in each of
three orthogonal planes with signature $(0,2)$, $(2,0)$ and $(2,0)$
with a double null rotation in two orthogonal $(1,2)$-planes.

There are no indecomposable blocks of the relevant signatures and of
size $>6$.

\begin{table}[h!]
  \centering
  \setlength{\extrarowheight}{3pt}
  \renewcommand{\arraystretch}{1.3}
    \begin{tabular}{|>{$}l<{$}|>{$}l<{$}|>{$}l<{$}|}\hline
      \multicolumn{1}{|c|}{Name} & \multicolumn{1}{c|}{Minimal Polynomial} &
      \multicolumn{1}{c|}{Bilinear Form} \\
      \hline\hline
      B^{(2,3)} & x^5 & \begin{bmatrix}0 & 1 & -1 & 0 & -1 \\
        -1 & 0 & 0 & 1 & 0 \\ 1 & 0 & 0 & -1 & 0 \\
        0 & -1 & 1 & 0 & -1  \\ 1 & 0 & 0 & 1 & 0
      \end{bmatrix}\\[5pt]
      B^{(2,4)}_\pm (\varphi\neq0) & (x-i\varphi)^3 (x+i\varphi)^3 &
      \begin{bmatrix}
        0 & \mp \varphi & 0 & 0 & -1 & 0 \\
        \pm \varphi & 0 & 0 & 0 & 0 & \mp 1 \\
        0 & 0 & 0 & \varphi & -1 & 0 \\
        0 & 0 & -\varphi & 0 & 0 & -1 \\
        1 & 0 & 1 & 0 & 0 & \varphi\\
        0 & \pm 1 & 0 & 1 & -\varphi & 0
      \end{bmatrix}\\
      \hline
    \end{tabular}
  \vspace{8pt}
  \caption{Elementary blocks of sizes $5$ and $6$ relative to an
    ordered frame.}
  \label{tab:blocks56}
\end{table}

Finally we must check whether any of the above blocks are related by
an orthogonal transformation which does not preserve the orientation.
That is, whether two $(m,n)$ blocks which are not $\SO(m,n)$-related
are $\O(m,n)$-related.  This can only happen when $m+n$ is even, since
when $m+n$ is odd, $-S$ preserves orientation whenever $S$ reverses
it, yet clearly conjugation by $S$ and $-S$ is the same; whence if two
endomorphisms are conjugate under $\O(m,n)$ they are also conjugate
under $\SO(m,n)$.  When $m+n$ is even, it is easy to investigate how
the above blocks change under an orientation-reversing orthogonal
transformation.  One finds the following relations:
\begin{itemize}
\item $B^{(0,2)}(\varphi) \sim B^{(0,2)}(-\varphi)$
\item $B^{(1,1)}(\beta) \sim B^{(1,1)}(-\beta)$
\item $B^{(2,0)}(\varphi) \sim B^{(2,0)}(-\varphi)$
\item $B^{(2,2)}_+ \sim B^{(2,2)}_-$
\item $B^{(2,2)}_+(\varphi) \sim B^{(2,2)}_-(\varphi)$
\item $B^{(2,2)}_+(\beta) \sim B^{(2,2)}_-(\beta)$
\item $B^{(2,2)}_+(\beta,\varphi) \sim B^{(2,2)}_-(\beta,\varphi)$, and
\item $B^{(2,4)}_+(\varphi) \sim B^{(2,4)}_-(\varphi)$.
\end{itemize}
In summary, ignoring orientation, we may ignore the $\pm$ sign in the
blocks $B^{(m,n)}_\pm$ with or without parameters, whereas for blocks
with parameters but without $\pm$ signs, we may take the parameters to
be positive.

It is convenient to write the bilinear forms appearing in each of the
elementary blocks in terms of the usual basis for $\Lambda^2 \RR^{m,n}
\cong \fso(m,n)$ consisting of wedge products of the elements of the
ordered frame.  This information is displayed in
Table~\ref{tab:blocksSO} where we employ the following notation :
$\be_1,\be_2$ will denote the timelike elements of an ordered frame
and $\be_3,\dots$ will denote the spacelike elements. In general, the
two-form is given by $\half \sum_{i,j} B^{ij} \be_i \wedge \be_j$,
where $B^{ij}$ is obtained by raising the indices of $B_{ij}$ with the
metric.  This explains the apparent discrepancies in signs from some
of the entries in Tables \ref{tab:blocks123}, \ref{tab:blocks4},
\ref{tab:blocks56} and in Table~\ref{tab:blocksSO}.

\begin{table}[h!]
  \centering
  \setlength{\extrarowheight}{3pt}
  \renewcommand{\arraystretch}{1.3}
    \begin{tabular}{|>{$}l<{$}|>{$}l<{$}|}\hline
      \multicolumn{1}{|c|}{Name} & \multicolumn{1}{c|}{Two-Form} \\
      \hline\hline
      B^{(0,1)} & 0 \\
      B^{(1,0)} & 0 \\
      B^{(0,2)}(\varphi) & \varphi \be_{34} \\
      B^{(1,1)}(\beta) & \beta \be_{13} \\ 
      B^{(2,0)}(\varphi) & \varphi \be_{12} \\
      B^{(1,2)} & \be_{13} - \be_{34} \\
      B^{(2,1)} & \be_{12} - \be_{23} \\
      B^{(2,2)}_\pm & \mp\be_{12} - \be_{13} \pm \be_{24} + \be_{34} \\
      B^{(2,2)}_\pm(\beta) & \mp\be_{12} - \be_{13} \pm \be_{24} + \be_{34}
      + \beta ( \be_{14} \mp \be_{23} ) \\
      B^{(2,2)}_\pm(\varphi) & \mp\be_{12} - \be_{13} \pm \be_{24} + \be_{34}
      + \varphi ( \pm\be_{12} + \be_{34} ) \\
      B^{(2,2)}_\pm(\beta,\varphi) & \varphi (\pm \be_{12} - \be_{34} )
      + \beta ( \be_{14} \mp \be_{23} ) \\
      B^{(2,3)} & \be_{12} + \be_{13} + \be_{15} - \be_{24} - \be_{34}
      - \be_{45} \\
      B^{(2,4)}_\pm (\varphi) & \be_{15} - \be_{35} \pm\be_{26} - \be_{46}
      + \varphi ( \mp\be_{12} + \be_{34} + \be_{56} )\\
      \hline
    \end{tabular}
  \vspace{8pt}
  \caption{The elementary blocks as two-forms.}
  \label{tab:blocksSO}
\end{table}

We are now ready to construct all the one-parameter subgroups of
$\SO(2,p)$ for any $p$.  We simply play \emph{Lego} with these
building blocks: assembling all the possible $(2,p)$ blocks out of
them, taking care to identify the resulting blocks under $\SO(2,p)$,
which may require identifying some of the building blocks under
$\O(m,n)$.  Finally, to obtain the subgroups, we may projectivise the
resulting normal forms; although doing so here would be premature
because we are ultimately interested in backgrounds of the form
$\AdS_{p+1} \times S^q$, whence the projectivisation comes after
adding a possible element of $\fso(q+1)$.

\section{One-parameter subgroups of $\SO(2,p)$}
\label{sec:so2p}

In this section we will classify the one-parameter subgroups of
$\SO(2,p)$ for $p=2,3,4,6$.  The general case is no harder in
principle, albeit of growing complexity as we will see already in low
dimension.

\subsection{One-parameter subgroups of $\SO(2,2)$}
\label{sec:so22}

\subsubsection{Adjoint orbits of $\fso(2,2)$}

The following decompositions are possible of a $(2,2)$-signature
space:
\begin{itemize}
\item $(2,2)$
\item $(2,1) \oplus (0,1)$
\item $(1,2) \oplus (1,0)$
\item $(2,0)\oplus (0,2)$
\item $(2,0) \oplus 2 (0,1)$
\item $2(1,1)$
\item $(1,1) \oplus (1,0) \oplus (0,1)$
\item $(0,2) \oplus 2 (1,0)$
\item $2(0,1) \oplus 2 (1,0)$
\end{itemize}
To each such decomposition there corresponds a block-diagonal
decomposition of the endomorphism $B$ and we are instructed to make
all possible block-diagonal decompositions and then make sure that no
two decompositions can be related by an $\SO(2,2)$ transformation.
There is always included in these decompositions the trivial case
$B=0$, which in the above list occurs at the end.  We will discard
this case henceforth.  Whenever there is a trivial factor in the
decomposition, e.g., $(2,1) \oplus (0,1)$, there is at least one
vector $\bv$ in the kernel of $B$ which is either timelike or
spacelike.  The orthogonal transformation $\bv \mapsto -\bv$ does not
change $B$ yet reverses orientation.  This means that we can use it to
compensate an orientation-reversing change of basis in the nontrivial
block(s), here $(2,1)$.  In other words, the possible endomorphisms
$B$ with a block-diagonal decomposition $(2,1) \oplus (0,1)$ up to the
action of $\SO(2,2)$ are in one-to-one correspondence with the
possible $(2,1)$-blocks up to the action of $\O(2,1)$.  We will write
this as $(2,1)_{\O}$.  More generally, the notation $(m,n)_{\O}$ will
denote all possible $(m,n)$-blocks ignoring orientation.  Similarly
the notation $[(m,n)\oplus(m',n')]_{\SO}$ means that we must take into
account all possible combinations of $(m,n)$- and $(m',n')$-blocks and
then identify them by orientation-preserving transformations of the
resulting $(m+m',n+n')$-block which however do not restrict to
orientation-preserving transformation in each of the sub-blocks.  With
this notation, and discarding the trivial case $B=0$, we can write
down the possible block-diagonal decompositions of $B$ up to
$\SO(2,2)$:
\begin{itemize}
\item $(2,2)$
\item $(2,1)_{\O}\oplus (0,1)$
\item $(1,2)_{\O}\oplus (1,0)$
\item $[(2,0)\oplus (0,2)]_{\SO}$
\item $(2,0)_{\O} \oplus 2 (0,1)$
\item $[2(1,1)]_{\SO}$
\item $(1,1)_{\O} \oplus (1,0) \oplus (0,1)$
\item $(0,2)_{\O} \oplus 2 (1,0)$
\end{itemize}

We now need to work out the two cases: $[(2,0)\oplus (0,2)]_{\SO}$ and
$[2(1,1)]_{\SO}$, since they are not given simply in terms of the
elementary blocks we have already classified, but will involve a
further restriction of parameters.

For $[(2,0)\oplus (0,2)]_{\SO}$ we have a $(2,2)$-block with bilinear
form
\begin{equation*}
  \begin{bmatrix}
    0 &  \varphi_1 & 0 & 0 \\
    -\varphi_1 & 0 & 0 & 0\\
    0 & 0 & 0 & \varphi_2 \\
    0 & 0 & -\varphi_2 & 0
  \end{bmatrix} \qquad \text{where $\varphi_1\varphi_2 \neq 0$.}
\end{equation*}
Under an $\SO(2,2)$ transformation not in $\SO(2,0) \times \SO(0,2)$
we can change the signs of the $\varphi_i$ simultaneously; which allows
us to choose $\varphi_1 >0$, say.

Similarly for $[2(1,1)]_{\SO}$ we have a $(2,2)$-block with bilinear
form
\begin{equation*}
  \begin{bmatrix}
    0 & 0 & -\beta_1 & 0\\
    0 & 0 & 0 & -\beta_2\\
    \beta_1 & 0 & 0 & 0\\
    0 & \beta_2 & 0 & 0
  \end{bmatrix} \qquad \text{where $\beta_1\beta_2 \neq 0$,}
\end{equation*}
which does not look block-diagonal because we had to reorder the basis
in order to make the timelike directions appear first.  Under
$\SO(2,2)$ transformations not in $\SO(1,1) \times \SO(1,1)$ we can
exchange $\beta_1 \leftrightarrow \beta_2$ and also change their signs
simultaneously.  Hence we may choose $\beta_1 \geq |\beta_2| > 0$.

With these considerations behind us, and taking into account Tables
\ref{tab:blocks123}, \ref{tab:blocks4} and \ref{tab:blocks56} of
elementary blocks, we can finally list all the possible elements of
$\fso(2,2)$ modulo the action of $\SO(2,2)$:
\begin{enumerate}
\item $B^{(2,0)}(\varphi>0) \oplus 2 B^{(0,1)}$
\item $B^{(1,1)}(\beta>0) \oplus B^{(1,0)} \oplus B^{(0,1)}$
\item $B^{(2,1)} \oplus B^{(0,1)}$
\item $B^{(0,2)}(\varphi>0) \oplus 2 B^{(1,0)}$
\item $B^{(1,2)} \oplus B^{(1,0)}$
\item $B^{(2,2)}_\pm$
\item $B^{(2,2)}_\pm(\beta>0)$
\item $B^{(2,2)}_\pm(\varphi\neq0)$
\item $B^{(2,2)}_\pm(\beta>0,\varphi>0)$
\item $B^{(0,2)}(\varphi_1>0) \oplus B^{(2,0)}(\varphi_2\neq0)$
\item $B^{(1,1)}(\beta_1) \oplus B^{(1,1)}(\beta_2)$, with $\beta_1
  \geq |\beta_2| > 0$
\end{enumerate}
We have reordered them to more easily reflect  the natural embeddings
$\fso(2,2) \subset \fso(2,3) \subset \cdots$ later on.

It is convenient to rewrite this list in a more traditional notation.
Let $\{\be_1,\dots,\be_4\}$ be an ordered frame with $\be_1$ and $\be_2$
timelike and the rest spacelike.  As usual we will write $\be_{ij} =
\be_i \wedge \be_j \in \Lambda^2 \RR^{2,2} \cong \fso(2,2)$.  Then,
using Table~\ref{tab:blocksSO}, the above elements can be written as
follows:
\begin{enumerate}
\item $\varphi \be_{12}$, ($\varphi>0$);
\item $\beta \be_{13}$, ($\beta>0$).
\item $\be_{12} - \be_{23}$
\item $\varphi \be_{34}$, ($\varphi>0$);
\item $\be_{13} - \be_{34}$
\item $\mp \be_{12} - \be_{13} \pm \be_{24} + \be_{34}$;
\item $\mp \be_{12} - \be_{13} \pm \be_{24} + \be_{34} + \beta
  (\be_{14} \mp \be_{23})$, ($\beta>0$);
\item $\mp \be_{12} - \be_{13} \pm \be_{24} + \be_{34} + \varphi (\pm
  \be_{12} + \be_{34})$, ($\varphi\neq0$);
\item $\varphi(\pm \be_{12} - \be_{34} ) + \beta ( \be_{14} \mp
  \be_{23} )$, ($\beta>0$, $\varphi>0$);
\item $\varphi_1 \be_{12} + \varphi_2 \be_{34}$, ($\varphi_1>0$,
  $\varphi_2 \neq 0$);
\item $\beta_1 \be_{13} + \beta_2 \be_{24}$, ($\beta_1 \geq |\beta_2|
  > 0$); and
\end{enumerate}

Clearly by letting some of the parameters become $0$, we can subsume
some of these cases into others; but we prefer not to do this at this
stage.  The above list is in one-to-one correspondence with the
conjugacy classes of nonzero elements of $\fso(2,2)$ under
$\SO(2,2)$.

A direct comparison with the results of \cite{BHTZ} is now possible.
We do not think it relevant to perform this comparison in detail here.
Let us merely observe that if we were to further identify elements of
$\fso(2,2)$ under the action of $\O(2,2)$, then we would obtain the
results of \cite{BHTZ} except for the fact that the parameters in
their classes $\text{I}_a$, $\text{I}_b$ and $\text{I}_c$ should then
be constrained.

\subsubsection{Causal properties of orbits}

Next we determine the causal properties of the orbits in $\AdS_3$
under the one-parameter subgroups.  We will do this by computing the
norm of the vector field which generates the subgroup.  These are
easier to compute in the local model for $\AdS_3$ given by the quadric
in $\RR^{2,2}$, since the Killing vectors are the restriction to the
quadric of linear vector fields in $\RR^{2,2}$.  The Killing vectors
and also their norms lift to $\AdS_3$.  This is perhaps most
concretely realised by giving coordinates to $\AdS_3$ which are
adapted to the action of the fundamental group of the quadric and
noticing that the Killing vectors (and hence their norms) are
invariant under this action, whence they lift trivially.  In practical
terms, what this means is that if the action of the elementary group
is generated by translating a coordinate, $\tau$ say, by a some period
$T$, then the local coordinate expressions for the Killing vectors
will be periodic in $\tau$ with period $T$.  The same will be true, of
course, for anti-de~Sitter spaces of higher dimension.

We introduce flat coordinates $x^i$ for $\RR^{2,2}$ associated to the
ordered frame $\be_i$, so that a point in $\RR^2$ has coordinates $\bx
= \sum x^i \be_i$.  In these coordinates the metric takes the form
\begin{equation*}
  g = \eta_{ij} dx^i dx^j = - (dx^1)^2 - (dx^2)^2 + (dx^3)^2 +
  (dx^4)^2~.
\end{equation*}
The Killing vector associated to the Lie algebra element $\be_{ij} \in
\fso(2,2)$ is $x_i \d_j - x_j \d_i$, and hence if $X = \half  B^{ij}
\be_{ij}$, then the norm of the corresponding Killing vector 
$\xi_X = \half B^{ij} (x_i \d_j - x_j \d_i)$ is given by
\begin{equation}
  \label{eq:normKilling}
  |\xi_X|^2 = \eta_{ij} B_k{}^i B_\ell{}^j x^k x^\ell = \eta^{ij}
  B_{ik}x^k B_{j\ell} x^\ell~,
\end{equation}
which we then restrict to the quadric
\begin{equation*}
  - (x^1)^2 - (x^2)^2 + (x^3)^2 + (x^4)^2 = - R^2~.
\end{equation*}
Equation \eqref{eq:normKilling} is very easily implemented from the
explicit expressions of the bilinear forms in
Tables~\ref{tab:blocks123}, \ref{tab:blocks4} and \ref{tab:blocks56}:
we simply apply the bilinear form as a matrix to the (column) vector
$(x^1,x^2,x^3,x^4)=(-x_1, -x_2, x_3, x_4)$ and then compute the
Minkowski norm of the resulting vector.  Doing so, we quickly arrive
at the following result for the Killing vectors in the same order as
that given above:
\begin{enumerate}
\item $|\xi|^2 = -\varphi^2 (R^2 + x_3^2 + x_4^2)$, which is unbounded
  below;
\item $|\xi|^2 = \beta^2 ( R^2 - x_2^2 + x_4^2 )$, which is
  unbounded below;
\item $|\xi|^2 = - (x_3 + x_1)^2$, which is unbounded below;
\item $|\xi|^2 = \varphi^2 ( x_3^2 + x_4^2) \geq 0$;
\item $|\xi|^2 = (x_1 + x_4)^2 \geq 0$;
\item $|\xi|^2 = 0$;
\item $|\xi|^2 = \beta^2 R^2 + 4\beta (x_1+x_4) (x_3 \pm x_2)$,
  which is unbounded below;
\item $|\xi|^2 = -\varphi^2 R^2 + 2 \varphi( (x_1+x_4)^2 + (x_3 \pm
  x_2)^2)$, which is $> - \varphi^2 R^2$ if $\varphi>0$;
\item $|\xi|^2 = (\beta^2 - \varphi^2) R^2 - 4\beta \varphi (x_1x_3
  \pm x_2x_4)$, which is unbounded below;
\item $|\xi|^2 = -\varphi_2^2 R^2 + (\varphi_1^2 -
  \varphi_2^2)(x_3^2 + x_4^2)$, which is $\geq -\varphi_2^2 R^2$
  if $\varphi_1^2 \geq \varphi_2^2$; and
\item $|\xi|^2 = \beta_1^2 R^2 + (\beta_2^2 - \beta_1^2)(x_2^2 -
  x_4^2)$, which is positive if $\beta_1 = \pm \beta_2$ and
  unbounded below otherwise.
\end{enumerate}

Before going on to the next dimension, let us pause a moment to
explain how these norms will be used.  In future sections, we will be
interested in quotients of $\AdS_{p+1} \times S^q$ by everywhere
spacelike Killing vectors.  The sphere being compact and riemannian,
the norm of a Killing vector is non-negative and bounded above.
Moreover if $q$ is odd there are Killing vectors whose norm on the
sphere is positive and pinched away from zero.  This means that in
those cases we can allow Killing vectors whose norms are bounded below
on $\AdS_{p+1}$ but not necessarily positive.  Such reductions
appeared for the first time in \cite{FigSimBranes} in the context of
reductions of elementary M-branes and are known to have closed causal
curves \cite{Liatjoan}.  The same phenomenon will happen here whenever
the norm of the Killing vector acting on $\AdS_{p+1}$ is bounded from
below and negative in some regions \cite{FOMRS}.

Notice that the property of the norm of a Killing vector `not being
bounded below' in $\AdS$ is hereditary under the embeddings
$\fso(2,2) \subset \fso(2,3) \subset \cdots$ Hence if a certain
Killing vector cannot be used in $\AdS_{p+1}$ it cannot be used either
in a higher-dimensional $\AdS$ space, since $\AdS_n$ contains $\AdS_m$
subspaces for $m<n$ where the norm can already be arbitrarily
negative.  On the other hand, the norm of a Killing vector in
$\AdS_{p+1}$ which is bounded below may become unbounded upon
embedding in a higher-dimensional $\AdS$ space.

\subsection{One-parameter subgroups of $\SO(2,3)$}
\label{sec:so23}

\subsubsection{Adjoint orbits of $\fso(2,3)$}

The possible decompositions of a skew-symmetric endomorphism $B$ up to
conjugation by $\SO(2,3)$ are as follows:
\begin{itemize}
\item $(2,3)$
\item $(2,2)_{\O} \oplus (0,1)$
\item $[(2,1) \oplus (0,2)]_{\SO} = (2,1) \oplus (0,2)_{\O}$
\item $(2,1) \oplus 2 (0,1)$
\item $[(2,0) \oplus (0,2)]_{\O} \oplus (0,1)$
\item $(2,0)_{\O} \oplus 3 (0,1)$
\item $[(1,2) \oplus (1,1)]_{\SO} = (1,2) \oplus (1,1)_{\O}$
\item $(1,2) \oplus (1,0) \oplus (0,1)$
\item $[2(1,1)]_{\O} \oplus (0,1)$
\item $[(1,1) \oplus (0,2)]_{\O} \oplus (1,0)$
\item $(1,1)_{\O} \oplus 2 (0,1) \oplus (1,0)$
\item $(0,2)_{\O} \oplus 2 (1,0) \oplus (0,1)$
\end{itemize}
Here, as before, the notation $[2(m,n)]_{\O}$ denotes all possible
pairs of $(m,n)$-blocks identified under transformations of
$\O(2m,2n)$ which do not belong to $\O(m,n) \times \O(m,n)$; namely
the orthogonal transformation which interchanges the two blocks.  Some
of the above blocks are not elementary and must be separately
investigated.  For example, the block $[(2,0) \oplus (0,2)]_{\O}$ is
represented by the bilinear form
\begin{equation*}
  \begin{bmatrix}
    0 & \varphi_1 & 0 & 0\\
    -\varphi_1 & 0 & 0 & 0\\
    0 & 0 & 0 & \varphi_2\\
    0 & 0 & -\varphi_2 & 0
  \end{bmatrix} \qquad \text{where $\varphi_1\varphi_2 \neq 0$.}
\end{equation*}
Under $\O(2,2)$ we can change the signs of $\varphi_i$ independently,
whence we can choose them both positive.  Similarly the block
$[2(1,1)]_{\O}$ is represented by the bilinear form
\begin{equation*}
  \begin{bmatrix}
    0 & 0 & -\beta_1 & 0\\
    0 & 0 & 0 & -\beta_2\\
    \beta_1 & 0 & 0 & 0\\
    0 & \beta_2 & 0 & 0
  \end{bmatrix} \qquad \text{where $\beta_1\beta_2\neq 0$.}
\end{equation*}
Under $\O(2,2)$ we can interchange $\beta_1$ and $\beta_2$ and change
their signs independently; whence we can always choose $\beta_1 \geq
\beta_2 > 0$.  Finally we have $[(1,1) \oplus (0,2)]_{\O}$, which is
represented by the bilinear form
\begin{equation*}
  \begin{bmatrix}
    0 & -\beta & 0 & 0\\
    \beta & 0 & 0 & 0\\
    0 & 0 & 0 & \varphi \\
    0 & 0 & -\varphi & 0
  \end{bmatrix} \qquad \text{where $\beta \varphi \neq 0$.}
\end{equation*}
Under $\O(2,2)$ we can change the sign of $\beta$ and $\varphi$
independently, whence we can choose them positive.

We can now list all the possible elements of $\fso(2,3)$ up to the
action of $\SO(2,3)$:
\begin{enumerate}
\item $B^{(2,0)}(\varphi>0) \oplus 3 B^{(0,1)}$
\item $B^{(1,1)}(\beta>0) \oplus 2 B^{(0,1)} \oplus B^{(1,0)}$
\item $B^{(2,1)} \oplus 2 B^{(0,1)}$
\item $B^{(0,2)}(\varphi>0) \oplus 2 B^{(1,0)} \oplus B^{(0,1)}$
\item $B^{(1,2)} \oplus B^{(1,0)} \oplus B^{(0,1)}$
\item $B^{(2,2)}_+ \oplus B^{(0,1)}$
\item $B^{(2,2)}_+(\beta>0) \oplus B^{(0,1)}$
\item $B^{(2,2)}_+(\varphi\neq0) \oplus B^{(0,1)}$
\item $B^{(2,2)}_+(\beta>0,\varphi>0) \oplus B^{(0,1)}$
\item $B^{(2,0)}(\varphi_1>0) \oplus B^{(0,2)}(\varphi_2>0) \oplus B^{(0,1)}$
\item $B^{(1,1)}(\beta_1) \oplus B^{(1,1)}(\beta_2) \oplus B^{(0,1)}$, with $\beta_1 \geq \beta_2 > 0$
\item $B^{(1,1)}(\beta>0) \oplus B^{(0,2)}(\varphi>0) \oplus B^{(1,0)}$
\item $B^{(2,3)}$
\item $B^{(2,1)} \oplus B^{(0,2)}(\varphi>0)$
\item $B^{(1,2)} \oplus B^{(1,1)}(\beta>0)$
\end{enumerate}
where again we have reordered them in such a way that the numbering
makes the embedding $\fso(2,2) \subset \fso(2,3)$ manifest; that is,
each of the cases (1)-(11) is the embedding in $\fso(2,3)$ of the
corresponding case in $\fso(2,2)$; except that sometimes the extra
freedom in conjugating by the larger group $\SO(2,3)$ results in a
further constraint on the parameters.

Again we rewrite this list in a more traditional notation.  Let
$\{\be_1,\dots,\be_5\}$ be an ordered frame with $\be_1$ and $\be_2$
timelike and the rest spacelike.  Then, using
Table~\ref{tab:blocksSO}, the above elements can be written as
follows:
\begin{enumerate}
\item $\varphi \be_{12}$, ($\varphi>0$);
\item $\beta \be_{13}$, ($\beta>0$); and
\item $\be_{12} - \be_{23}$;
\item $\varphi \be_{34}$, ($\varphi>0$).
\item $\be_{13} - \be_{34}$;
\item $- \be_{12} - \be_{13} + \be_{24} + \be_{34}$
\item $- \be_{12} - \be_{13} + \be_{24} + \be_{34} + \beta (\be_{14} -
  \be_{23})$, ($\beta>0$);
\item $- \be_{12} - \be_{13} + \be_{24} + \be_{34} + \varphi (\be_{12}
  + \be_{34})$, ($\varphi\neq0$);
\item $\varphi(\be_{12} - \be_{34}) + \beta (\be_{14} - \be_{23})$,
  ($\beta>0$, $\varphi>0$);
\item $\varphi_1 \be_{12} + \varphi_2 \be_{34}$, ($\varphi_i>0$);
\item $\beta_1 \be_{13} + \beta_2 \be_{24}$, ($\beta_1 \geq \beta_2 >
  0$);
\item $\beta \be_{13} + \varphi \be_{45}$, ($\beta>0$, $\varphi>0$);
\item $\be_{12} + \be_{13} + \be_{15} - \be_{24} - \be_{34} -
  \be_{45}$
\item $\be_{12} - \be_{23} + \varphi \be_{45}$, ($\varphi>0$);
\item $\be_{13} - \be_{34} + \beta \be_{25}$, ($\beta>0$);
\end{enumerate}

Again by letting some of the parameters become $0$, we can subsume
some of these cases into others; but we prefer not to do this at this
stage.  The above list is in one-to-one correspondence with the
conjugacy classes of nonzero elements of $\fso(2,3)$ under 
$\SO(2,3)$.

A direct comparison with the results of \cite{HolstPeldan} is now also
possible and as in the case of $\fso(2,2)$, our results morally agree
with those of \cite{HolstPeldan} provided we identify elements under
the action of $\O(2,3)$.

\subsubsection{Causal properties of orbits}

We compute the norms of the Killing vectors as we did for
$\fso(2,2)$.  Notice that we can read off the norms of the Killing
vectors (1)-(11) coming from $\fso(2,2)$ simply by noticing that
whenever $R^2$ appears in the calculation of the norms for $\AdS_3$ we
now have $R^2 + x_5^2$.  Similarly we can read off the norms of
decomposable blocks by adding the norms of each of the blocks, taken
care that the coordinates should correspond.  These considerations and
a simple calculation yield immediately the following norms:
\begin{enumerate}
\item $|\xi|^2 = -\varphi^2 (R^2 + x_3^2 + x_4^2 + x_5^2)$, which is
  unbounded below;
\item $|\xi|^2 = \beta^2 ( R^2 - x_2^2 + x_4^2 + x_5^2 )$, which is
  unbounded below;
\item $|\xi|^2 = - (x_3 + x_1)^2$, which is unbounded below; 
\item $|\xi|^2 = \varphi^2 ( x_3^2 + x_4^2) \geq 0$;
\item $|\xi|^2 = (x_1 + x_4)^2 \geq 0$;
\item $|\xi|^2 = 0$;
\item $|\xi|^2 = \beta^2 (R^2+x_5^2) + 4\beta (x_1+x_4) (x_3 +
  x_2)$, which is unbounded below;
\item $|\xi|^2 = -\varphi^2 (R^2+x_5^2) + 2 \varphi( (x_1+x_4)^2 +
  (x_2 + x_3)^2)$, which is unbounded below because there are points
  in $\AdS_4$ with arbitrarily small $|x_1+x_4|$ and $|x_2 + x_3|$ and
  arbitrarily large $R^2 + x_5^2 = (x_1+x_4)(x_1-x_4) + (x_2+x_3)(x_2
  - x_3)$;
\item $|\xi|^2 = (\beta^2 - \varphi^2) (R^2+x_5^2) - 4\beta \varphi
  (x_1x_3 + x_2x_4)$, which is unbounded below;
\item $|\xi|^2 = -\varphi_2^2 (R^2+x_5^2) + (\varphi_1^2 -
  \varphi_2^2)(x_3^2 + x_4^2)$, which is unbounded below since in the
  subspace of $\AdS_4$ where $x_3=x_4=0$, $|x_5|$ is not bounded;
\item $|\xi|^2 = \beta_1^2 (R^2+x_5^2) + (\beta_2^2 - \beta_1^2)(x_2^2
  - x_4^2)$, which is $\geq \beta_1^2 R^2$ if $\beta_1 = \beta_2$ and
  unbounded below otherwise;
\item $|\xi|^2 = \beta^2 (x_1^2 - x_3^2) + \varphi^2 (x_4^2 + x_5^2)$,
  which is unbounded below;
\item $|\xi|^2 = (x_4 - x_1)^2 - 4 (x_2 + x_3) x_5$, which is
  unbounded below;
\item $|\xi|^2 = -(x_1+x_3)^2 + \varphi^2 (x_4^2 + x_5^2)$, which is
  unbounded below; and
\item $|\xi|^2 = (x_1+x_4)^2 + \beta^2 (x_2^2 - x_5^2)$, which is also
  unbounded below.
\end{enumerate}

Since we will be applying these results to the reductions of $\AdS_4
\times S^7$ and $S^7$ does possess Killing vectors without zeroes we
will be needing those Killing vectors on $\AdS_4$ whose norms are
bounded below.  From the above list we see immediately that these are
(4), (5), (6) and (11) for $\beta_1 = \beta_2$.

\subsection{One-parameter subgroups of $\SO(2,4)$}
\label{sec:so24}

\subsubsection{Adjoint orbits of $\fso(2,4)$}

Up to conjugation by $\SO(2,4)$ a skew-symmetric endomorphism $B$ 
can have the following block-diagonal decompositions:
\begin{itemize}
\item $(2,4)$
\item $(2,3) \oplus (0,1)$
\item $[(2,2)\oplus(0,2)]_{\SO} = (2,2)_{\O} \oplus (0,2)$
\item $(2,2)_{\O} \oplus 2 (0,1)$
\item $(2,1)\oplus(0,2)_{\O} \oplus (0,1)$
\item $(2,1) \oplus 3 (0,1)$
\item $2(1,2)$
\item $(1,2) \oplus (1,1)_{\O} \oplus (0,1)$
\item $(1,2) \oplus (0,2)_{\O} \oplus (1,0)$
\item $(1,2) \oplus (1,0) \oplus 2(0,1)$
\item $[(2,0)\oplus 2(0,2)]_{\SO} = (2,0) \oplus [2(0,2)]_{\O}$
\item $(2,0)_{\O} \oplus (0,2)_{\O} \oplus 2 (0,1)$
\item $(2,0)_{\O} \oplus 4(0,1)$
\item $[2(1,1) \oplus (0,2)]_{\SO} = [2(1,1)]_{\O} \oplus (0,2)$
\item $[2(1,1)]_{\O} \oplus 2(0,1)$
\item $(1,1)_{\O} \oplus (0,2)_{\O} \oplus (0,1) \oplus (1,0)$
\item $(1,1)_{\O} \oplus (1,0) \oplus 3 (0,1)$
\item $[2(0,2)]_{\O} \oplus 2 (1,0)$
\item $(0,2)_{\O} \oplus 2(1,0) \oplus 2(0,1)$
\end{itemize}
We still have to work out $[2(0,2)]_{\O}$ which is represented by a
bilinear form of the type
\begin{equation*}
  \begin{bmatrix}
    0 & \varphi_1 & 0 & 0\\
    -\varphi_1 & 0 & 0 & 0\\
    0 & 0 & 0 & \varphi_2\\
    0 & 0 & -\varphi_2 & 0
  \end{bmatrix} \qquad \text{where $\varphi_1\varphi_2 \neq 0$.}
\end{equation*}
Under $\O(2,2)$ we can interchange $\varphi_1$ and $\varphi_2$ and
change their signs independently; whence we choose $\varphi_1 \geq
\varphi_2 > 0$.

With these considerations behind us, we can now list all the possible
elements of $\fso(2,4)$ up to the action of $\SO(2,4)$, where we have
again ordered them in such a way that (1)-(15) correspond to the
embedding in $\fso(2,4)$ of the corresponding elements of $\fso(2,3)$:
\begin{enumerate}
\item $B^{(2,0)}(\varphi>0) \oplus 4 B^{(0,1)}$
\item $B^{(1,1)}(\beta>0) \oplus 3 B^{(0,1)} \oplus B^{(1,0)}$
\item $B^{(2,1)} \oplus 3 B^{(0,1)}$
\item $B^{(0,2)}(\varphi>0) \oplus 2 B^{(1,0)} \oplus 2 B^{(0,1)}$.
\item $B^{(1,2)} \oplus 2 B^{(0,1)} \oplus B^{(1,0)}$
\item $B^{(2,2)}_+ \oplus 2 B^{(0,1)}$
\item $B^{(2,2)}_+(\beta>0) \oplus 2 B^{(0,1)}$
\item $B^{(2,2)}_+(\varphi\neq0) \oplus 2 B^{(0,1)}$
\item $B^{(2,2)}_+(\beta>0,\varphi>0) \oplus 2 B^{(0,1)}$
\item $B^{(2,0)}(\varphi_1>0) \oplus B^{(0,2)}(\varphi_2>0) \oplus 2  B^{(0,1)}$
\item $B^{(1,1)}(\beta_1) \oplus B^{(1,1)}(\beta_2) \oplus  2B^{(0,1)}$, with $\beta_1 \geq \beta_2 > 0$
\item $B^{(1,1)}(\beta>0) \oplus B^{(0,2)}(\varphi>0) \oplus B^{(0,1)}  \oplus B^{(1,0)}$
\item $B^{(2,3)} \oplus B^{(0,1)}$
\item $B^{(2,1)} \oplus B^{(0,2)}(\varphi>0) \oplus B^{(0,1)}$
\item $B^{(1,2)} \oplus B^{(1,1)}(\beta>0) \oplus B^{(0,1)}$
\item $B^{(0,2)}(\varphi_1) \oplus B^{(0,2)}(\varphi_2) \oplus 2  B^{(1,0)}$, with $\varphi_1 \geq \varphi_2 > 0$
\item $B^{(1,2)} \oplus B^{(0,2)}(\varphi>0) \oplus B^{(1,0)}$
\item $B^{(2,4)}_\pm(\varphi\neq0)$
\item $B^{(2,2)}_+(\beta>0) \oplus B^{(0,2)}(\varphi\neq0)$
\item $B^{(2,2)}_+(\varphi_1\neq0) \oplus B^{(0,2)}(\varphi_2\neq 0)$
\item $B^{(2,2)}_+ \oplus B^{(0,2)}(\varphi\neq 0)$
\item $B^{(2,2)}_+(\beta>0,\varphi_1>0) \oplus B^{(0,2)}(\varphi_2\neq   0)$
\item $2B^{(1,2)}$
\item $B^{(2,0)}(\varphi_1\neq0) \oplus B^{(0,2)}(\varphi_2) \oplus  B^{(0,2)}(\varphi_3)$, with $\varphi_2 \geq \varphi_3 > 0$
\item $B^{(1,1)}(\beta_1) \oplus B^{(1,1)}(\beta_2) \oplus  B^{(0,2)}(\varphi\neq0)$, with $\beta_1 \geq \beta_2 > 0$
\end{enumerate}

Again, let $\{\be_1,\dots,\be_6\}$ be an ordered frame with $\be_1$
and $\be_2$ timelike and the rest spacelike.  Then, using
Table~\ref{tab:blocksSO}, the above elements can be written as
follows:
\begin{enumerate}
\item $\varphi \be_{12}$, ($\varphi>0$);
\item $\beta \be_{13}$, ($\beta>0$);
\item $\be_{12} - \be_{23}$;
\item $\varphi \be_{34}$, ($\varphi>0$)
\item $\be_{13} - \be_{34}$;
\item $- \be_{12} - \be_{13} + \be_{24} + \be_{34}$
\item $- \be_{12} - \be_{13} + \be_{24} + \be_{34} + \beta  (\be_{14} - \be_{23})$, ($\beta>0$);
\item $- \be_{12} - \be_{13} + \be_{24} + \be_{34} + \varphi (\be_{12}  + \be_{34} )$, ($\varphi \neq 0$);
\item $\varphi ( \be_{12} - \be_{34} ) + \beta ( \be_{14} -  \be_{23})$, ($\beta>0$, $\varphi>0$);
\item $\varphi_1 \be_{12} + \varphi_2 \be_{34}$, ($\varphi_i>0$);
\item $\beta_1 \be_{13} + \beta_2 \be_{24}$, ($\beta_1 \geq \beta_2 >  0$);
\item $\beta \be_{13} + \varphi \be_{45}$, ($\beta>0$, $\varphi>0$);
\item $\be_{12} + \be_{13} + \be_{15} - \be_{24} - \be_{34} -  \be_{45}$;
\item $\be_{12} - \be_{23} + \varphi \be_{45}$, ($\varphi>0$);
\item $\be_{13} - \be_{34} + \beta \be_{25}$, ($\beta>0$);
\item $\varphi_1 \be_{34} + \varphi_2 \be_{56}$, ($\varphi_1 \geq
  \varphi_2 > 0$);
\item $\be_{13} - \be_{34} + \varphi \be_{56}$, ($\varphi>0$);
\item $\varphi( \mp \be_{12} + \be_{34} + \be_{56} ) + \be_{15} -  \be_{35} \pm \be_{26} - \be_{46}$, ($\varphi\neq 0$)
\item $- \be_{12} - \be_{13} + \be_{24} + \be_{34} + \beta  (\be_{14} - \be_{23}) + \varphi \be_{56}$, ($\beta>0$,$\varphi  \neq 0$);
\item $- \be_{12} - \be_{13} + \be_{24} + \be_{34} + \varphi_1 ( \be_{12} + \be_{34} ) + \varphi_2 \be_{56}$, ($\varphi_i\neq 0$);
\item $- \be_{12} - \be_{13} + \be_{24} + \be_{34} + \varphi  \be_{56}$, ($\varphi\neq0$);
\item $\varphi_1 ( \be_{12} - \be_{34} ) + \beta ( \be_{14} -  \be_{23} ) + \varphi_2 \be_{56}$, ($\beta>0$, $\varphi_1>0$,  $\varphi_2\neq 0$);  
\item $\be_{13} - \be_{34} + \be_{25} - \be_{56}$;
\item $\varphi_1 \be_{12} + \varphi_2 \be_{34} + \varphi_3 \be_{56}$,
  ($\varphi_2 \geq \varphi_3 > 0$, $\varphi_1\neq 0$);
\item $\beta_1 \be_{13} + \beta_2 \be_{24} + \varphi \be_{56}$,
  ($\varphi\neq 0$, $\beta_1 \geq \beta_2 > 0$); 
\end{enumerate}

Clearly by letting some of the parameters become $0$, we can subsume
some of these cases into others; but we prefer not to do this at this
stage.  The above list is in one-to-one correspondence with the
conjugacy classes of nonzero elements of $\fso(2,4)$ under
$\SO(2,4)$.

\subsubsection{Causal properties of orbits}

Similar considerations as those explained in the previous section on
norms allow us to immediately write the norms of the first 15 Killing
vector fields with the proviso that $R^2$ becomes $R^2 + x_6^2$ with
respect to the $\AdS_4$ norms.  All cases but (18) involve no new
computations, just adding results of previous computations.  At the
end of the day, one obtains the following norms:
\begin{enumerate}
\item $|\xi|^2 = -\varphi^2 (R^2 + x_3^2 + x_4^2 + x_5^2 + x_6^2)$,
  which is unbounded below;
\item $|\xi|^2 = \beta^2 ( R^2 - x_2^2 + x_4^2 + x_5^2 +x_6^2 )$,
  which is unbounded below;
\item $|\xi|^2 = - (x_3 + x_1)^2$, which is unbounded below;
\item $|\xi|^2 = \varphi^2 ( x_3^2 + x_4^2) \geq 0$;
\item $|\xi|^2 = (x_1 + x_4)^2 \geq 0$;
\item $|\xi|^2 = 0$;
\item $|\xi|^2 = \beta^2 (R^2+x_5^2+x_6^2) + 4\beta (x_1+x_4) (x_3 +
  x_2)$, which is unbounded below;
\item $|\xi|^2 = -\varphi^2 (R^2+x_5^2+x_6^2) + 2 \varphi( (x_1+x_4)^2 +
  (x_3 + x_2)^2)$, which is unbounded below;
\item $|\xi|^2 = (\beta^2 - \varphi^2) (R^2+x_5^2+x_6^2) - 4\beta \varphi
  (x_1x_3 + x_2x_4)$, which is unbounded below;
\item $|\xi|^2 = -\varphi_2^2 (R^2+x_5^2+x_6^2) + (\varphi_1^2 -
  \varphi_2^2)(x_3^2 + x_4^2)$, which is unbounded below;
\item $|\xi|^2 = \beta_1^2 (R^2+x_5^2+x_6^2) + (\beta_2^2 -
  \beta_1^2)(x_2^2 - x_4^2)$, which is $\geq \beta_1^2 R^2$ if
  $\beta_1 = \beta_2$ and unbounded below otherwise;
\item $|\xi|^2 = \beta^2 (x_1^2 - x_3^2) + \varphi^2 (x_4^2 + x_5^2)$,
  which is unbounded below;
\item $|\xi|^2 = (x_1 - x_4)^2 - 4 (x_2 + x_3) x_5$, which is
  unbounded below;
\item $|\xi|^2 = -(x_1+x_3)^2 + \varphi^2 (x_4^2 + x_5^2)$, which is
  unbounded below;
\item $|\xi|^2 = (x_1+x_4)^2 + \beta^2 (x_2^2 - x_5^2)$, which is
  unbounded below;
\item $|\xi|^2 = \varphi_1^2 (x_3^2 + x_4^2) + \varphi_2^2 (x_5^2 +
  x_6^2) \geq 0$;
\item $|\xi|^2 = (x_1+x_4)^2 + \varphi^2 (x_5^2 + x_6^2) \geq 0$;
\item $|\xi|^2 = -\varphi^2 R^2 + (x_1-x_3)^2 + (x_4 \mp x_2)^2 -
  4\varphi ((x_4 \mp x_2) x_5 + (x_1 - x_3) x_6)$, which is unbounded
  below, since in the subspace of $\AdS_5$ where $x_4 \mp x_2=0$ and
  $x_1 - x_3 =-2\varphi x_6$ we can take $|x_6|$ as large as desired;
\item $|\xi|^2 = \beta^2 R^2 + 4\beta (x_1+x_4)(x_2+x_3) + (\varphi^2
  + \beta^2) (x_5^2 + x_6^2)$, which is unbounded below;
\item $|\xi|^2 = -\varphi_1^2 R^2 + 2 \varphi_1 ((x_1+x_4)^2 + (x_2 +
  x_3)^2)) + (\varphi_2^2 -\varphi_1^2)(x_5^2 + x_6^2)$, which is
  $\geq -\varphi_1^2 R^2$ provided that $|\varphi_2| \geq \varphi_1 >
  0$ and unbounded below otherwise;
\item $|\xi|^2 = \varphi^2 (x_5^2 + x_6^2) \geq 0$;
\item $|\xi|^2 = (\beta^2-\varphi_1^2) R^2 + (\beta^2 - \varphi_1^2 +
  \varphi_2^2) (x_5^2 + x_6^2) - 4\beta \varphi_1 (x_1 x_3 + x_2
  x_4)$, which is unbounded below;
\item $|\xi|^2 = (x_1 + x_4)^2 + (x_2 + x_6)^2 >0$;
\item $|\xi|^2 = -\varphi_1^2 R^2 + (\varphi_2^2 - \varphi_1^2) (x_3^2
  + x_4^2) + (\varphi_3^2 - \varphi_1^2) (x_5^2 + x_6^2)$, which is
  $\geq -\varphi_1^2 R^2$ provided that $\varphi_3 \geq \varphi_2 \geq
  |\varphi_1|>0$; and
\item $|\xi|^2 = \beta_1^2 R^2 + (\beta_2^2 - \beta_1^2) (x_2^2 -
  x_4^2) + (\varphi^2 + \beta_1^2)(x_5^2 + x_6^2)$, which is $\geq
  \beta_1^2 R^2$ if $\beta_2 = \beta_1$ and unbounded below otherwise.
\end{enumerate}

Since we will be applying these results to the reductions of $\AdS_5
\times S^5$ and $S^5$ does possess Killing vectors without zeroes, we
will be needing those Killing vectors on $\AdS_5$ whose norms  are
bounded below.  From the above list we see immediately that these 
are (4), (5), (6), (11) for $\beta_1 = \beta_2$, (16), (17), (20) for
$|\varphi_2| \geq \varphi_1 > 0$, (21), (23), (24) for $\varphi_2 \geq
\varphi_3 \geq |\varphi_1|>0$, and (25) for $\beta_1 = \beta_2$.

\subsection{One-parameter subgroups of $\SO(2,6)$}
\label{sec:so26}

\subsubsection{Adjoint orbits of $\fso(2,6)$}

The following decompositions are possible for a skew-symmetric
endomorphism $B$ up to conjugation by $\SO(2,6)$:
\begin{itemize}
\item $[(2,4)\oplus (0,2)]_{\SO} = (2,4)_{\O} \oplus (0,2)$
\item $(2,4)_{\O} \oplus 2(0,1)$
\item $(2,3) \oplus (0,2)_{\O} \oplus (0,1)$
\item $(2,3) \oplus 3(0,1)$
\item $[(2,2) \oplus 2(0,2)]_{\SO} = (2,2)_{\O} \oplus [2(0,2)]_{\SO}$
\item $(2,2)_{\O} \oplus (0,2) \oplus 2(0,1)$
\item $(2,2)_{\O} \oplus 4(0,1)$
\item $(2,1) \oplus [2(0,2)]_{\O} \oplus (0,1)$
\item $(2,1) \oplus (0,2)_{\O} \oplus 3(0,1)$
\item $(2,1) \oplus 5 (0,1)$
\item $2(1,2) \oplus (0,2)_{\O}$
\item $2(1,2) \oplus 2(0,1)$
\item $(1,2) \oplus [2(0,2)]_{\O} \oplus (1,0)$
\item $(1,2) \oplus (1,1)_{\O} \oplus (0,2)_{\O} \oplus (0,1)$
\item $(1,2) \oplus (1,1)_{\O} \oplus 3 (0,1)$
\item $(1,2) \oplus (0,2)_{\O} \oplus (1,0) \oplus 2(0,1)$
\item $(1,2) \oplus (1,0) \oplus 4 (0,1)$
\item $(2,0) \oplus [3(0,2)]_{\O}$
\item $(2,0)_{\O} \oplus [2(0,2)]_{\O} \oplus 2(0,1)$
\item $(2,0)_{\O} \oplus (0,2)_{\O} \oplus 4 (0,1)$
\item $(2,0)_{\O} \oplus 6(0,1)$
\item $[2(1,1) \oplus 2(0,2)]_{\SO}$
\item $[2(1,1)]_{\O} \oplus 4 (0,1)$
\item $[2(1,1) \oplus (0,2)]_{\O} \oplus 2 (0,1)$
\item $[(1,1) \oplus 2(0,2)]_{\O} \oplus (1,0) \oplus (0,1)$
\item $[(1,1) \oplus (0,2)]_{\O} \oplus (1,0) \oplus 3(0,1)$
\item $(1,1)_{\O} \oplus (1,0) \oplus 5(0,1)$
\item $[3(0,2)]_{\O} \oplus 2 (1,0)$
\item $[2(0,2)]_{\O} \oplus 2 (1,0) \oplus 2(0,1)$
\item $(0,2)_{\O} \oplus 2(1,0) \oplus 4(0,1)$
\end{itemize}

We still have to work out a few of the blocks which are not
elementary, namely $[2(0,2)]_{\SO}$, $[3(0,2)]_{\O}$, $[2(1,1) \oplus
2(0,2)]_{\SO}$, $[2(1,1) \oplus (0,2)]_{\O}$, $[(1,1) \oplus
2(0,2)]_{\O}$ and $[(1,1) \oplus (0,2)]_{\O}$.  We will simply state
the results, which are easily verified as was done in previous cases
already treated in detail:
\begin{itemize}
\item $[2(0,2)]_{\SO}$ is represented by $B^{(0,2)}(\varphi_1) \oplus
  B^{(0,2)}(\varphi_2)$, where $\varphi_1 \geq |\varphi_2| > 0$;
\item $[3(0,2)]_{\O}$ is represented by $B^{(0,2)}(\varphi_1) \oplus
  B^{(0,2)}(\varphi_2) \oplus B^{(0,2)}(\varphi_3)$, where $\varphi_1
  \geq \varphi_2 \geq \varphi_3 > 0$;
\item $[2(1,1) \oplus 2(0,2)]_{\SO}$ is represented by
  $B^{(1,1)}(\beta_1) \oplus B^{(1,1)}(\beta_2) \oplus
  B^{(0,2)}(\varphi_1) \oplus B^{(0,2)}(\varphi_2)$, where $\beta_1
  \geq \beta_2 > 0$ and $\varphi_1 \geq |\varphi_2| > 0$;
\item $[2(1,1) \oplus (0,2)]_{\O}$ is represented by
  $B^{(1,1)}(\beta_1) \oplus B^{(1,1)}(\beta_2) \oplus
  B^{(0,2)}(\varphi)$, where $\beta_1 \geq \beta_2 > 0$ and $\varphi >
  0$;
\item $[(1,1) \oplus 2(0,2)]_{\O}$ is represented by
  $B^{(1,1)}(\beta) \oplus B^{(0,2)}(\varphi_1) \oplus 
  B^{(0,2)}(\varphi_2)$, where $\beta > 0$ and $\varphi_1 \geq
  \varphi_2 > 0$; and
\item $[(1,1) \oplus (0,2)]_{\O}$ is represented by
  $B^{(1,1)}(\beta) \oplus B^{(0,2)}(\varphi)$, where $\beta > 0$ and
  $\varphi > 0$.
\end{itemize}

We are now able to finally list all the possible elements of
$\fso(2,6)$ up to the action of $\SO(2,6)$:
\begin{enumerate}
\item $B^{(2,0)}(\varphi_1>0) \oplus 6 B^{(0,1)}$
\item $B^{(1,1)}(\beta>0) \oplus 5 B^{(0,1)} \oplus B^{(1,0)}$
\item $B^{(2,1)} \oplus 5 B^{(0,1)}$
\item $B^{(0,2)}(\varphi>0) \oplus 2 B^{(1,0)} \oplus 4 B^{(0,1)}$
\item $B^{(1,2)} \oplus B^{(1,0)} \oplus 4 B^{(0,1)}$
\item $B^{(2,2)}_+ \oplus 4 B^{(0,1)}$
\item $B^{(2,2)}_+(\beta>0) \oplus 4 B^{(0,1)}$
\item $B^{(2,2)}_+(\varphi\neq0) \oplus 4 B^{(0,1)}$
\item $B^{(2,2)}_+(\beta>0,\varphi>0) \oplus 4 B^{(0,1)}$
\item $B^{(2,0)}(\varphi_1>0) \oplus B^{(0,2)}(\varphi_2>0) \oplus 4
  B^{(0,1)}$
\item $B^{(1,1)}(\beta_1) \oplus B^{(1,1)}(\beta_2) \oplus 4
  B^{(0,1)}$ with $\beta_1 \geq \beta_2 > 0$
\item $B^{(1,1)}(\beta>0) \oplus B^{(0,2)}(\varphi>0) \oplus 3
  B^{(0,1)}  \oplus B^{(1,0)}$
\item $B^{(2,3)} \oplus 3 B^{0,1}$
\item $B^{(2,1)} \oplus B^{(0,2)}(\varphi>0) \oplus 3 B^{(0,1)}$
\item $B^{(1,2)} \oplus B^{(1,1)}(\beta>0) \oplus 3 B^{(0,1)}$
\item $B^{(0,2)}(\varphi_1) \oplus B^{(0,2)}(\varphi_2) \oplus  2
  B^{(0,1)} \oplus 2 B^{(1,0)}$, where $\varphi_1 \geq \varphi_2 >  0$
\item $B^{(1,2)} \oplus B^{(0,2)}(\varphi>0) \oplus B^{(1,0)} \oplus 2
  B^{(0,1)}$
\item $B^{(2,4)}_+(\varphi\neq0) \oplus 2B^{(0,1)}$
\item $B^{(2,2)}_+(\beta>0) \oplus B^{(0,2)}(\varphi\neq0) \oplus
  2B^{(0,1)}$
\item $B^{(2,2)}_+(\varphi_1\neq0) \oplus B^{(0,2)}(\varphi_2\neq0)
  \oplus  2B^{(0,1)}$
\item $B^{(2,2)}_+ \oplus B^{(0,2)}(\varphi\neq0) \oplus   2B^{(0,1)}$
\item $B^{(2,2)}_+(\beta>0,\varphi_1>0) \oplus B^{(0,2)}(\varphi_2\neq0)
  \oplus   2B^{(0,1)}$
\item $2B^{(1,2)} \oplus 2 B^{(0,1)}$
\item $B^{(2,0)}(\varphi_1>0) \oplus B^{(0,2)}(\varphi_2) \oplus
  B^{(0,2)}(\varphi_3) \oplus 2 B^{(0,1)}$ with $\varphi_2 \geq
  \varphi_3 > 0$
\item $B^{(1,1)}(\beta_1) \oplus B^{(1,1)}(\beta_2) \oplus
  B^{(0,2)}(\varphi>0) \oplus 2 B^{(0,1)}$, with $\beta_1 \geq \beta_2
  > 0$
\item $B^{(1,1)}(\beta>0) \oplus B^{(0,2)}(\varphi_1) \oplus
  B^{(0,2)}(\varphi_2) \oplus B^{(0,1)} \oplus B^{(1,0)}$, with
  $\varphi_1 \geq \varphi_2>0$
\item $B^{(2,3)} \oplus B^{(0,2)}(\varphi>0) \oplus B^{0,1}$
\item $B^{(2,1)} \oplus B^{(0,2)}(\varphi_1) \oplus
  B^{(0,2)}(\varphi_2) \oplus B^{(0,1)}$ with $\varphi_1 \geq
  \varphi_2 > 0$
\item $B^{(1,2)} \oplus B^{(1,1)}(\beta>0) \oplus B^{(0,2)}(\varphi>0)
  \oplus B^{(0,1)}$
\item $B^{(0,2)}(\varphi_1) \oplus B^{(0,2)}(\varphi_2) \oplus
  B^{(0,2)}(\varphi_3) \oplus 2 B^{(1,0)}$, where $\varphi_1 \geq
  \varphi_2 \geq \varphi_3 > 0$
\item $B^{(1,2)} \oplus B^{(0,2)}(\varphi_1) \oplus
  B^{(0,2)}(\varphi_2) \oplus B^{(1,0)}$, with $\varphi_1 \geq
  \varphi_2 > 0$
\item $B^{(2,4)}_+(\varphi_1\neq0) \oplus B^{(0,2)}(\varphi_2\neq0)$
\item $B^{(2,2)}_+ \oplus B^{(0,2)}(\varphi_1) \oplus
  B^{(0,2)}(\varphi_2)$ with $\varphi_1 \geq |\varphi_2| > 0$
\item $B^{(2,2)}_+(\beta>0) \oplus B^{(0,2)}(\varphi_1) \oplus
  B^{(0,2)}(\varphi_2)$ with $\varphi_1 \geq |\varphi_2| > 0$
\item $B^{(2,2)}_+(\varphi_1\neq0) \oplus B^{(0,2)}(\varphi_2) \oplus
  B^{(0,2)}(\varphi_3)$ with $\varphi_2 \geq |\varphi_3| > 0$
\item $B^{(2,2)}_+(\beta>0,\varphi_1>0) \oplus B^{(0,2)}(\varphi_2)
  \oplus  B^{(0,2)}(\varphi_3)$ with $\varphi_2 \geq |\varphi_3| > 0$
\item $2B^{(1,2)} \oplus B^{(0,2)}(\varphi>0)$
\item $B^{(2,0)}(\varphi_1\neq0) \oplus B^{(0,2)}(\varphi_2) \oplus
  B^{(0,2)}(\varphi_3) \oplus B^{(0,2)}(\varphi_4)$ with $\varphi_2
  \geq \varphi_3 \geq \varphi_4 > 0$
\item $B^{(1,1)}(\beta_1) \oplus B^{(1,1)}(\beta_2) \oplus
  B^{(0,2)}(\varphi_1) \oplus B^{(0,2)}(\varphi_2)$, with $\beta_1
  \geq \beta_2 > 0$ and $\varphi_1 \geq |\varphi_2|>0$
\end{enumerate}
where we have again reordered them in such a way that the embeddings
$\fso(2,2) \subset \fso(2,3) \subset \fso(2,4) \subset \fso(2,6)$ are
compatible with the labelling; in particular, cases (1)-(25)
correspond to the embedding in $\fso(2,6)$ of the corresponding
elements in $\fso(2,4)$.

Again we rewrite this list in a more traditional notation.  Let
$\{\be_1,\dots,\be_8\}$ be an ordered frame with $\be_1$ and $\be_2$
timelike and the rest spacelike.  Then, using
Table~\ref{tab:blocksSO}, the above elements can be written as
follows:
\begin{enumerate}
\item $\varphi \be_{12}$, $(\varphi>0$);
\item $\beta \be_{13}$, ($\beta>0$);
\item $\be_{12} - \be_{23}$;
\item $\varphi \be_{34}$, ($\varphi>0$);
\item $\be_{13} - \be_{34}$;
\item $- \be_{12} - \be_{13} + \be_{24} + \be_{34}$;
\item $- \be_{12} - \be_{13} + \be_{24} + \be_{34} + \beta  (\be_{14}
  - \be_{23})$, ($\beta>0$);
\item $- \be_{12} - \be_{13} + \be_{24} + \be_{34} + \varphi (
  \be_{12} + \be_{34} )$ ($\varphi\neq0$);
\item $\varphi ( \be_{12} - \be_{34} ) + \beta (\be_{14} - \be_{23})$,
  ($\beta>0$, $\varphi>0$);
\item $\varphi_1 \be_{12} + \varphi_2 \be_{34}$, ($\varphi_i>0$);
\item $\beta_1 \be_{13} + \beta_2 \be_{24}$, ($\beta_1 \geq \beta_2 >
  0$);
\item $\beta \be_{13} + \varphi \be_{45}$, ($\varphi > 0$, $\beta >  0$);
\item $\be_{12} + \be_{13} + \be_{15} - \be_{24} - \be_{34} - \be_{45}$;
\item $\be_{12} - \be_{23} + \varphi \be_{45}$, ($\varphi > 0$);
\item $\be_{13} - \be_{34} + \beta \be_{25}$, ($\beta>0$);
\item $\varphi_1 \be_{34} + \varphi_2 \be_{56}$, ($\varphi_1 \geq
  \varphi_2 > 0$);
\item $\be_{13} - \be_{34} + \varphi \be_{56}$, ($\varphi>0$);
\item $\varphi( -\be_{12} + \be_{34} + \be_{56} ) + \be_{15} -
  \be_{35} + \be_{26} - \be_{46}$, ($\varphi\neq 0$);
\item $- \be_{12} - \be_{13} + \be_{24} + \be_{34} + \beta  (\be_{14}
  - \be_{23}) + \varphi \be_{56}$, ($\beta>0$, $\varphi \neq  0$);
\item $- \be_{12} - \be_{13} + \be_{24} + \be_{34} + \varphi_1 (
  \be_{12} + \be_{34} ) + \varphi_2 \be_{56}$, $(\varphi_i\neq0$);
\item $- \be_{12} - \be_{13} + \be_{24} + \be_{34} + \varphi
  \be_{56}$, ($\varphi\neq 0$);
\item $\varphi_1 ( \be_{12} - \be_{34} ) + \beta ( \be_{14} - \be_{23} )
  + \varphi_2 \be_{56}$, ($\beta>0$, $\varphi_1>0$, $\varphi_2 \neq 0$);
\item $\be_{13} - \be_{34} + \be_{25} - \be_{56}$;
\item $\varphi_1 \be_{12} + \varphi_2 \be_{34} + \varphi_3 \be_{56}$,
  ($\varphi_1>0$, $\varphi_2 \geq   \varphi_3 > 0$);
\item $\beta_1 \be_{13} + \beta_2 \be_{24} + \varphi \be_{56}$,
  ($\varphi > 0$, $\beta_1 \geq \beta_2 > 0$);
\item $\beta \be_{13} + \varphi_1 \be_{56} + \varphi_2 \be_{78}$,
  ($\varphi_1 \geq \varphi_2 > 0$, $\beta > 0$);
\item $\be_{12} + \be_{13} + \be_{15} - \be_{24} - \be_{34} - \be_{45}
  + \varphi \be_{78}$, ($\varphi>0$);
\item $\be_{12} - \be_{23} + \varphi_1 \be_{45} + \varphi_2 \be_{67}$,
  ($\varphi_1 \geq \varphi_2 > 0$);
\item $\be_{13} - \be_{34} + \beta \be_{25} + \varphi \be_{67}$,
  ($\beta>0$, $\varphi>0$);
\item $\varphi_1 \be_{34} + \varphi_2 \be_{56} + \varphi_3 \be_{78}$,
  ($\varphi_1 \geq \varphi_2 \geq \varphi_3 > 0$);
\item $\be_{13} - \be_{34} + \varphi_1 \be_{56} + \varphi_2 \be_{78}$,
  ($\varphi_1 \geq \varphi_2 > 0$);
\item $\varphi_1 ( -\be_{12} + \be_{34} + \be_{56} ) + \be_{15} -
  \be_{35} + \be_{26} - \be_{46} + \varphi_2 \be_{78}$,
  ($\varphi_i\neq 0$);
\item $- \be_{12} - \be_{13} + \be_{24} + \be_{34} + \varphi_1
  \be_{56} + \varphi_2 \be_{78}$, ($\varphi_1 \geq |\varphi_2| > 0$);
\item $- \be_{12} - \be_{13} + \be_{24} + \be_{34} + \beta  (\be_{14}
  - \be_{23}) + \varphi_1 \be_{56} + \varphi_2 \be_{78}$,  ($\beta>0$,
  $\varphi_1 \geq |\varphi_2| > 0$);
\item $- \be_{12} - \be_{13} + \be_{24} + \be_{34} + \varphi_1 (
  \be_{12} + \be_{34} ) + \varphi_2 \be_{56} + \varphi_3 \be_{78}$,
  ($\varphi_1\neq 0$, $\varphi_2 \geq |\varphi_3| > 0$);
\item $\varphi_1 ( \be_{12} - \be_{34} ) + \beta ( \be_{14} -  \be_{23}
  ) + \varphi_2 \be_{56} + \varphi_3 \be_{78}$, ($\beta>0$,
  $\varphi_1>0$, $\varphi_2 \geq |\varphi_3| > 0$);
\item $\be_{13} - \be_{34} + \be_{25} - \be_{56} + \varphi \be_{78}$,
  ($\varphi>0$);
\item $\varphi_1 \be_{12} + \varphi_2 \be_{34} + \varphi_3 \be_{56} +
  \varphi_4 \be_{78}$, ($\varphi_1\neq0$, $\varphi_2 \geq \varphi_3 \geq
  \varphi_4 > 0$);
\item $\beta_1 \be_{13} + \beta_2 \be_{24} + \varphi_1 \be_{56} +
  \varphi_2 \be_{78}$, ($\varphi_1 \geq |\varphi_2| > 0$, $\beta_1
  \geq \beta_2 > 0$);
\end{enumerate}

Again it is clear that by letting some of the parameters become $0$,
we can subsume some of these cases into others; but we prefer not to
do this at this stage.  The above list is in one-to-one correspondence
with the conjugacy classes of nonzero elements of $\fso(2,6)$ under
$\SO(2,6)$.

\subsubsection{Causal properties of orbits}

Similar considerations as those explained in the previous section on
norms allow us to immediately write the norms of the first 25 Killing
vector fields with the proviso that $R^2$ becomes $R^2 + x_7^2 +
x_8^2$ with respect to the $\AdS_5$ norms.  All other cases involve no
new computations, just adding results of previous computations.  At
the end of the day (or night!), one obtains the following norms:
\begin{enumerate}
\item $|\xi|^2 = -\varphi^2 (R^2 + x_3^2 + x_4^2 + x_5^2 + x_6^2 +
  x_7^2 + x_8^2 )$, which is unbounded below;
\item $|\xi|^2 = \beta^2 ( R^2 - x_2^2 + x_4^2 + x_5^2 +x_6^2 + x_7^2
  + x_8^2)$, which is unbounded below;
\item $|\xi|^2 = - (x_3 + x_1)^2$, which is unbounded below;
\item $|\xi|^2 = \varphi^2 ( x_3^2 + x_4^2) \geq 0$;
\item $|\xi|^2 = (x_1 + x_4)^2 \geq 0$;
\item $|\xi|^2 = 0$;
\item $|\xi|^2 = \beta^2 (R^2+x_5^2+x_6^2 + x_7^2 + x_8^2) + 4\beta
  (x_1+x_4) (x_3 + x_2)$, which is unbounded below;
\item $|\xi|^2 = -\varphi^2 (R^2+x_5^2+x_6^2 + x_7^2 + x_8^2) + 2
  \varphi( (x_1+x_4)^2 + (x_3 + x_2)^2)$, which is unbounded below;
\item $|\xi|^2 = (\beta^2 - \varphi^2) (R^2+x_5^2+x_6^2+ x_7^2 +
  x_8^2) - 4\beta \varphi (x_1x_3 + x_2x_4)$, which is unbounded
  below;
\item $|\xi|^2 = -\varphi_2^2 (R^2+x_5^2+x_6^2+ x_7^2 + x_8^2) +
  (\varphi_1^2 - \varphi_2^2)(x_3^2 + x_4^2)$, which is unbounded
  below;
\item $|\xi|^2 = \beta_1^2 (R^2+x_5^2+x_6^2+ x_7^2 + x_8^2) +
  (\beta_2^2 - \beta_1^2)(x_2^2 - x_4^2)$, which is $\geq \beta_1^2
  R^2$ if $\beta_1 = \beta_2$ and unbounded below otherwise;
\item $|\xi|^2 = \beta^2 (x_1^2 - x_3^2) + \varphi^2 (x_4^2 + x_5^2)$,
  which is unbounded below;
\item $|\xi|^2 = (x_4 - x_1)^2 - 4 (x_2 + x_3) x_5$, which is
  unbounded below;
\item $|\xi|^2 = -(x_1+x_3)^2 + \varphi^2 (x_4^2 + x_5^2)$, which is
  unbounded below;
\item $|\xi|^2 = (x_1+x_4)^2 + \beta^2 (x_2^2 - x_5^2)$, which is
  unbounded below;
\item $|\xi|^2 = \varphi_1^2 (x_3^2 + x_4^2) + \varphi_2^2 (x_5^2 +
  x_6^2) \geq 0$;
\item $|\xi|^2 = (x_1+x_4)^2 + \varphi^2 (x_5^2 + x_6^2) \geq 0$;
\item $|\xi|^2 = -\varphi^2 (R^2 + x_7^2 + x_8^2) + (x_1-x_3)^2 + (x_4
  - x_2)^2 - 4\varphi ((x_4 - x_2) x_5 + (x_1 - x_3) x_6)$, which
  is unbounded below;
\item $|\xi|^2 = \beta^2 (R^2 + x_7^2 + x_8^2) + 4\beta
  (x_1+x_4)(x_2+x_3) + (\varphi^2 + \beta^2) (x_5^2 + x_6^2)$, which
  is unbounded below;
\item $|\xi|^2 = -\varphi_1^2 (R^2 + x_7^2 + x_8^2) + 2 \varphi_1
  ((x_1+x_4)^2 + (x_2 + x_3)^2)) + (\varphi_2^2 -\varphi_1^2)(x_5^2 +
  x_6^2)$, which is unbounded below;
\item $|\xi|^2 = \varphi^2 (x_5^2 + x_6^2) \geq 0$;
\item $|\xi|^2 = (\beta^2-\varphi_1^2) (R^2 + x_7^2 + x_8^2) +
  (\beta^2 - \varphi_1^2 + \varphi_2^2) (x_5^2 + x_6^2) - 4\beta
  \varphi_1 (x_1 x_3 + x_2 x_4)$, which is unbounded below;
\item $|\xi|^2 = (x_1 + x_4)^2 + (x_2 + x_6)^2 >0$;
\item $|\xi|^2 = -\varphi_1^2 (R^2 + x_7^2 + x_8^2) + (\varphi_2^2 -
  \varphi_1^2) (x_3^2 + x_4^2) + (\varphi_3^2 - \varphi_1^2) (x_5^2 +
  x_6^2)$, which is $\geq -\varphi_1^2 R^2$ provided that $\varphi_2
  \geq \varphi_3 \geq |\varphi_1|>0$;
\item $|\xi|^2 = \beta_1^2 (R^2 + x_7^2 + x_8^2) + (\beta_2^2 -
  \beta_1^2) (x_2^2 - x_4^2) + (\varphi^2 + \beta_1^2)(x_5^2 +
  x_6^2)$, which is $\geq \beta_1^2 R^2$ if $\beta_2 = \beta_1$ and
  unbounded below otherwise;
\item $|\xi|^2 = \beta^2 (R^2 - x_2^2 + x_4^2) + (\varphi_1^2 +
  \beta^2) (x_5^2 + x_6^2) + (\varphi_2^2 + \beta^2) (x_7^2 +
  x_8^2)$, which is unbounded below;
\item $|\xi|^2 = (x_4-x_1)^2 - 4 x_5(x_2 + x_3) + \varphi^2 (x_7^2 +
  x_8^2)$, which is unbounded below;
\item $|\xi|^2 = -(x_1+x_3)^2 + \varphi_1 (x_4^2 + x_5^2) + \varphi_2
  (x_6^2 + x_7^2)$, which is unbounded below;
\item $|\xi|^2 = (x_1+x_4)^2 + \beta^2 (x_2^2 - x_5^2) + \varphi^2
  (x_6^2 + x_7^2)$, which is unbounded below;
\item $|\xi|^2 = \varphi_1^2 (x_3^2 + x_4^2) + \varphi_2^2 (x_5^2 +
  x_6^2) + \varphi_3^2 (x_7^2 + x_8^2) \geq 0$;
\item $|\xi|^2 = (x_1+x_4)^2 + \varphi_1^2 (x_5^2 + x_6^2) +
  \varphi_2^2 (x_7^2 + x_8^2) \geq 0$;
\item $|\xi|^2 = -\varphi_1^2 (R^2 + x_7^2 + x_8^2) + (x_1-x_3)^2 + (x_4
  - x_2)^2 - 4\varphi_1 ((x_4 - x_2) x_5 + (x_1 - x_3) x_6) +
  \varphi_2^2 (x_7^8 + x_8^2)$, which is unbounded below;
\item $|\xi|^2 = \varphi_1^2 (x_5^2 + x_6^2) + \varphi_2^2 (x_7^2 +
  x_8^2) \geq 0 $;
\item $|\xi|^2 = \beta^2 R^2 + 4\beta (x_1 + x_4)(x_2 + x_3) +
  (\varphi_1^2 + \beta^2) (x_5^2 + x_6^2) + (\varphi_2^2 + \beta^2)
  (x_7^2 + x_8^2)$, which is unbounded below;
\item $|\xi|^2 = -\varphi_1^2 R^2 + 2 \varphi_1 ((x_1+x_4)^2 + (x_2 +
  x_3)^2) + (\varphi_2^2 - \varphi_1^2)(x_5^2 + x_6^2) + (\varphi_3^2
  - \varphi_1^2)(x_7^2 + x_8^2)$, which is $\geq -\varphi_1^2 R^2$
  provided that $\varphi_2 \geq |\varphi_3| \geq \varphi_1 > 0$ and
  unbounded below otherwise;
\item $|\xi|^2 = (\beta^2 - \varphi_1^2) R^2 + (\beta^2 + \varphi_2^2
  - \varphi_1^2) (x_5^2 + x_6^2) + (\beta^2 + \varphi_3^2 -
  \varphi_1^2) (x_7^2 + x_8^2) - 4\beta\varphi_1 (x_1 x_3 + x_2
  x_4)$, which is unbounded below;
\item $|\xi|^2 = (x_1+x_4)^2 + (x_2 + x_6)^2 + \varphi^2 (x_7^2 +
  x_8^2) > 0$;
\item $|\xi|^2 = -\varphi_1^2 R^2 + (\varphi_2^2 - \varphi_1^2)(x_3^2
  + x_4^2) + (\varphi_3^2 - \varphi_1^2)(x_5^2 + x_6^2) + (\varphi_4^2
  - \varphi_1^2)(x_7^2 + x_8^2)$, which is $\geq -\varphi^2 R^2$ for
  $\varphi_2 \geq \varphi_3 \geq \varphi_4 \geq |\varphi_1| > 0$; and
\item $|\xi|^2 = \beta_1^2 R^2 + (\beta_2^2 - \beta_1^2)(x_2^2 -
  x_4^2) + (\varphi_1^2 + \beta_1^2)(x_5^2 + x_6^2) + (\varphi_2^2 +
  \beta_1^2)(x_7^2 + x_8^2)$, which is $\geq \beta_1^2 R^2$ provided
  that $\beta_2 = \beta_1$ and unbounded below otherwise.
\end{enumerate}

In this case, since we will be applying these results to the
reductions of $\AdS_7 \times S^4$ and the hair on $S^4$ cannot be
combed, we will only be needing those Killing vectors on $\AdS_7$
whose norms are positive. From the above list we see immediately 
that these are (11), (25) and (39) all three with $\beta_1 = \beta_2$, 
(23) and (37).

\section{Supersymmetry}
\label{sec:susy}

In this section we discuss the conditions under which a particular
reduction of a Freund--Rubin background of the form $\AdS_{p+1} \times
S^q$ preserves any of the (maximal) supersymmetry.  This is a subtle
issue, for which we have to distinguish between M-theory and
supergravity.  Supersymmetry in the supergravity limit is realised
geometrically in terms of Killing spinors.  In M-theory this cannot be
the full story.  In fact, as illustrated in
\cite{DLP,DuffLuPopeUnt,PSS} and lucidly explained more recently in
\cite{HullHolonomy}, a background such as the Freund--Rubin vacuum
$\AdS_5 \times S^5$ in type IIB string theory, whose supersymmetry is
fully realised geometrically, is T-dual (and hence equivalent) to a
type IIA string background $\AdS_5 \times \CP^2 \times S^1$, which does
not even have a spin structure.

At the supergravity level, T-duality defines a correspondence between
certain backgrounds of type IIA and type IIB supergravities: two
backgrounds being T-dual if their Kaluza--Klein reductions to $d{=}9$
$N{=}2$ supergravity coincide:
\begin{equation*}
  \xymatrix{d{=}10~\text{IIB} \ar[dr]_{\text{KK}}
    \ar@{<.>}[rr]^{\text{T-duality}} &
    & d{=}10~\text{IIA} \ar[dl]^{\text{KK}}\\
    & d{=}9~N{=}2~;}
\end{equation*}
for instance,
\begin{equation*}
  \xymatrix{\AdS_5 \times S^5 \ar[dr] \ar@{<.>}[rr] &
    & \AdS_5 \times \CP^2 \times S^1 \ar[dl]\\
    & \AdS_5 \times \CP^2 & }
\end{equation*}
In string or M-theory we must keep all the Kaluza--Klein modes, and
not just the zero modes.  Supersymmetries which would
appear broken from the supergravity point of view due to the
non-invariance of the Killing spinors, manifest themselves among the
nonzero Kaluza--Klein modes.  Under T-duality these in turn become
winding modes of the dual background.  Indeed, the spinors in $\AdS_5
\times S^5$ transform nontrivially under the circle subgroup along
which the T-duality is being performed and hence the supersymmetry in
the T-dual picture is manifested in winding modes.

From the M-theory point of view the existence of a spin structure in
the quotient is therefore somewhat of a red herring, since the
Killing spinors of the original background will descend in some
fashion to the Kaluza--Klein reduction, even if not necessarily as
spinors.  From a supergravity perspective, however, it is only those
Killing spinors which are invariant which do descend to Killing
spinors of the reduced background and as we will see a necessary
condition for the existence of invariant Killing spinors is the
existence of a spin structure in the quotient.

Our primary focus in this paper being the supergravity backgrounds, we
will take the point of view that when we talk about the supersymmetry
preserved in a background, we will mean the geometrically realised
supersymmetry of the supergravity background and hence it will be that
problem which we will study.  This section is thus divided into two
parts.  In the first part we analyse, for the geometries of interest,
the problem of whether a quotient of a spin manifold is again spin.
We will arrive at a criterion which can be tested by a simple
calculation in a Clifford algebra.  In the second part we analyse the
problem of whether the background (before reduction) allows invariant
Killing spinors.

\subsection{Spin structures}
\label{sec:spin}

Given a Freund--Rubin background $M$, let $\Gamma$ be the
one-parameter subgroup of isometries by which we are reducing.  When
is $M/\Gamma$ spin?

We will generalise and consider $(M^{1+n},g)$ a simply-connected
lorentzian spin manifold and $\Gamma$ a one-parameter subgroup of
spacelike isometries, and ask under what conditions the quotient
(assumed smooth) is spin.

Let $P_{\SO}(M) \to M$ denote the bundle of oriented orthonormal
frames on $M$.  It is a principal bundle with structure group
$\SO(1,n)$.  Since $M$ is spin and simply-connected, it has a unique
spin structure $\theta: P_{\Spin}(M) \to P_{\SO}(M)$.  The bundle
$P_{\Spin}(M) \to M$ is a principal bundle with group $\Spin(1,n)$ and
the bundle map $\theta$ above is a double cover which agrees on each
fibre with the standard double-cover $\theta_0 : \Spin(1,n) \to
\SO(1,n)$.

Let $\xi$ denote the Killing vector which generates the action of
$\Gamma$.  By assumption, it is everywhere spacelike, with norm
$|\xi|>0$.  Let $P \subset P_{\SO}(M)$ be the sub-bundle consisting
of those frames which have $\xi/|\xi|$ as the first vector.  This is
again a principal bundle, but now with structure group $\SO(1,n-1)$,
this being isomorphic to the isotropy subgroup of the spacelike vector
$\xi/|\xi|$ at any given point.

The action of $\Gamma$ on $M$ induces an action on $P_{\SO}(M)$, an
element $\gamma \in \Gamma$ taking an oriented orthonormal frame at
the point $x \in M$ to an oriented orthonormal frame at the point
$\gamma \cdot x \in M$.  If the frame at $x$ is in $P$, so that it has
the normalised Killing vector as the first vector, then so will be the
transformed frame.  This is because the Killing vector $\xi$ is
$\Gamma$-invariant.  The action of $\Gamma$ on $P_{\SO}(M)$ commutes
with the natural action of $\SO(1,n)$, whence we have a free action of
$\SO(1,n) \times \Gamma$ on $P_{\SO}(M)$.  Restricting to the
sub-bundle $P$ we have a free action of $\SO(1,n-1) \times \Gamma$.
We can therefore take the quotient $P/\Gamma$ to obtain a principal
$\SO(1,n-1)$-bundle over $M/\Gamma$.  Since oriented orthonormal
frames of $M/\Gamma$ are in one-to-one correspondence with oriented
orthonormal frames on $M$ with $\xi/|\xi|$ as first vector, we see
that indeed $P/\Gamma \to M/\Gamma$ is the bundle of oriented
orthonormal frames on $M/\Gamma$.

Now consider the bundle $\theta^{-1}P \to M$, which is a sub-bundle of
$P_{\Spin}(M)$ with structure group $\Spin(1,n-1) = \theta_0^{-1}
\SO(1,n-1)$.  If the action of $\Gamma$ on $P_{\SO}(M)$ lifts to
$P_{\Spin}(M)$ in such a way that $\theta$ is $\Gamma$-equivariant,
then it preserves $\theta^{-1}P$ (by equivariance) and the quotient
$\theta^{-1}P/\Gamma$ is a principal $\Spin(1,n-1)$ bundle over
$M/\Gamma$---indeed, it is the spin lift of $P/\Gamma$.  In other
words, it is a spin bundle on $M/\Gamma$.  Conversely, any spin bundle
$P_{\Spin}(M/\Gamma)$ on $M/\Gamma$ pulls back via the projection $\pi
: M \to M/\Gamma$ to a spin bundle
\begin{equation*}
  \pi^* P_{\Spin}(M/\Gamma) \times_{\Spin(1,n-1)} \Spin(1,n) \to M
\end{equation*}
on $M$ after enlarging the structure group to $\Spin(1,n)$, with an
equivariant $\Gamma$ action.

In other words, $M/\Gamma$ is spin if and only if the action of
$\Gamma$ lifts to an action on the spin bundle $P_{\Spin}(M)$.
If it does, the spin bundle on $M$ is called \emph{projectable}, a
term introduced in \cite{MoroianuThesis} and discussed in more detail
in \cite{AmmannBaer}.

There are two possible topologies for the one-parameter group
$\Gamma$: either $\RR$ or $S^1$.  In the former case, $\Gamma$ is
simply-connected and one obtains an action of $\Gamma$ on
$P_{\Spin}(M)$ simply by integrating the infinitesimal action.  In
this case the spin bundle is always projectable.  However when
$\Gamma \cong S^1$ it may be that the infinitesimal action only lifts
to an action of a double cover of $\Gamma$.  Therefore the question of
$M/\Gamma$ being spin only arises for reductions by circle subgroups.

If the spin bundle $P_{\Spin}(M)$ is not projectable, then the action
of $\Gamma$ ($\cong S^1$) on the frame bundle does not lift.  Instead
the spin bundle admits an action of a double cover $\widetilde\Gamma
\cong S^1$ of $\Gamma$:
\begin{equation*}
  \begin{CD}
    1 @>>> \ZZ_2 @>>> \widetilde\Gamma @>\sigma>> \Gamma @>>> 1~,
  \end{CD}
\end{equation*}
defined so that if $\tilde p \in P_{\Spin}(M)$, then we have
\begin{equation}
  \label{eq:liftedaction}
  \theta(\tilde\gamma \cdot \tilde p) = \sigma(\tilde \gamma) \cdot
  \theta(\tilde p)
\end{equation}
for every $\tilde\gamma \in \widetilde\Gamma$.  This shows that the
action is free, since so is the action of $\Gamma$ on $P_{\SO}(M)$.

Let $P \subset P_{\SO}(M)$ be as above and consider the
$\Spin(1,n-1)$-bundle $\theta^{-1}P$.  Since $\Gamma$ preserves $P$,
it follows from \eqref{eq:liftedaction} that $\widetilde\Gamma$
preserves $\theta^{-1} P$.

Therefore $\theta^{-1}P$ admits a free action of $\Spin(1,n-1)$ on the
right and a commuting free action of $\widetilde\Gamma$ on the left.
This action is not effective, however, and there is a normal $\ZZ_2$
subgroup which acts trivially.  To see this, notice that this subgroup
must belong to the kernel of the homomorphism
\begin{equation*}
  \begin{CD}
    \Spin(1,n-1) \times \widetilde\Gamma @>\theta_0 \times \sigma>>
    \SO(1,n-1) \times \Gamma~,
  \end{CD}
\end{equation*}
which is isomorphic to $\ZZ_2 \times \ZZ_2$.  Indeed, if $\tilde\gamma
\in \widetilde\Gamma$, $\tilde p \in \theta^{-1} P$ and $\tilde g \in
\Spin(1,n-1)$, then
\begin{equation*}
  \theta(\tilde\gamma \cdot \tilde p \cdot \tilde g) =
  \sigma(\tilde\gamma) \cdot \theta(\tilde p) \cdot \theta_0(\tilde
  g)~;
\end{equation*}
whence if $\tilde \gamma\cdot \tilde p \cdot \tilde g = \tilde p$,
then $\sigma(\tilde\gamma) \cdot \theta(\tilde p) \cdot
\theta_0(\tilde g) = \theta(\tilde p)$, which implies that
$\sigma(\tilde \gamma) = 1$ and $\theta_0(\tilde g) = 1$, as
$\SO(1,n-1) \times \Gamma$ acts freely on $P$.  Conversely, if
$(\tilde g, \tilde \gamma)$ belongs to the kernel of $\theta_0 \times
\sigma$ then $\theta(\tilde \gamma \cdot \tilde p \cdot \tilde g) =
\theta(\tilde p)$, whence $\tilde \gamma \cdot \tilde p \cdot \tilde
g$ and $\tilde p$ are in the same fibre of $\theta^{-1} P \to P$.
Since $\theta$ is a double cover, this fibre consists of two points.
Clearly $\ZZ_2 \times \ZZ_2$, a group of order $4$, cannot act
effectively in a set with two points.  The kernel of this action
cannot contain $(-1,1)$ or $(1,-1)$ since this would imply that either
$\Spin(1,n-1)$ or $\widetilde \Gamma$ did not act effectively, which
they do.  On the other hand, being a two-point set, the actions of
both $(-1,1)$ and $(1,-1)$ on any point $\tilde p$ must yield the
other point $\tilde p'$ in the same fibre as $\tilde p$; whence their
product $(-1,-1)$ preserves all points $\tilde p$.  In summary,
$\theta^{-1} P$ admits a free effective action of the group
\begin{equation*}
  \Spin(1,n-1)^c := \Spin(1,n-1)\times_{\ZZ_2} \widetilde\Gamma
\end{equation*}
obtained by quotienting $\Spin(1,n-1)\times \widetilde\Gamma$ by the
$\ZZ_2$ subgroup generated by $(-1,-1)$.

In summary, the bundle $\theta^{-1} P$ can be viewed as a
$\Spin(1,n-1)^c$-bundle over $M/\Gamma$; that is, it yields a
spin$^c$-structure in the quotient.

We shall not attempt here to give general conditions for a spin bundle
on a manifold $M$ to yield a spin or a spin$^c$ structure in the
quotient by a circle action.  Instead we shall concentrate on the
geometries of interest, namely a product of space forms.  Let us start
with the spheres.

\subsubsection{Spin structures on sphere quotients}

The $q$-sphere $S^q \cong \SO(q+1)/\SO(q)$ is a homogeneous space of
$\SO(q+1)$ and (the total space of) its oriented orthonormal frame
bundle can be identified with $\SO(q+1)$ itself.  Let $\vartheta:
\Spin(q+1) \to \SO(q+1)$ be the two-fold covering map.  It is a bundle
map over $S^{q+1}$ and either composing these two maps or noticing
that $S^q \cong \Spin(q+1)/\Spin(q)$ is also a homogeneous space for
the Spin group, we see that $\Spin(q+1)$ can be identified with the
total space of the spin bundle over $S^q$.

The usefulness of this description is that the action of isometries on
the frame bundle is very natural; namely, the action of $\SO(q+1)$ on
$S^q$ is induced from left multiplication on $\SO(q+1)$.  This action
is well defined on $S^q = \SO(q+1)/\SO(q)$ because left and right
multiplications commute.  Therefore the action of a subgroup $\Gamma
\subset \SO(q+1)$ on $S^q$ is induced by left multiplication by
$\Gamma$ itself in $\SO(q+1)$.  A lift of this action to the spin
bundle is then simply a subgroup $\widehat\Gamma \subset \Spin(q+1)$
acting on $\Spin(q+1)$ by left multiplication and projecting
\emph{isomorphically} to $\Gamma$ via $\vartheta$.  In summary, spin
structures over $S^q/\Gamma$ are in one-to-one correspondence with
subgroups $\widehat\Gamma \subset \Spin(q+1)$ such that $\vartheta
(\widehat\Gamma) = \Gamma$ is an isomorphism.  The one-parameter
groups $\Gamma$ we will consider are obtained by applying the
exponential map to a given line $\RR X \subset \fso(q+1)$.  By
embedding the Lie algebra $\fso(q+1) \subset \Cl(q+1)$ in the Clifford
algebra and exponentiating $t X$ there, we obtain a subgroup
$\widehat\Gamma \subset \Spin(q+1)$.  It is then a simple matter to
check whether or not $\vartheta: \widehat\Gamma \to \Gamma$ is an
isomorphism.  By construction it is a local diffeomorphism, so the
question is whether $\widehat\Gamma$ covers $\Gamma$ precisely once.

Let us illustrate this for the complex projective spaces.  Let
$S^{2n-1} \subset \CC^n$ be the unit sphere and let $\Gamma \subset
\SO(2n)$ be the circle subgroup acting diagonally on $\CC^n$ via
\begin{equation*}
  (z_1,\dots,z_n) \mapsto (e^{it} z_1, \dots, e^{it} z_n)~.
\end{equation*}
This action is clearly free and the resulting quotient is the complex
projective space $\CP^{n-1}$.  We would like to know whether there
exists a subgroup $\widehat\Gamma \subset \Spin(2n)$ isomorphic to
$\Gamma$ and such that $\vartheta(\widehat\Gamma) = \Gamma$.  We will
answer this by working in the Clifford algebra $\Cl(2n)$.  The
subgroup $\Gamma$ is generated infinitesimally by
\begin{equation*}
  X = \Gamma_{12} + \dots + \Gamma_{2n-1,2n} \in \fso(2n) \subset \Cl(2n)~.
\end{equation*}
Exponentiating the line containing $X$ in $\Cl(2n)$ we arrive at a
subgroup $\widehat\Gamma \subset \Spin(2n) \subset \Cl(2n)$
consisting of elements
\begin{equation*}
  \widehat{g}(t) := \exp(t/2 X) = (\id \cos \tfrac{t}2 + \Gamma_{12}
  \sin\tfrac{t}2) \cdots (\id \cos \tfrac{t}2 + \Gamma_{2n-1,2n}
  \sin\tfrac{t}2)~,
\end{equation*}
which clearly projects to $\Gamma \subset \SO(2n)$: $\widehat{g}(t)
\mapsto g(t)$, where
\begin{equation*}
  g(t) = R_{12}(t) \cdots R_{2n-1,2n}(t)~,
\end{equation*}
where $R_{ij}(t)$ is the rotation by an angle $t$ in the $(ij)$-plane.
The question is whether the map $\widehat{g}(t) \mapsto g(t)$ is an
isomorphism.  Locally it is clearly an isomorphism, but the question
is whether it is one-to-one or two-to-one.  Equivalently, since the
pre-image of the identity in $\SO(2n)$ consists of $\pm \id$, the
question is whether the circle $\widehat{g}(t)$ passes or not
through the point $-\id$.  Clearly this may only happen at $t=2\pi$
where
\begin{equation*}
  \widehat{g}(2\pi) = (-1)^n \id~,
\end{equation*}
whence $\widehat\Gamma$ contains $-\id$ if and only if $n$ is odd.
Therefore if $n$ is even, $\CP^{n-1}$ admits a spin structure, whereas
if $n$ is odd, it only admits a spin$^c$ structure.

\subsubsection{Spin structures on quotients of $\AdS$}
\label{sec:sotilde}

Now we turn our attention to the $\AdS$ spaces.  The situation here is
complicated by the fact that the isometry group of $\AdS$ space is not
a matrix group; that is, it has no faithful finite-dimensional linear
representations.  Recall that in this paper $\AdS$ is
simply-connected, whereas the quadric $Q_{p+1} \subset \RR^{2,p}$ is a
quotient by the action of the fundamental group, which is isomorphic
to $\ZZ$ for $p>2$ and to $\ZZ \oplus \ZZ$ for $p=2$.  The quadric
$Q_{p+1}$ is a homogeneous space for the group $\SO(2,p)$, whereas
$\AdS_{p+1}$ is a homogeneous space for the group
$\widetilde{\SO(2,p)}$ which is the central extension of $\SO(2,p)$ by
the fundamental group of the quadric.  Let $\tau \in
\widetilde{\SO(2,p)}$ denote the generator of the fundamental group of
$Q_{p+1}$, for $p>2$ and let $C_p = \left<\tau\right> \cong \ZZ$ be
the infinite cyclic subgroup of $\widetilde{\SO(2,p)}$ generated by
$\tau$, for $p>2$.  For $p=2$ we have two generators $\tau_i \in
\widetilde{\SO(2,2)}$ for $i=1,2$ and we let $C_2 =
\left<\tau_1,\tau_2\right> \cong \ZZ \oplus \ZZ$ denote the abelian
subgroup generated by $\tau_1,\tau_2$.  Then we have the following
exact sequence of groups
\begin{equation*}
  \begin{CD}
    0 @>>> C_p @>>> \widetilde{\SO(2,p)} @>\pi>> \SO(2,p) @>>> 1~,
  \end{CD}
\end{equation*}
making $\widetilde{\SO(2,p)}$ into a central extension by $C_p$ of
$\SO(2,p)$.  The above sequence does not split; that is, there is no
subgroup $\SO(2,p)$ of $\widetilde{\SO(2,p)}$ which is a section of
$\pi$.  Therefore $\widetilde{\SO(2,p)}$ is a nontrivial central
extension and is thus characterised by a group 2-cocycle $\gamma$ of
$\SO(2,p)$ with values in $C_p$.  Group elements in
$\widetilde{\SO(2,p)}$ are given by pairs $(c,g)$ where $c\in C_p$ and
$g \in \SO(2,p)$ with product
\begin{equation*}
  (c_1,g_1) (c_2,g_2) = (c_1 + c_2 + \gamma(g_1,g_2), g_1 g_2)~,
\end{equation*}
where the cocycle $\gamma$ is normalised so that $\gamma(e,g) =
\gamma(g,e) = 0$ for $e \in \SO(2,p)$ the identity element.
Associativity of the above multiplication is equivalent to the cocycle
condition for $\gamma$.  It should be possible to derive an explicit
formula for this cocycle in local coordinates adapted to an Iwasawa
decomposition of $\SO(2,p)$, but we do not believe it to be
particularly useful for our purposes.

The isotropy of a point of the quadric $Q_{p+1}$ is isomorphic to
$\SO(1,p)$ and, as we will now show, the same is true for $\AdS_{p+1}$.
Let $x\in\AdS_{p+1}$ be a point and let $G_x$ denote the corresponding
isotropy subgroup of $\widetilde{\SO(2,p)}$.  Now let $\bar x \in
Q_{p+1}$ denote the corresponding point on the quadric and $G_{\bar x}
\subset \SO(2,p)$ its isotropy group.  The projection $\pi:
\widetilde{\SO(2,p)} \to \SO(2,p)$ restricts to a covering map
$\pi: G_x \to G_{\bar x}$.  We wish to show that this map is actually
one-to-one.  To see this notice that if $g_1,g_2 \in G_x$ project to
the same element of $G_{\bar x}$, then $g_1 g_2^{-1}$ belongs to the
kernel of $\pi$, which is the central subgroup $C_p$.  However no
nontrivial element of $C_p$ fixes the point $x$: the fundamental group
acts via deck transformations which act freely and properly
discontinuous.  Therefore $g_1 g_2^{-1} = e$ and hence $g_1 = g_2$.

This means that we have two principal $\SO(1,p)$ fibrations, shown by
the vertical arrows of the following commutative diagram:
\begin{equation*}
  \begin{CD}
    @. \\
    \widetilde{\SO(2,p)} @>{\pi}>>  \SO(2,p)\\
    @VVV          @VVV \\
    \AdS_{p+1} @>>> Q_{p+1}\\
    @.
  \end{CD}
\end{equation*}
Indeed, as in the case of the sphere, these fibrations define the
oriented orthonormal frame bundles of $\AdS_{p+1}$ and $Q_{p+1}$,
respectively.  Any subgroup $\Gamma \subset \widetilde{\SO(2,p)}$
acting on $\AdS_{p+1}$ acts naturally on the frame bundle by left
multiplication in the group $\widetilde{\SO(2,p)}$.

As in the case of the sphere, the spin bundles are similarly described
by composing the above fibrations with the spin double covers
$\widetilde\vartheta: \widetilde{\Spin(2,p)} \to \widetilde{\SO(2,p)}$
and $\vartheta: \Spin(2,p) \to \SO(2,p)$, yielding the following
commutative diagram
\begin{equation*}
  \begin{CD}
    @. \\
    \widetilde{\Spin(2,p)}  @>>> \Spin(2,p)\\
    @V{\widetilde\vartheta}VV  @VV{\vartheta}V \\
    \widetilde{\SO(2,p)} @>{\pi}>>  \SO(2,p)\\
    @VVV  @VVV \\
    \AdS_{p+1} @>>> Q_{p+1}\\
    @.
  \end{CD}
\end{equation*}

The action of a subgroup $\Gamma \subset \widetilde{\SO(2,p)}$ on
$\AdS_{p+1}$ lifts to the spin bundle if and only if there is a
subgroup $\widehat\Gamma \subset \widetilde{\Spin(2,p)}$ such that
$\widetilde\vartheta$ maps $\widehat\Gamma$ isomorphically to
$\Gamma$.  The difficulty in testing for the existence of such a
subgroup is that, unlike $\Spin(2,p)$, the spin group
$\widetilde{\Spin(2,p)}$ admits no finite-dimensional faithful
representations, whence in particular it cannot be embedded in, say, a
Clifford algebra.  We will show, however, that the subgroups of
interest in this paper are actually subgroups of $\Spin(2,p)$ and
hence we can (and will) work with the Clifford algebra.

\subsubsection{Spin structures on quotients of Freund--Rubin
  backgrounds}

Finally let us consider the product $\AdS_{p+1} \times S^q$.  The
oriented orthonormal frame bundle of this product, which is a
principal $\SO(1,q+p)$ bundle, admits a reduction to an $\SO(1,p)
\times \SO(q)$ bundle, whose total space is the Lie group
$\widetilde{\SO(2,p)} \times \SO(q+1)$.  Similarly the spin bundle
admits a reduction to the subgroup of $\Spin(1,p+q)$ obtained as the
image of $\Spin(1,p) \times \Spin(q)$ under the canonical lift
(denoted by the dotted line)
\begin{equation}
  \label{eq:lift}
  \xymatrix{ & & \Spin(1,p+q) \ar[d] \\
    \Spin(1,p) \times \Spin(q) \ar@{.>}[urr] \ar[r]
    & \SO(1,p) \times \SO(q) \ar[r] &  \SO(1,p+q) }
\end{equation}
of the map obtained by composing the natural projection to $\SO(1,p)
\times \SO(q)$ with the natural inclusion of this group into
$\SO(1,p+q)$.  The lift exists because $\Spin(1,p) \times \Spin(q)$ is
simply-connected (for the values of $p$ and $q$ that we are interested
in and restricting implicitly to the identity component) and hence any
map to a space---here $\SO(1,p+q)$---lifts to its universal cover.
The lift \eqref{eq:lift} is not an embedding however; but actually
factors through a $\ZZ_2$-quotient.  This can be seen more
transparently in the Clifford algebra.  We can embed $\Spin(1,p)$ and
$\Spin(q)$ into the Clifford algebras $\Cl(1,p)$ and $\Cl(q)$,
respectively.  The lift \eqref{eq:lift} is then induced by the natural
isomorphism $\Cl(1,p) \widehat\otimes \Cl(q) \cong \Cl(1,p+q)$, where
$\widehat\otimes$ is the $\ZZ_2$-graded tensor product.  (This just
means that the $\gamma$-matrices of $\Cl(1,p)$ anticommute with those
of $\Cl(q)$.)  Under this embedding we see that the element
$(-\id,-\id)$ in $\Spin(1,p) \times \Spin(q)$ gets mapped to the
identity in $\Spin(1,p+q)$.  Therefore the lift \eqref{eq:lift} factors
through the group
\begin{equation*}
  \Spin(1,p)\times_{\ZZ_2} \Spin(q)
\end{equation*}
where $\ZZ_2$ is the subgroup generated by $(-\id,-\id)$.

The total space of the reduction of the spin bundle to the above group
is then precisely the quotient of $\widetilde{\Spin(2,p)} \times
\Spin(q+1)$ by the $\ZZ_2$-subgroup of
$\Spin(1,p) \times \Spin(q)$ generated by $(-\id,-\id)$, now embedded
in $\widetilde{\Spin(2,p)} \times \Spin(q+1)$.

A subgroup $\Gamma \subset \widetilde{\SO(2,p)} \times \SO(q+1)$ acts
on the frame bundle by left multiplication and its action lifts to the
spin bundle if there is a subgroup $\widehat\Gamma \subset
\widetilde{\Spin(2,p)} \times_{\ZZ_2} \Spin(q+1)$ which projects
isomorphically to $\Gamma$ under the spin covering map
\begin{equation*}
  \begin{CD}
   \widetilde{\Spin(2,p)} \times_{\ZZ_2}
     \Spin(q+1)  @>>> \widetilde{\SO(2,p)} \times \SO(q+1)~,
  \end{CD}
\end{equation*}
which is a double cover.

We are interested in one-parameter subgroups $\Gamma$ and these are
obtained by exponentiating a one-dimensional subspace of the Lie
algebra.  There are two possible topologies for $\Gamma$: either the
real line or the circle.  If $\Gamma$ is diffeomorphic to the real
line, then the lift $\widehat\Gamma$ is also diffeomorphic to the real
line, since it has to be a connected cover of the real line, which is
simply-connected; hence $\widehat\Gamma \cong \Gamma$.  Only when
$\Gamma$ is diffeomorphic to the circle, can we have an obstruction.

Circle subgroups of $\widetilde{\SO(2,p)} \times \SO(q+1)$ are
contained in a maximal compact subgroup, which is isomorphic to
$\SO(p) \times \SO(q+1)$.  Moreover, the $\SO(p)$ component is
contained in the isotropy $\SO(1,p)$ of a point in $\AdS_{p+1}$, which
as we showed above projects isomorphically to the isotropy of a point
in $Q_{p+1}$.  In other words, the maximal compact subgroups $\SO(p)
\times \SO(q+1)$ are diffeomorphic to their image under the projection
$\pi$ and this means that we can effectively work in the quotient
group $\SO(2,p) \times \SO(q+1)$.  This simplifies the analysis
considerably because $\Spin(2,p) \times_{\ZZ_2} \Spin(q+1)$ can be
embedded in the Clifford algebra $\Cl(2,p+q+1)$.  A circle subgroup
$\Gamma \subset \SO(2,p) \times \SO(q+1)$ is obtained by
exponentiating a line $\RR X \subset \fso(2,p) \oplus \fso(q+1)$ in
the Lie algebra, and exponentiating this same line in the Clifford
algebra $\Cl(2,p+q+1)$ yields a circle subgroup $\widehat\Gamma
\subset \Spin(2,p) \times_{\ZZ_2} \Spin(q+1)$, which covers $\Gamma$.
We must determine if $\widehat\Gamma$ is isomorphic to $\Gamma$ or if,
on the contrary, it is a double cover.  This will only be the case if
$\widehat\Gamma$ contains the point other than the identity in
$\Spin(2,p) \times_{\ZZ_2} \Spin(q+1)$ which projects to the identity
in $\SO(2,p) \times \SO(q+1)$; equivalently, if $\widehat\Gamma
\subset \Cl(2,p+q+1)$ contains the point $-\id$.  This can then be
checked in the same way as was illustrated earlier for the complex
projective spaces.

\subsection{Killing spinors}
\label{sec:killing}

Let $S$ denote the representation of $\Spin(1,n)$ corresponding to the
spinors in the supergravity theory on $M$.  The spinor bundle
associated to this representation is defined to be
\begin{equation*}
  S(M) := P_{\Spin}(M) \times_{\Spin(1,n)} S~.
\end{equation*}
If the spin bundle is projectable, then $\Gamma$ acts naturally on
$S(M)$, since it acts on $P_{\Spin}(M)$ and commutes with the action
of $\Spin(1,n)$.  The supergravity Killing spinors on $M$ are sections
of $S(M)$ which are parallel with respect to a connection
\begin{equation*}
  D : S(M) \to T^*M \otimes S(M)
\end{equation*}
which is $\Gamma$-invariant because $D$ depends on the bosonic fields
of the background which are left invariant by the action of $\Gamma$.
This means that the space of Killing spinors is a representation of
$\Gamma$.  The $\Gamma$-invariant Killing spinors are the Killing
spinors of the quotient.

Let us consider the case of a circle action, for definiteness.
Associated to the principal circle bundle $M \to M/\Gamma$ there is a
complex line bundle
\begin{equation*}
  \begin{CD}
    L = M \times_\Gamma \CC \\
    @VVV\\
    M/\Gamma~.
  \end{CD}
\end{equation*}
This line bundle is equipped with a natural connection coming from the
connection on $M \to M/\Gamma$ which declares those vectors orthogonal
to the Killing vector $\xi$ as horizontal.  In the case where $M$ is a
background of eleven-dimensional supergravity, the connection one-form
on the IIA background $M/\Gamma$ is the Ramond-Ramond one-form
potential.

Sections of $L$ are identified with the first Kaluza--Klein mode of a
complex valued function $f:M \to \CC$.  In other words, a function
which obeys $\xi f = i f$.  More generally, any function $f:M \to
\CC$ admits a decomposition into Kaluza--Klein modes
\begin{equation*}
  f = \sum_{n\in\ZZ} f_n
\end{equation*}
where $\xi f_n = i n f_n$.  This is nothing but the Fourier
decomposition of the function $f$ along the orbits of $\Gamma$.  From
the point of view of the quotient, $f_n$ defines a section through
$L^n$, where $L^{-1} = L^*$ is the dual bundle and $L^0$ is the
trivial bundle.  Similarly we can decompose a complex spinor $\psi$
into its Kaluza--Klein modes
\begin{equation*}
  \psi = \sum_{n\in\ZZ} \psi_n
\end{equation*}
where $\eL_\xi \psi_n = i n \psi_n$, with $\eL_\xi$ the spinorial Lie
derivative.  From the point of view of the quotient, $\psi_n$ is a
section of the bundle $S(M/\Gamma) \otimes L^n$, where $S(M/\Gamma)$
is the bundle $\theta^{-1} P \times_{\Spin(1,n-1)} S$.  The
Kaluza--Klein zero modes are spinors of $M/\Gamma$, whereas the
nonzero modes are the `charged spinors' of \cite{HullHolonomy}.

The covariant derivative $D$ defining the Killing spinors commutes
with the action of $\Gamma$, whence it preserves the Fourier
decomposition.  This means that we will be able to write the
(complexified) Killing spinors as a finite linear combination of
Fourier modes.  The zero modes will be Killing spinors on the
quotient, which correspond to the geometrically realised
supersymmetries.  It is not clear to us whether there is any
supergravity interpretation of the nonzero modes.  In M/string-theory
and under T-duality, these modes become winding (or brane) modes in
the dual background and are the ones responsible for the
`supersymmetry without supersymmetry' of \cite{DLP}.

If the spin bundle on $M$ is not projectable we have a similar
situation with the difference that the infinitesimal action of $\xi$
on spinors integrates to an action of the double cover
$\widetilde\Gamma$.  In terms of the eigenvalues of $\eL_\xi$ this means
that they are now $i \lambda$ with $\lambda \in \ZZ + \half$.  It may
be surprising at first that there are no integral eigenvalues---after
all this is still a representation of $\widetilde\Gamma$, albeit one
with a kernel.  The reason is topological.  A spinor on $M$ with
eigenvalue $i n$ with $n\in \ZZ$ defines a section through a bundle
$S'\otimes L^n$, where $S'$ would be a spinor bundle on $M/\Gamma$,
but this bundle does not exist.  Instead a spinor with eigenvalue
$i(n+\half)$ defines a section through $(S'\otimes L^{1/2}) \otimes
L^n$, where the bundles $S'$ and $L^{1/2}$ do not exist individually
but their tensor product $S^c(M/\Gamma) = S'\otimes L^{1/2}$
does---indeed, it is the bundle of spin$^c$ spinors, as the notation
suggests.  Killing spinors on $M$ decompose into Fourier modes to give
rise to sections of $S^c(M/\Gamma) \otimes L^n$ for some $n$.  We see
therefore that even when $n=0$, the sections of $S^c(M/\Gamma)$ carry
(fractional) charge under $\Gamma$, and hence there are no zero modes.
The supergravity interpretation of these charged spinors is also not
clear, although some speculations are presented in the paper
\cite{HullHolonomy}.

These remarks give another way to test for the existence of a spin
structure in the quotient in the case of a circle action, which is the
only case where there is a possible obstruction in the groups we are
considering.  We simply analyse the representation of the circle group
(normalised to $\RR/2\pi\ZZ$) on the Killing spinors and check whether
the charges are integral, in which case there is a spin structure
in the quotient, or half-integral, in which case there is not and only
a spin$^c$ structure.

It should be clear that this analysis requires determining the action
of isometries on the Killing spinors of the backgrounds in question.
It is not always appreciated that this question can be answered
without ever having to compute a Killing spinor explicitly and,
although doing so might provide a concrete check, the problem is
simply one of group theory.  The upshot of the analysis, to be
presented in detail in \cite{FigLeiSim} as part of a general study of
smooth Freund--Rubin backgrounds, is that Killing spinors on
$\AdS_{p+1} \times S^q$ are in one-to-one correspondence with the
tensor product of half-spin representations of $\Spin(2,p)$ and
$\Spin(q+1)$ (and possibly the R-symmetry group) and that this
correspondence is equivariant with respect to the action of the
isometry group \cite{JMFKilling} and the R-symmetry group.

Let us be more precise.  As a representation of the symmetry (i.e.,
isometry $\times$ R-symmetry) group, the space of Killing spinors of
the different Freund--Rubin vacua considered in this paper is
isomorphic to the ones in Table~\ref{tab:Killing}.  Let us explain the
notation in the table.  The notation $\Delta^{m,n}$ stands for the
spinorial representation of $\Spin(m,n)$, and $\Delta^{m,n}_\pm$ for
the chiral spinor representations, whenever they exist.  Therefore,
$\Delta^{2,2}_-$ is the real $2$-dimensional representation of
$\Spin(2,2)$ consisting of negative chirality spinors, whereas
$\Delta^{4,0}_-$ denotes the negative chirality spinor representation
of $\Spin(4)$, which is quaternionic of (complex) dimension $2$.
$\Delta$ is the fundamental representation of $\Sp(1)$, which is also
quaternionic and of (complex) dimension $2$.  $\Delta^{2,3}$ is the
$4$-dimensional spinorial representation of $\Spin(2,3)$ and
$\Delta^{8,0}_-$ is one of the two $8$-dimensional spinorial
representations of $\Spin(8)$.  Both representations are real.
Similarly, $\Delta^{5,0}$ is the fundamental representation of $\Sp(2)
\cong \Spin(5)$, whence it is quaternionic and of complex dimension
$4$.  $\Delta_-^{2,6}$, which is also quaternionic but of complex
dimension $8$, is one of the two spinorial representations of
$\Spin(2,6)$.  The tensor product of two quaternionic representations
has a real structure and the notation $[R]$ denotes the underlying
real representation.  Finally, $\Delta^{2,4}$ is one of the
$4$-dimensional complex spinor representations of $\Spin(2,4)$ and
$\Delta^{6,0}$ is the fundamental representation of $\SU(4) \cong
\Spin(6)$, which is also complex and $4$-dimensional.  The notation
$[\![R]\!]$ means the underlying real representation of $R \oplus \bar
R$, which has a natural real structure.  We remind the reader that
$\dim_\RR [R] = \dim_\CC R$ whereas $\dim_\RR [\![R]\!] = 2 \dim_\CC
R$.

\begin{table}[h!]
  \centering
  \setlength{\extrarowheight}{3pt}
  \renewcommand{\arraystretch}{1.3}
    \begin{tabular}{|>{$}l<{$}|>{$}l<{$}|>{$}l<{$}|}\hline
      \multicolumn{1}{|c|}{Vacuum} &   \multicolumn{1}{c|}{Symmetry
        group} & \multicolumn{1}{c|}{Killing spinors} \\
      \hline\hline
      \AdS_3 \times S^3 & \Spin(2,2) \times \Spin(4) \times \Sp(1)& 
      \Delta^{2,2}_- \otimes [ \Delta^{4,0}_- \otimes \Delta ] \\
      \AdS_4 \times S^7 & \Spin(2,3) \times \Spin(8) & \Delta^{2,3}
      \otimes \Delta^{8,0}_-\\
      \AdS_5 \times S^5 & \Spin(2,4) \times \Spin(6) &
      [\![\Delta^{2,4} \otimes \Delta^{6,0} ]\!]\\
      \AdS_7 \times S^4 & \Spin(2,6) \times \Spin(5) & [
      \Delta^{2,6}_- \otimes \Delta^{5,0}]\\[3pt]
      \hline
    \end{tabular}
  \vspace{8pt}
  \caption{Killing spinors of Freund--Rubin vacua as representations
    of the symmetry group. (See the main text for an explanation of
    the notation.)}
  \label{tab:Killing}
\end{table}

\section{Supersymmetric quotients of Freund--Rubin backgrounds}
\label{sec:squots}

In this section we bring together the technology developed in the
previous sections and classify the supersymmetric smooth $S^1$ or
$\RR$ quotients of the maximally supersymmetric Freund--Rubin vacua of
type IIB superstring and M-theory, as well as (the trivial lift to IIB
supergravity of) the Freund--Rubin vacuum of $(1,0)$ six-dimensional
supergravity.  The discrete version of these and other quotients will
be studied in more detail in a separate publication \cite{FOMRS}.

To this end we would like to classify all the (connected)
one-parameter subgroups $\Gamma\subset \SO(2,p) \times \SO(q+1)$ such
that $(\AdS_{p+1}\times S^q)/\Gamma$ is a smooth supersymmetric
background of the relevant supergravity theory in $p+q+1$ dimensions,
for $(p,q)=(2,3),(3,7),(4,5),(6,4)$.  Since $\Gamma$ is generated by a
Killing vector $\xi$, these conditions on the quotient become
conditions on $\xi$ and on its integral curves.  First of all, we
demand that $\xi$ be everywhere spacelike: this implies in particular
that it never vanishes, whence the action of $\Gamma$ is locally free.
For the quotient to be smooth, we demand in addition that every point
should have trivial stabiliser.  For the quotient to be a supergravity
background we require that it be spin, which as discussed in the
previous section requires the spin structure in $\AdS_{p+1} \times
S^q$ to be projectable, so that it is acted upon by $\Gamma$.
Finally, supersymmetry requires that some of the Killing vectors in
$\AdS_{p+1} \times S^q$ should be $\Gamma$-invariant.

Following the discussion in \cite{FigSimFlat,FigSimBranes,FigSimGrav},
the amount of supersymmetry preserved by the reduction by a
one-parameter subgroup $\Gamma$ of isometries generated by a Killing
vector $\xi$ is the dimension of the vector space of solutions to the
equation
\begin{equation}
  \eL_\xi \varepsilon = \nabla_\xi\varepsilon +
  \frac{1}{4}d\xi^\flat\cdot\varepsilon = 0~,
 \label{eq:susyeq}
\end{equation}
where $\varepsilon$ is a Killing spinor and $\eL_\xi$ is the Lie
derivative operator along the Killing vector field $\xi$ introduced in
\cite{Kosmann}.  For the backgrounds under consideration, this is
simply the action of the element $\xi$ in the Lie algebra of the
isometry group on the relevant spinorial representation.  These
representations appear in Table~\ref{tab:Killing} and the action of
$\xi$ on them follows in many cases by a simple weight analysis.  The
method has been explained in detail in \cite[Appendix A]{FigSimBranes}
and we will simply apply it here to the cases at hand.

Finally, we stress again that we shall not discuss whether the
everywhere spacelike quotients we classify have closed causal curves,
even though we are aware that some of them do.  Such considerations
will appear in the forthcoming paper \cite{FOMRS}.

\subsection{Supersymmetric quotients of $\AdS_3 \times S^3 (\times
  \RR^4)$}
\label{sec:ads3s3}

The $\AdS_3 \times S^3$ vacuum of six-dimensional $(1,0)$
supergravity corresponds geometrically to a simply-connected
parallelised Lie group: $\widetilde{\SL}(2,\RR) \times \SU(2)$, where
$\widetilde{\SL}(2,\RR)$ is the universal covering group of
$\SL(2,\RR)$.  Both factors have equal radii of curvature, whence
their scalar curvatures have equal magnitude but opposite sign.
Letting $s>0$ denote the scalar curvature of the sphere, the $3$-form
is then given by
\begin{equation*}
  H = \sqrt{\tfrac23 s} \left( \dvol_A - \dvol_S \right)~,
\end{equation*}
where $\dvol_A$ and $\dvol_S$ are the volume forms of $\AdS_3$ and
$S^3$, respectively.  This vacuum solution has eight supersymmetries.
It is also a vacuum solution of the $(2,0)$ supergravity, this time
with 16 supersymmetries \cite{CFOSchiral}.  We remark here, once and
for all, that in this and other Freund--Rubin backgrounds we will
study in this paper, the volume forms $\dvol_A$ and/or $\dvol_S$ are
the volume forms of the actual metrics which appear in the
supergravity solution.  In particular, they are \emph{not} normalised
to unit radius of curvature.

This background can be lifted to IIB supergravity giving rise to
non-dilatonic backgrounds of the form $\AdS_3 \times S^3 \times X$,
where supersymmetry implies that $X$ admits parallel spinors.  We will
focus on the simply-connected half-BPS background where $X = \RR^4$.
The space of Killing spinors for this background is isomorphic to
\begin{equation}
  \label{eq:KS334}
    \left( \Delta^{2,2}_+ \otimes \left[ \Delta^{4,0}_+ \otimes
        \Delta^{0,4}_+ \right] \right) \oplus \left( \Delta^{2,2}_-
      \otimes \left[ \Delta^{4,0}_- \otimes \Delta^{0,4}_+ \right]
    \right)
\end{equation}
as a representation of $\Spin(2,2) \times \Spin(4) \times \Spin(4)$,
in the notation introduced and explained at the end of the previous
section.  We will moreover restrict our attention to quotients of this
IIB background by one-parameter subgroups of the isometries of the
$\AdS_3 \times S^3$ geometry.  The general case can be treated along
similar lines without any additional difficulty, but it falls somewhat
outside the scope of this paper thematically.  Restricting to these
quotients has the added virtue that we can then simply read off the
supersymmetric quotients of the $\AdS_3 \times S^3$ vacuum of $(1,0)$
supergravity in six dimensions, whose Killing spinors are in
one-to-one correspondence with the second factor in the above
decomposition, after identifying the $\Sp(1) \subset \Spin(4)$
subgroup acting nontrivially on $\Delta^{0,4}_+$ with the R-symmetry
group of the six-dimensional theory.  Similarly we can work in $(2,0)$
supergravity by substituting $\Sp(1)$ for $\Sp(2)$ and essentially
doubling the number of supersymmetries.  This means that the
supersymmetry fraction of the quotient is the same for $(1,0)$ and
$(2,0)$ supergravities.

The Killing vector $\xi$ generating the action of $\Gamma$ decomposes
as
\begin{equation*}
  \xi = \xi_A + \xi_S~,
\end{equation*}
where $\xi_A$ is the component on $\AdS_3$ and $\xi_S$ is the
component on $S^3$.   Moreover its norm is the sum of the norms of
these two components. 

Let $S^3$ denote the sphere of radius $R_S$ in $\RR^4$.  Up to
conjugation by $\SO(4)$, the most general Killing vector can be
written as
\begin{equation*}
  \xi_S = \theta_1 R_{12} + \theta_2 R_{34}~,
\end{equation*}
where $\theta_i$ are real parameters and where $R_{ij}$ is the
infinitesimal generator of rotations in the $ij$-plane.  Strictly
speaking we still have to quotient by the Weyl group, which acts on
the $\theta_i$ by permuting them and changing both their signs
simultaneously.  This means that we can always arrange the $\theta_i$
in such a way that $\theta_1 \geq |\theta_2|$.  We will take this
ordering to hold from now on.  Notice that $\theta_2$ need not be
non-negative.

The norm of the Killing vector $\xi_S$ is bounded both
above and below in the sphere:
\begin{equation*}
  \theta_2^2 R^2_S \leq |\xi_S|^2 \leq \theta_1^2 R^2_S~.
\end{equation*}
It is not hard to see that these bounds are sharp.  As a consequence,
the norm of $\xi = \xi_A + \xi_S$ obeys the bounds
\begin{equation*}
    |\xi|^2 \geq  |\xi_A|^2 + \theta_2^2 R^2_S~,
\end{equation*}
whence $\xi$ will be always spacelike provided that
\begin{equation*}
  |\xi_A|^2 > - \theta_2^2 R^2_S~.
\end{equation*}
From the results in Section~\ref{sec:so22}, the only Killing vectors
$\xi_A$ on $\AdS_3$ which satisfy this condition without any
requirement are the ones labelled (4), (5), (6) and (11) with $\beta_1
= \pm \beta_2$.  Recall that for this vacuum solution, $R_S = R_A =
R$.  Thus, the cases labelled (8) and (10), which are bounded from
below whenever $\varphi>0$ and $\varphi_1^2\geq\varphi_2^2$,
respectively, will also satisfy the above condition if
$\theta_2^2>\varphi^2$ and $\theta_2^2>\varphi_2^2$, respectively. The
resulting Killing vectors are everywhere spacelike with norms bounded
below by $\theta_2^2 R^2$ for cases (4), (5) and (6); by $(\theta_2^2
+ \beta_1^2) R^2$ for case (11); by $(\theta_2^2-\varphi^2)R^2$
$(\varphi>0)$ for case (8); and by $(\theta_2^2-\varphi_2^2)R^2$
(where $\varphi_1^2\geq\varphi_2^2$) for case (10).  We now proceed to
analyse each case in more detail.

Notice that in principle given a Killing vector field $\xi$, what we
have is a pencil of Killing vectors: a linear combination of a Killing
vector on $\AdS_3$ and a Killing vector on $S^3$ which is then
projectivised.  This means that one or the other of the two Killing
vectors (but not both) may vanish.  However since we are only
interested in those Killing vectors being spacelike everywhere, and
the norm of $\xi_S$ is bounded from above, it is clear that we shall
only consider the subset of $\xi_A$ being bounded from below. In
particular, the sphere component must always be present in cases (4),
(5), (6), (8) and (10).  As explained above, in cases (8) and (10)
even the infinitesimal parameters in $\xi_S$ can no longer be
arbitrary but have to satisfy some requirements. On the other hand,
this sphere component may be absent in case (11) with
$\beta_1=\pm\beta_2$.  We start by analysing the case when only the
spherical component is present.

\subsubsection{$\xi=\xi_S$}

For $\xi$ to be everywhere nonvanishing (and hence spacelike) we
require $\theta_2>0$.  This guarantees that the action of $\Gamma$ is
locally free.  To have a smooth quotient, the stabilisers must be
trivial.  To check when this is the case it is convenient to think of
$S^3$ as embedded in $\CC^2$ with standard coordinates $z_i$, $a=1,2$
as the quadric
\begin{equation}
  \label{eq:s3quadric}
  |z_1|^2 + |z_2|^2 = R^2~.
\end{equation}
The action of $\xi$ integrates to
\begin{equation*}
  (z_1,z_2) \mapsto (e^{i\theta_1 t} z_1, e^{i\theta_2 t} z_2)~.
\end{equation*}
The quotient will not be Hausdorff unless the integral curves are
periodic (compare with the discussion in
\cite[Section~3.2]{FigSimGrav}), whence there is a smallest positive
number $T$ for which
\begin{equation*}
  (e^{i\theta_1 T} z_1, e^{i\theta_2 T} z_2) = (z_1,z_2)~,
\end{equation*}
for all $(z_1,z_2)$.  This implies that there are integers
$p_i$ such that $\theta_i T = 2\pi p_i$ and since $T$ is the smallest
such positive integer, the $p_i$ are coprime.  Using the freedom
to rescale the Killing vector, we can take
\begin{equation*}
  \xi = p_1 R_{12} + p_2 R_{34} 
  \quad\text{with $\gcd(p_1,p_2)=1$},
\end{equation*}
which integrates to a circle action
\begin{equation*}
  (z_1,z_2) \mapsto (e^{ip_1 t} z_1, e^{ip_2 t} z_2)~,
\end{equation*}
with period $2\pi$.  The action has trivial stabilisers if for all
$(z_1,z_2)$ obeying \eqref{eq:s3quadric}, the equation
\begin{equation*}
  (z_1,z_2) = (e^{ip_1 t} z_1, e^{ip_2 t} z_2)
\end{equation*}
holds only for $t \in 2\pi\ZZ$.  By looking at the points $(2R,0)$,
$(0,2R)$, we see that $p_i$ have to be $\pm 1$ giving $2^2 = 4$
possibilities which, taking into account the action of the Weyl group,
are reduced to only two: $p_1=1$ and $p_2=\pm1$.  In either case the
resulting quotient is $\CP^1$ (with a different orientation for each
quotient), which as we saw above, admits a spin structure.

Let us analyse whether supersymmetry is preserved when taking this
quotient.  As discussed above, the Killing spinors of this background
are in a representation \eqref{eq:KS334} of $\Spin(2,2)\times \Spin(4)
\times \Spin(4)$.  Complexifying this representation and computing the
weights of the element in the first $\fso(4)$ corresponding to
$\xi^\pm = R_{12} \pm R_{34}$, we see that
\begin{equation*}
    \Delta^{2,2}_\pm \otimes \Delta^{4,0}_\pm \otimes \Delta^{0,4}_+
\end{equation*}
has all eight weights equal to $0$ for $\xi^\pm$, and has weights
$(1,-1)$, each with multiplicity $4$, for $\xi^\mp$.  We conclude that
either quotient preserves one half of the sixteen supersymmetries of
the original background.  In other words, these quotients have eight
supersymmetries, or equivalently, they  preserve a fraction $\nu =
\frac14$ of the IIB supersymmetry.

As backgrounds of $(1,0)$ and $(2,0)$ $d{=}6$ supergravities, the
quotient by $\xi^+$ breaks all supersymmetry, whereas the one by
$\xi^-$ preserves all.  The existence of nontrivial quotients
preserving all supersymmetry was explained in \cite{CFOSchiral}, and
is intimately linked to the fact that these vacua can be interpreted
as anti-selfdual lorentzian Lie groups.  The resulting
five-dimensional backgrounds were also classified in \cite{CFOSchiral}
and further elucidated in \cite{FHLT}.

We now turn our attention to the four possible families of reductions
where the anti-de~Sitter component of the Killing vector is nonzero.

\subsubsection{$\xi=\xi_A^{(4)} + \xi_S$}

This case is very similar to the previous one, with the possibility of
adding a spatial rotation $\xi_A^{(4)} = \varphi \be_{34}$ to $\xi$.
The same analysis as before shows that $\varphi T = 2\pi n$ for some
integer $n$ which is coprime to the $p_i$, whence after rescaling the
Killing vector it becomes
\begin{equation*}
  \xi = n \be_{34} +  p_1 R_{12} + p_2 R_{34} \quad\text{with 
  $\gcd(n, p_1,p_2)=1$}.
\end{equation*}
As before, the triviality of the stabilisers demands that $p_i=\pm 1$,
and hence up to rescaling $\xi$ and conjugating by the Weyl group of
$\SO(4)$ one has
\begin{equation*}
  \xi^\pm = n \be_{34} +  R_{12} \pm R_{34}
  \quad\text{with $n=1,2,\dots$.}
\end{equation*}
Let us employ the convention that $\gamma_i$ denote the
$\gamma$-matrices for $\Cl(2,2)$ and $\Gamma_a$ denote those of
$\Cl(4)$.  Then the element in $\fso(2,2) \oplus \fso(4) \subset
\Cl(2,6)$ corresponding to $\xi^\pm$ is given by
\begin{equation*}
  X^\pm = n \gamma_{34} + \Gamma_{12} \pm \Gamma_{34}~.
\end{equation*}
Exponentiating this element we get
\begin{multline*}
  \widehat g(t) = \exp\frac{tX^\pm}2 = \left(\id \cos\frac{nt}2 +
    \gamma_{34} \sin \frac{nt}2\right) \left(\id \cos\frac{t}2 +
    \Gamma_{12} \sin\frac{t}2\right)\\
  \times \left(\id \cos\frac{t}2 \pm \Gamma_{34} \sin\frac{t}2\right)
\end{multline*}
and this can only possibly equal $-\id$ at $t=2\pi$, where
\begin{equation*}
  \widehat g(2\pi) = (-1)^n \id~;
\end{equation*}
whence only for $n$ even do we have a spin structure in the quotient.

Computing the weights under $\xi^\pm$ of the (complexified) Killing
spinors we obtain from
\begin{equation*}
  \Delta^{2,2}_+ \otimes \Delta^{4,0}_+ \otimes \Delta^{0,4}_+
\end{equation*}
$\pm \frac{n}2$, each with multiplicity $4$, for $\xi^+$ and
$\pm \frac{n}2 \pm 1$ with uncorrelated signs, each with multiplicity
$2$, for $\xi^-$; whereas from
\begin{equation*}
  \Delta^{2,2}_- \otimes \Delta^{4,0}_- \otimes \Delta^{0,4}_+
\end{equation*}
$\pm\frac{n}2 \pm 1$ with uncorrelated signs, each with multiplicity
$2$, for $\xi^+$ and $\pm \frac{n}2$, each with multiplicity $4$, for
$\xi^-$.  Since $n$ is positive and even, we see that we only have
zero weights when $n=2$, in which case there are four for either
quotient.  Therefore the quotient preserves one fourth of the
supersymmetry of the background, or a fraction $\nu = \frac18$ of the
IIB supersymmetry.

As six-dimensional supergravity backgrounds, the $\xi^+$ quotient
is a half-BPS background for $(1,0)$ and $(2,0)$ supergravities,
whereas the $\xi^-$ quotient is not supersymmetric.

\subsubsection{$\xi=\xi_A^{(5)} + \xi_S$}

Here the Killing vector field is
\begin{equation*}
  \xi = \be_{13} - \be_{34} + \theta_1 R_{12} + \theta_2 R_{34}~.
\end{equation*}
This vector fields integrates to the action of $\Gamma \cong \RR$.
Its orbits are generically non-periodic, except at those points
where the action of $\xi_A$ vanishes, that is, at $x_1=x_3=x_4=0$,
which corresponds to the two points $x_2 = \pm R$, where we must
be more careful. In these points, the orbits are either not periodic, 
in which case they would be dense in a submanifold of dimension at least 
$2$ and hence the quotient would not be Hausdorff, or else they are periodic 
in which case their points have nontrivial stabiliser, namely the subgroup
$\Gamma_0 \subset \Gamma$ consisting of the periods.  In either case
the quotient is not smooth.  Nevertheless see Section~\ref{sec:N+SI}
for a discussion on how such a singular quotient can still preserve
some supersymmetry.

\subsubsection{$\xi=\xi_A^{(6)} + \xi_S$}
\label{sec:nilprod}

In this case the Killing vector is
\begin{equation*}
  \xi^\pm = \mp\be_{12} - \be_{13} \pm \be_{24} + \be_{34} + \theta_1
  R_{12} + \theta_2 R_{34}~,
\end{equation*}
where the $\theta_i$ are nonzero for $\xi$ to be everywhere
spacelike.  Notice that the anti-de~Sitter component is never zero,
although its norm vanishes.  It integrates to the following action on
$\RR^{2,2}$:
\begin{equation*}
  \begin{pmatrix}
    x^1\\x^2\\x^3\\x^4
  \end{pmatrix}
  \mapsto
  \begin{pmatrix}
    x^1 + t (x^3 \mp x^2)\\
    x^2 \pm t (x^1 - x^4)\\
    x^3 + t (x^1 - x^4)\\
    x^4 + t (x^3 \mp x^2)
  \end{pmatrix}~,
\end{equation*}
whose orbits are straight lines ruling the quadric.  Thus the
combined Killing vector $\xi^\pm$ integrates to an action of $\Gamma
\cong \RR$.  It is easy to see that it has trivial stabilisers, since
it does on $\AdS_3$.  Therefore the resulting quotient is smooth and,
since $\Gamma \cong \RR$, also spin.

To determine the amount of preserved supersymmetry, we have to solve
the algebraic equation \eqref{eq:susyeq} for this particular Killing
vector field $\xi$.  The action of $\xi$ on the Killing spinors
defines an algebraic operator which splits into a nilpotent piece
\begin{equation*}
  N^\pm = \mp\gamma_{12}-\gamma_{13} \pm \gamma_{24}+\gamma_{34}\,,
\end{equation*} 
and a semisimple piece
\begin{equation*}
  S=\theta_1\,\Gamma_{12} + \theta_2\,\Gamma_{34}~,
\end{equation*} 
which moreover commute with each other.  Therefore the equation
\eqref{eq:susyeq} splits into two:
\begin{equation*}
  N^\pm \,\varepsilon = 0 \qquad\text{and}\qquad S\,\varepsilon = 0~.
\end{equation*}
Thus, the invariant Killing spinors correspond to the intersection of 
the kernels of $N^\pm$ and $S$ acting on the space
\begin{equation*}
  \left( \Delta^{2,2}_+ \otimes \Delta^{4,0}_+ \otimes \Delta^{0,4}_+
  \right) \oplus \left( \Delta^{2,2}_- \otimes \Delta^{4,0}_- \otimes
    \Delta^{0,4}_+ \right)
\end{equation*}
of (complexified) Killing spinors.

First of all, let us determine the dimension of the kernel of the
nilpotent operators $N^\pm$. To analyse this question we can simply
notice that $N^\pm = N_1^\pm N_2$ where $N_1^\pm=
\pm\gamma_2+\gamma_3$  and $N_2 = \gamma_1 + \gamma_4$ are commuting
nilpotent operators.  The kernel of $N^\pm$ is therefore the subspace
generated by the kernels of $N_1$ and $N_2$, which has $\frac34$ the
dimension of the space.

Alternatively, the equation $N^\pm \varepsilon = 0$ is equivalent to
\begin{equation*}
  \left(\gamma_{14} + \gamma_{23} - \gamma_{1234}\right)\varepsilon =
  \varepsilon~.
\end{equation*}
The solution to this eigenvalue problem shows that the action on
$\AdS_3$ preserves $\frac{3}{4}$ of the supersymmetry.

It is convenient for the full discussion of the supersymmetry
preserved by these quotients to determine how the nilpotent operators
act in each of the subspaces of (complexified) Killing spinors.
We first observe that $N^\pm$ acts trivially on $\Delta^{2,2}_\mp$.
Indeed, a simple calculation in $\Cl(2,2)$ shows that
\begin{equation*}
  N^\pm (\id \mp \gamma_{1234}) = 0~.
\end{equation*}
Since the kernel of $N^\pm$ in $\Delta^{2,2}_+ \oplus
\Delta^{2,2}_-$ is three-dimensional, we learn that $N^\pm$ has a
one-dimensional kernel on $\Delta^{2,2}_\pm$.

To analyse the equation $S\varepsilon = 0$, let us observe that $S$
has weights $\pm \half (\theta_1 - \theta_2)$ in $\Delta^{4,0}_+$ and
$\pm \half (\theta_1 + \theta_2)$ in $\Delta^{4,0}_-$.  Therefore
there are only zero weights when $\theta_1 = \pm \theta_2$ in
$\Delta^{4,0}_\pm$ and none otherwise.  

In summary, we see that the quotient generated by $\xi^+$ preserves
$\nu=\frac{1}{8}$ of the original supersymmetries in type IIB when
$\theta_1=\theta_2$.  That is, it has four supercharges which live in
\begin{equation*}
  \left(\ker
    N^+ \cap \Delta^{2,2}_+\right) \otimes
  [\Delta^{4,0}_+\otimes\Delta^{0,4}_+]~.
\end{equation*}
Furthermore, it preserves $\nu=\frac{1}{4}$ in type IIB when
$\theta_1=-\theta_2$. These are all eight supercharges living in 
\begin{equation*}
  \Delta^{2,2}_- \otimes [\Delta^{4,0}_- \otimes \Delta^{0,4}_+]~.
\end{equation*}
It has no supercharges whenever $\theta_1\neq \pm\theta_2$.

On the other hand, the quotient generated by $\xi^-$ preserves
$\nu=\frac{1}{8}$ of the original supersymmetries in type IIB when
$\theta_1=-\theta_2$.  That is, it has four supercharges which live in
\begin{equation*}
  \left(\ker N^- \cap \Delta^{2,2}_- \right) \otimes
  [\Delta^{4,0}_-\otimes\Delta^{0,4}_+]~.
\end{equation*}
Furthermore, it preserves $\nu=\frac{1}{4}$ in type IIB when
$\theta_1=\theta_2$.  These are all eight supercharges living in 
\begin{equation*}
  \Delta^{2,2}_+\otimes[\Delta^{4,0}_+\otimes\Delta^{0,4}_+]~.
\end{equation*}
It has no supercharges whenever $\theta_1\neq \pm\theta_2$.

In six-dimensional supergravity, $\xi^\pm$ break all supersymmetry
when $\theta_1\neq -\theta_2$, whereas if $\theta_1=-\theta_2$,
$\xi^+$ preserves all supersymmetry and $\xi^-$ preserves
$\nu=\frac{1}{2}$.  As shown in \cite{FHLT} and conjectured in
\cite{CFOSchiral}, the quotient by $\xi^+$ is the near-horizon
geometry \cite{GMT1} of the critically over-rotating black hole
\cite{BMPV,KRW}.

\subsubsection{$\xi=\xi_A^{(8)} + \xi_S$}
\label{sec:N=N1N2}

In this case, the Killing vector field is
\begin{equation*}
  \xi^\pm = \mp\be_{12} - \be_{13} \pm\be_{24} + \be_{34} + 
  \varphi\left(\be_{34}\pm\be_{12}\right) +
  \theta_1 R_{12} + \theta_2 R_{34}~,
\end{equation*}
where $\theta_1 \geq |\theta_2| > \varphi >0$, to ensure that
$\xi^\pm$ is everywhere spacelike.  It integrates to the following
free action of $\Gamma \cong \RR$ on $\RR^{2,2}$:
\begin{equation*}
  \begin{pmatrix}
    x^1\\x^2\\x^3\\x^4
  \end{pmatrix}
  \mapsto
  \begin{pmatrix}
    x^1 \cos\varphi t  \pm x^2 \sin\varphi t +
    t\left(\left(x^1-x^4\right)\,\sin\varphi t
      + \left(\mp x^2 + x^3\right)\,\cos\varphi t\right) \\
    x^2 \cos\varphi t  \mp x^1 \sin\varphi t \pm
    t\left(\left(x^1-x^4\right)\,\cos\varphi t +
      \left(\pm x^2-x^3\right)\,\sin\varphi t\right) \\
    x^3 \cos\varphi t  - x^4 \sin\varphi t +
    t\left(\left(x^1-x^4\right)\,\cos\varphi t +
      \left(\pm x^2-x^3\right)\,\sin\varphi t\right) \\
    x^4 \cos\varphi t  + x^3 \sin\varphi t +
    t\left(\left(x^1-x^4\right)\,\sin\varphi t +
      \left(\mp x^2 + x^3\right)\,\cos\varphi t\right)
  \end{pmatrix}~,
\end{equation*}
which lifts to a free action on $\AdS_3$.  The resulting quotient is
smooth and, since $\Gamma\cong \RR$, is also spin.

The analysis of supersymmetry follows closely the one in the previous
section.  We first observe that the action of $\xi$ on Killing spinors
is given by the sum of two commuting operators: the same nilpotent
operator as in the previous case,
\begin{equation*}
  N^\pm = \mp\gamma_{12} - \gamma_{13} \pm \gamma_{24} + \gamma_{34}~,
\end{equation*}
and a semisimple one
\begin{equation*}
  S^\pm = \varphi\left(\gamma_{34}\pm\gamma_{12}\right) + \theta_1
  \Gamma_{12} + \theta_2 \Gamma_{34}~.
\end{equation*}
Once again, the invariant Killing spinors correspond to the
intersection of the kernels of $N^\pm$ and $S^\pm$ acting on the space
\begin{equation*}
  \left( \Delta^{2,2}_+ \otimes \Delta^{4,0}_+ \otimes \Delta^{0,4}_+
  \right) \oplus \left( \Delta^{2,2}_- \otimes \Delta^{4,0}_- \otimes
    \Delta^{0,4}_+ \right)
\end{equation*}
of (complexified) Killing spinors.

Since we already know the kernel of $N^\pm$ from the previous section,
we summarise this discussion as follows.  To find the amount of
supersymmetry preserved by the quotient along $\xi^\pm$, we need to find
the invariants of
\begin{equation*}
  \theta_1 \Gamma_{12} + \theta_2 \Gamma_{34} \quad\text{acting on
    $\Delta^{4,0}_\pm \otimes \Delta^{0,4}_+$}
\end{equation*}
and those of
\begin{equation*}
  \varphi(\pm \gamma_{12} + \gamma_{34}) + \theta_1 \Gamma_{12} +
  \theta_2 \Gamma_{34} \quad\text{acting on $\Delta^{2,2}_\mp \otimes
    \Delta^{4,0}_\mp \otimes \Delta^{0,4}_+$.}
\end{equation*}

Let us consider $\xi^+$ first.  The weights of $\theta_1 \Gamma_{12} +
\theta_2 \Gamma_{34}$ on $\Delta^{4,0}_+$ are $\pm \half (\theta_1 -
\theta_2)$, whereas those of $\varphi(\gamma_{12} + \gamma_{34}) +
\theta_1 \Gamma_{12} + \theta_2 \Gamma_{34}$ on $\Delta^{2,2}_-
\otimes \Delta^{4,0}_-$ are $\pm\varphi \pm \half (\theta_1 +
\theta_2)$.  The zero weights occur when $\theta_1 = \theta_2 >
\varphi > 0$ with multiplicity $4$ and when $2\varphi = \theta_1 +
\theta_2$ with $\theta_1 > - \theta_2 > \varphi > 0$, also with
multiplicity $4$.  In either case there are $4$ supersymmetries and
hence the quotient preserves a fraction $\nu=\tfrac18$ of the IIB
supersymmetry .  No further enhancement of supersymmetry is possible
given the range of parameters defined by the Weyl group action, the
$\SO(2,2)$ action and the causality constraint on $\xi^+$. Notice that
the first supersymmetric example $(\theta_1=\theta_2>\varphi>0\,\,)$
provides an example of the phenomenon of turning on a deformation
parameter $(\varphi)$ without decreasing the supersymmetry preserved
by the quotient.

For the six-dimensional theories, the quotient along $\xi^+$ is
half-BPS whenever $2\varphi = \theta_1 + \theta_2$ and
non-supersymmetric otherwise.

Going on to $\xi^-$, the weights of $\theta_1 \Gamma_{12} + \theta_2
\Gamma_{34}$ on $\Delta^{4,0}_-$ are $\pm \half (\theta_1 +
\theta_2)$, whereas those of $\varphi(-\gamma_{12} + \gamma_{34}) +
\theta_1 \Gamma_{12} + \theta_2 \Gamma_{34}$ on $\Delta^{2,2}_+
\otimes \Delta^{4,0}_+$ are $\pm\varphi \pm \half (\theta_1 -
\theta_2)$.  The zero weights occur when $\theta_1 = - \theta_2 >
\varphi > 0$ with multiplicity $4$, and when $2\varphi = \theta_1 -
\theta_2$ with $\theta_1 > \theta_2 > \varphi > 0$, also with
multiplicity $4$.  In either case there are $4$ supersymmetries and
hence the quotient preserves a fraction $\nu=\frac18$ of the IIB
supersymmetry.  No further supersymmetry enhancement is allowed in
this case either. The same comment on the phenomenon of free turning
of deformation parameter, supersymmetry wise, applies to
$\theta_1=-\theta_2$.

Again for the six-dimensional theories, the quotient along $\xi^-$ is
half-BPS whenever $\theta_1 = - \theta_2$ and non-supersymmetric
otherwise.

It should be remarked that for the IIB quotients there exists an
enhancement of supersymmetry to $\nu = \frac14$ by allowing the norm
of $\xi$ to vanish at places.

\subsubsection{$\xi=\xi_A^{(10)} + \xi_S$}

In this case, the Killing vector field is
\begin{equation*}
  \xi = \varphi_1\be_{12} + \varphi_2\be_{34} +
  \theta_1 R_{12} + \theta_2 R_{34}~,
\end{equation*}
where $\varphi_1\geq |\varphi_2|\geq 0$, otherwise $\xi_A^{(10)}$
would not be bounded from below and $\theta_1 \geq |\theta_2| >
|\varphi_2|$, so that $\xi$ is everywhere spacelike.  By rescaling
$\xi$ we may further set $\varphi_1 =1$.  The action of $\xi_A^{(10)}$
integrates on $\RR^{2,2}$ to
\begin{equation*}
  \begin{pmatrix}
    x^1\\x^2\\x^3\\x^4
  \end{pmatrix}
  \mapsto
  \begin{pmatrix}
    x^1 \cos\varphi_1 t  + x^2 \sin\varphi_1 t \\
    x^2 \cos\varphi_1 t  - x^1 \sin\varphi_1 t \\
    x^3 \cos\varphi_2 t  - x^4 \sin\varphi_2 t \\
    x^4 \cos\varphi_2 t  + x^3\sinh\varphi_2 t 
  \end{pmatrix}~.
\end{equation*}
However in the simply-connected $\AdS_3$ these orbits are not periodic
and hence $\xi$ integrates to an action of $\Gamma\cong\RR$.  There
are clearly no stabilisers and hence the quotient is smooth, and since
$\Gamma \cong \RR$ it is also spin.  (This would be different in the
quadric in $\RR^{2,2}$, where the analysis of smoothness would be much
more delicate.)

Let us move to discuss the supersymmetry of these quotients, noting
that $\varphi_2$ is allowed to vanish.  The weights of $\xi$ in the
(complexified) Killing spinor representation \eqref{eq:KS334} are
given by
\begin{equation*}
  \pm\half \left(\varphi_1 - \varphi_2 \pm (\theta_1 - \theta_2)
  \right)\qquad 
  \pm\half \left(\varphi_1 + \varphi_2 \pm (\theta_1 +
    \theta_2)\right)~,
\end{equation*}
respectively, with uncorrelated signs and each with multiplicity two.
There are two possible conditions for zero weights consistent with our
range of parameters, each giving rise to $4$ supersymmetries and
hence preserving a fraction $\nu=\frac18$ of the IIB supersymmetry:
\begin{itemize}
\item $\varphi_1 - \varphi_2 = \theta_1 - \theta_2$; or
\item $\varphi_1 + \varphi_2 = \theta_1 + \theta_2$.
\end{itemize}
Even though these two conditions cannot be satisfied simultaneously and still
comply with the conditions on the parameters coming from demanding
that the orbits be everywhere spacelike, it is still possible to have
enhancement of supersymmetry for the two cases :
\begin{itemize}
\item $\varphi=\varphi_2$ and $\theta_1=\theta_2$; or
\item $\varphi_1=-\varphi_2$ and $\theta_1=-\theta_2$.
\end{itemize}
They preserve $\nu=\frac14$.

For both of the two six-dimensional theories under consideration, the
condition $\varphi_1 + \varphi_2 = \theta_1 + \theta_2$ gives rise to
half-BPS quotients, whereas $\varphi_1=-\varphi_2$,
$\theta_1=-\theta_2$ preserves all the supersymmetry of the vacuum.
All other cases listed above break all supersymmetry.  As conjectured
in \cite{CFOSchiral} and shown recently in \cite{FHLT}, the maximally
supersymmetric quotient is the near-horizon geometry \cite{GMT1} of
the over-rotating black hole \cite{BMPV,KRW}.

\subsubsection{$\xi=\xi_A^{(11)} + \xi_S$}

We let $\beta_1 = \pm\beta_2 = \beta$ in case (11) to arrive at the
Killing vector
\begin{equation*}
  \xi^\pm = \beta( \be_{13} \pm \be_{24}) + \theta_1 R_{12} +
  \theta_2 R_{34}~,
\end{equation*}
where now the $\theta_i$ are allowed to be zero provided that $\beta$
is not.  Since we already discussed the case $\beta=0$, we will now
concentrate on $\beta\neq 0$, whence we can rescale $\xi^\pm$ such
that $\beta=1$.  The anti-de~Sitter component integrates to the
following action on $\RR^{2,2}$:
\begin{equation*}
  \begin{pmatrix}
    x^1\\x^2\\x^3\\x^4
  \end{pmatrix}
  \mapsto
  \begin{pmatrix}
    x^1 \cosh t  - x^3 \sinh t \\
    x^2 \cosh t  \mp x^4 \sinh t \\
    x^3 \cosh t  - x^1 \sinh t \\
    x^4 \cosh t  \mp x^2 \sinh t
  \end{pmatrix}
\end{equation*}
which is clearly a free action of $\Gamma \cong \RR$.  This also
lifts to a free action of $\Gamma$ on $\AdS_3$ and hence on the
background for all values of $\theta_i$.  The resulting quotient is
therefore smooth and, since $\Gamma \cong \RR$, also spin.

The Killing vector defines an action on the (complexified) Killing
spinor representation which decomposes into two commuting semisimple
pieces:
\begin{equation*}
  S_1^\pm = \gamma_{13} \pm \gamma_{24}
\end{equation*}
and
\begin{equation*}
  S_2 = \theta_1 \Gamma_{12} + \theta_2 \Gamma_{34}~.
\end{equation*}
On a complex vector space, they are both diagonalisable, but whereas
$S_1$ has real eigenvalues, those of $S_2$ are imaginary.  Therefore
the supersymmetries of the quotient consist of Killing spinors in the
kernel of both.

We observe that $S_1^\pm \varepsilon = 0$ if and only if $\varepsilon
\in \Delta^{2,2}_\pm \otimes [\Delta^{4,0}_\pm \otimes
\Delta^{0,4}_+]$.  This means that $S_2 \varepsilon = 0$ if and only
if $\theta_1 = \pm \theta_2$.  Hence either quotient preserves a
fraction $\nu=\frac14$ of the IIB supersymmetry.  This is still the
case if $\theta_i=0$, which corresponds to the selfdual orbifolds
introduced in \cite{CH} and studied recently in \cite{BalNaqSim}.
Thus, these are examples of quotients
in which by turning on the $\theta_i$ in an appropriate way no
supersymmetry is lost.  Note that this phenomenon originates in the
way the Killing vector acts on the Killing spinors and on the space in
which the latter live.  Therefore this phenomenon depends on dimension
and will not be a feature of all $\AdS_{p+1}$ quotients.

Finally we observe that whereas $\xi^+$ preserves no supersymmetry in
the six-dimensional theories, $\xi^-$ preserves all.  Indeed, the
quotient by $\xi^-$ is the near-horizon geometry \cite{GMT1} of the
rotating black hole of \cite{BMPV,KRW}.

\subsubsection{Summary}

The smooth spacelike supersymmetric quotients of $\AdS_3 \times S^3$
are summarised in Table~\ref{tab:AdS3S3}, where we have also indicated
the fraction of supersymmetry preserved by each of the theories under
consideration: type IIB and six-dimensional $(1,0)$ and $(2,0)$
supergravities.  The $(2,0)$ fraction is the same as the $(1,0)$
fraction and is hence not written explicitly.

\begin{table}[h!]
  \centering
  \setlength{\extrarowheight}{3pt}
  \renewcommand{\arraystretch}{1.3}
  \begin{tabular}{|>{$}l<{$}|>{$}l<{$}|>{$}l<{$}|>{$}c<{$}|>{$}c<{$}|}\hline
    & \multicolumn{2}{c|}{Conditions for} & 
      \multicolumn{2}{c|}{$\nu$ in}\\ 
      \multicolumn{1}{|c|}{Killing vector} &
      \multicolumn{1}{c|}{Causality/Weyl} &
      \multicolumn{1}{c|}{Supersymmetry} &
      \multicolumn{1}{c|}{IIB} &
      \multicolumn{1}{c|}{(1,0)} \\
      \hline\hline
      \xi_S = \theta_1 R_{12} + \theta_2 R_{34} & \theta_1 \geq
      |\theta_2| > 0 & \theta_i = 1 & \frac14 & 0\\
      & & \theta_i = (-1)^i & \frac14 & 1\\ \hline
      \varphi \be_{34} + \xi_S & \varphi > 0 & 
      \text{$\varphi=2$, $\theta_i = 1$} & \frac18 & \frac14 \\
      & \theta_1 \geq |\theta_2| > 0 & \text{$\varphi=2$, $\theta_i =
        (-1)^i$} & \frac18 & 0 \\ \hline
      -\be_{12} - \be_{13} & \theta_1
      \geq |\theta_2| > 0 & \theta_1 = \theta_2 & \frac1{8} & 0\\
      \hfill  + \be_{24} + \be_{34} + \xi_S & & \theta_1 = -\theta_2 &
      \frac1{4} & 1 \\ \hline
      \be_{12} - \be_{13} & \theta_1
      \geq |\theta_2| > 0 & \theta_1 = \theta_2 & \frac1{4} & 0\\
      \hfill  -\be_{24} + \be_{34} + \xi_S & & \theta_1 = -\theta_2 &
      \frac1{8} & \frac12 \\ \hline
      -\be_{12} - \be_{13} +\be_{24} + \be_{34} &
      \theta_1 \geq  |\theta_2| > \varphi > 0 & \theta_1 = \theta_2 &
      \frac18 & 0\\
      \hfill  + \varphi\left(\be_{34}+\be_{12}\right) + \xi_S & &
      \varphi = \half(\theta_1 + \theta_2) & \frac18 & \half\\ \hline
      \be_{12} - \be_{13} - \be_{24} + \be_{34} &
      \theta_1 \geq  |\theta_2| > \varphi > 0 & \theta_1 = -\theta_2 &
      \frac18 & \half\\
      \hfill  + \varphi\left(\be_{34} - \be_{12}\right) + \xi_S & & 
      \varphi = \half(\theta_1 - \theta_2) & \frac18 & 0\\ \hline
      \be_{12} + \varphi\be_{34} + \xi_S & 1 \geq |\varphi| &
      1 - \varphi = \theta_1 -\theta_2 & \frac18 & 0\\
      & \theta_1 \geq  |\theta_2| > |\varphi| & 1 +
      \varphi = \theta_1 + \theta_2 & \frac18 & \half\\ 
      & & \varphi=1\,,\theta_1=\theta_2 & \frac{1}{4} & 0 \\
      & & \varphi=-1\,,\theta_1=-\theta_2 & \frac{1}{4} & 1\\ \hline
      \be_{13} + \be_{24} + \xi_S & \theta_1 \geq  |\theta_2| \geq 0 &
      \theta_1 = \theta_2 & \frac14& 0\\ \hline
      \be_{13} - \be_{24} + \xi_S & \theta_1 \geq  |\theta_2| \geq 0 &
      \theta_1 = -\theta_2 & \frac14 & 1\\ \hline 
    \end{tabular}
  \vspace{8pt}
  \caption{Smooth spacelike supersymmetric quotients of $\AdS_3 \times
    S^3$}
  \label{tab:AdS3S3}
\end{table}

\subsection{Supersymmetric quotients of $\AdS_4 \times S^7$}
\label{sec:ads4s7}

The $\AdS_4 \times S^7$ vacuum of eleven-dimensional supergravity
is such that $\AdS_4$ has radius of curvature $R_A = R$ and $S^7$ has
radius of curvature $R_S = 2R$.  Letting $s>0$ denote the scalar
curvature of the sphere, the $4$-form is given by
\begin{equation*}
  F = \sqrt{\tfrac67 s} \dvol_A ~,
\end{equation*}
where $\dvol_A$ is the volume form of $\AdS_4$.  This vacuum solution
has thirty-two supersymmetries.  In this section we will study the
smooth supersymmetric quotients.

The situation here is analogous to the one in the previous section.
The spherical component $\xi_S$ of the Killing vector is given by
\begin{equation*}
  \xi_S = \theta_1 R_{12} + \theta_2 R_{34} + \theta_3 R_{56} +
  \theta_4 R_{78}~,
\end{equation*}
whose norm is bounded above and below by
\begin{equation*}
    m^2 R^2_S \leq |\xi_S|^2 \leq M^2 R^2_S~,
\end{equation*}
where
\begin{equation*}
  m^2 = \min_i \theta_i^2 \qquad\text{and}\qquad M^2 = \max_i
  \theta_i^2~.
\end{equation*}
It is easy to see that these bounds are sharp.  The Weyl group acts by
permuting the $\theta_i$ and changing an even number of their signs
(see, e.g., \cite[Section 12.1]{Humphreys}).  This means that we can
order them so that $\theta_1 \geq \theta_2 \geq \theta_3 \geq
|\theta_4|$.  Clearly $m^2$ and $M^2$ are Weyl-invariant.

Provided that $m^2>0$, we will be able to use Killing vectors in
$\AdS_4$ which are not necessarily everywhere spacelike, provided
their norms are bounded below.  From the results in
Section~\ref{sec:so23}, the only Killing vectors $\xi_A$ on $\AdS_4$
which satisfy this condition are the ones labelled (4), (5), (6) and
(11) with $\beta_1 = \beta_2$.  The resulting Killing vectors are
everywhere spacelike with norms bounded below by $m^2 R^2_S$ for cases
(4), (5) and (6); and by $m^2 R^2_S + \beta_1^2 R_A^2$ for case (11).

\subsubsection{$\xi = \xi_S$}

For $\xi$ to be everywhere nonvanishing (and hence spacelike) we
require all $\theta_i$ to be nonzero.  This guarantees that the action
of $\Gamma$ is locally free.  To analyse the stabilisers it is
convenient to think of $S^7$ an embedded in $\CC^4$ with standard
coordinates $z_i$, $i=1,2,3,4$ as the quadric
\begin{equation}
  \label{eq:s7quadric}
  |z_1|^2 + |z_2|^2 + |z_3|^2 + |z_4|^2  = 4R^2~.
\end{equation}
The action of $\xi$ integrates to
\begin{equation*}
  (z_1,z_2,z_3,z_4) \mapsto (e^{i\theta_1 t} z_1, e^{i\theta_2 t} z_2,
  e^{i\theta_3 t} z_3, e^{i\theta_4 t} z_4)~.
\end{equation*}
The quotient will not be Hausdorff unless the integral curves are
periodic (compare with the discussion in
\cite[Section~3.2]{FigSimGrav}), whence there is a smallest positive
number $T$ for which
\begin{equation*}
  (e^{i\theta_1 T} z_1, e^{i\theta_2 T} z_2, e^{i\theta_3 T} z_3,
  e^{i\theta_4 T} z_4) = (z_1,z_2,z_3,z_4)~,
\end{equation*}
for all $(z_1,z_2,z_3,z_4)$.  This implies that there are integers
$p_i$ such that $\theta_i T = 2\pi p_i$ and since $T$ is the smallest
such positive integer, the $p_i$ are coprime.  Using the freedom
to rescale the Killing vector, we can take
\begin{equation*}
  \xi = p_1 R_{12} + p_2 R_{34} + p_3 R_{56} + p_4 R_{78}
  \quad\text{with $\gcd(p_1,p_2,p_3,p_4)=1$},
\end{equation*}
which integrates to a circle action
\begin{equation*}
  (z_1,z_2,z_3,z_4) \mapsto (e^{ip_1 t} z_1, e^{ip_2 t} z_2,
  e^{ip_3 t} z_3, e^{ip_4 t} z_4)~,
\end{equation*}
with period $2\pi$.  The action has trivial stabilisers
if for all $(z_1,z_2,z_3,z_4)$ obeying \eqref{eq:s7quadric}, the
equation
\begin{equation*}
  (z_1,z_2,z_3,z_4) = (e^{ip_1 t} z_1, e^{ip_2 t} z_2,
  e^{ip_3 t} z_3, e^{ip_4 t} z_4)
\end{equation*}
holds only for $t \in 2\pi\ZZ$.  By looking at the points
$(2R,0,0,0)$, $(0,2R,0,0)$, etc we see that $p_i$ have to be $\pm 1$
giving $2^4 = 16$ possibilities which, taking into account the action
of the Weyl group, are reduced to only two: $p_1=p_2=p_3=1$ and $p_4 =
\pm 1$.  For $p_4 =1$, the resulting quotient is $\CP^3$, which as we
saw above, admits a spin structure.  For $p_4=-1$, we also obtain
$\CP^3$ but with the opposite orientation.  Indeed, notice that the
two cases are related by an orientation-reversing isometry in $\O(8)$.
In either case we have a quotient with a spin structure. We shall now
discuss in which cases $\AdS_4 \times \CP^3$ backgrounds do preserve
some amount of supersymmetry \cite{DLP,DuffLuPopeUnt}.

Let
\begin{equation*}
  \xi^\pm  = R_{12} + R_{34} + R_{56} \pm R_{78}
\end{equation*}
generate a one-parameter subgroup $\Gamma^\pm \subset \Spin(8)$.  The
Killing spinors in $S^7$ are in the representation $\Delta^{8,0}_-$ of
$\Spin(8)$.  The weights of $\xi^+$ on $\Delta^{8,0}_-$ are $\pm1$
with multiplicity $4$, whereas those of $\xi^-$ are $(0,\pm2)$ with
multiplicities $6$ and $1$, respectively.  Therefore quotienting by
$\Gamma^+$ breaks all supersymmetry, whereas quotienting by $\Gamma^-$
preserves a fraction $\nu= \frac34$.

Alternatively, we can solve \eqref{eq:susyeq} explicitly and arrive at
the same conclusion.  Given the one-to-one correspondence between
Killing spinors on $S^7$ and negative chirality spinors on $\RR^8$, we
can manifestly evaluate \eqref{eq:susyeq} for the
particular case of $\xi=\xi_S$ being considered now.  This gives rise
to
\begin{equation*}
  \left(\Gamma_{12} + \Gamma_{34} + \Gamma_{56} \pm
    \Gamma_{78}\right)\varepsilon= 0~.
\end{equation*}
Since $\Gamma_{12}$ in invertible, the above equation is equivalent to
\begin{equation}
  \left(\Gamma_{1234} + \Gamma_{1256} \pm
  \Gamma_{1278}\right)\varepsilon= \varepsilon~,
 \label{eq:algs7}
\end{equation}
where the square of all antisymmetric combinations of gamma matrices
equals one. The solution to this kind of eigenvalue problem was
described in detail in \cite{MolSim}. There is a subtlety though,
which is worth mentioning. The solution $\varepsilon$ to equation
\eqref{eq:algs7} has to be found in the subspace of negative chirality
spinors. In this subspace, the three operators appearing in the left
hand side of equation \eqref{eq:algs7} are no longer independent.  In
particular, the third one can be written as a product of the first two
ones, that is, $\Gamma_{1278} = \Gamma_{1234}\Gamma_{1256}$. Under
these circumstances, as pointed out in \cite{HullGauntlett} and
generalised in \cite{MolSim}, preservation of an exotic amount of
supersymmetry is allowed.  An explicit matrix realisation for these
operators
\begin{equation*}
  \begin{aligned}[m]
    \Gamma_{1234} &= \text{diag}(1,\,1,\,-1,\,-1)\otimes \id \\
    \Gamma_{1256} &= \text{diag}(1,\,-1,\,1,\,-1)\otimes \id \\
    \Gamma_{1278} &= \text{diag}(1,\,-1,\,-1,\,1)\otimes \id~,
  \end{aligned}
\end{equation*}
defines an eigenvalue problem given by
\begin{equation*}
  \text{diag}(M-\lambda_1^\pm,\,M-\lambda_2^\pm,\,M-\lambda_3^\pm,\,
  M-\lambda_4^\pm)\otimes \id\, \varepsilon = 0~,
\end{equation*}
where $M=1$, as can be seen from \eqref{eq:algs7}, and $\lambda_i^\pm$
are the diagonal elements of the matrix appearing in the left hand
side of equation \eqref{eq:algs7} in the realisation given above. It
is straightforward to check that for $p_4=1$, non of the diagonal
elements of the above eigenvalue problem vanish. Thus, the
corresponding spinor $\varepsilon$ vanishes.  Therefore, $\CP^3$
breaks all the supersymmetry. On the other hand, when $p_4=-1$,
$\lambda_i^-=1$ $i=1,2,3$. Thus, there exists a non-trivial spinor
solving the eigenvalue problem. The amount of supersymmetry preserved
is, indeed, given by $\nu=3/4$, since three out of four of the
eigenvalues vanish.

We now turn our attention to the four possible families of reductions
where the anti-de~Sitter component of the Killing vector is nonzero.

\subsubsection{$\xi = \xi_A^{(4)} + \xi_S$}

This case is very similar to the previous one, with the possibility of
adding a spatial rotation $\xi_A^{(4)} = \varphi \be_{34}$ to $\xi$.
The same analysis as before shows that $\varphi T = 2\pi n$ for some
integer $n$ which is coprime to the $p_i$, whence after rescaling the
Killing vector it becomes
\begin{equation*}
  \xi = n \be_{34} +  p_1 R_{12} + p_2 R_{34} + p_3 R_{56} + p_4
  R_{78} \quad\text{with $\gcd(n, p_1,p_2,p_3,p_4)=1$}.
\end{equation*}
As before, the triviality of the stabilisers demands that $p_i=\pm 1$,
and hence up to rescaling $\xi$ and conjugating by the Weyl group of
$\SO(8)$ one has
\begin{equation*}
  \xi = n \be_{34} +  R_{12} + R_{34} + R_{56}  \pm  R_{78}
  \quad\text{with $n=1,2,\dots$.}
\end{equation*}
Let us employ the convention that $\gamma_i$ denote the
$\gamma$-matrices for $\Cl(2,3)$ and $\Gamma_a$ denote those of
$\Cl(8)$.  Then the element in $\fso(2,3) \oplus \fso(8) \subset
\Cl(2,11)$ corresponding to $\xi$ is given by
\begin{equation*}
  X = n \gamma_{34} + \Gamma_{12} + \Gamma_{34} + \Gamma_{56}  \pm
  \Gamma_{78}~.
\end{equation*}
Exponentiating this element we get
\begin{multline*}
  \widehat g(t) = \exp\frac{tX}2 = \left(\id \cos\frac{nt}2 +
    \gamma_{34} \sin \frac{nt}2\right) \left(\id \cos\frac{t}2 +
    \Gamma_{12} \sin\frac{t}2\right)\\
  \times \left(\id \cos\frac{t}2 + \Gamma_{34} \sin\frac{t}2\right)
         \left(\id \cos\frac{t}2 + \Gamma_{56} \sin\frac{t}2\right)
         \left(\id \cos\frac{t}2 \pm \Gamma_{78} \sin\frac{t}2\right)
\end{multline*}
and this can only possibly equal $-\id$ at $t=2\pi$, where
\begin{equation*}
  \widehat g(2\pi) = (-1)^n \id~;
\end{equation*}
whence only for $n$ even do we have a spin structure in the quotient.

The analysis of supersymmetry is analogous to the ones given in the
previous subsection.  The weights of the element
\begin{equation*}
  n \gamma_{34} + \Gamma_{12} + \Gamma_{34} + \Gamma_{56}  +
  \Gamma_{78}
\end{equation*}
on the representation carried by the Killing spinors are $\pm
\frac{n}2 \pm 1$ with uncorrelated signs and multiplicity $8$.  There
are $16$ zero weights when $n=2$ and none otherwise.  The resulting
supersymmetric quotient is therefore half-BPS.  The element
\begin{equation*}
    n \gamma_{34} + \Gamma_{12} + \Gamma_{34} + \Gamma_{56}  -
    \Gamma_{78}
\end{equation*}
has weights $\pm \frac{n}2$ with multiplicity $12$ and $\pm \frac{n}2
\pm 2$ with uncorrelated signs and multiplicity $2$.  Since $n\neq 0$,
there are $4$ zero weights when $n=4$ and none otherwise.  The
resulting supersymmetric quotient preserves a fraction $\nu = \frac18$
of the supersymmetry of the original background.

\subsubsection{$\xi = \xi_A^{(5)} + \xi_S$}

Here the Killing vector field is
\begin{equation*}
  \xi = \be_{13} - \be_{34} + \theta_1 R_{12} + \theta_2 R_{34} +
  \theta_3 R_{56} + \theta_4 R_{78}~.
\end{equation*}
This vector fields integrates to the action of $\Gamma \cong \RR$.
The orbits are generically non-periodic, but on the hyperbola in
$\AdS_4$ consisting of points $x_1=x_3=x_4=0$ and $x_2^2 = R^2 +
x_5^2$, the orbits are either not periodic, in which case they would
be dense in a submanifold of dimension at least $2$ and hence the
quotient would not be Hausdorff, or else they are periodic in which
case their points have nontrivial stabiliser, namely the subgroup
$\Gamma_0 \subset \Gamma$ consisting of the periods.  In either case
the quotient is not smooth.

\subsubsection{$\xi = \xi_A^{(6)} + \xi_S$}

In this case the Killing vector is
\begin{equation*}
  \xi = -\be_{12} - \be_{13} + \be_{24} + \be_{34} + \theta_1 R_{12} +
  \theta_2 R_{34} + \theta_3 R_{56} + \theta_4 R_{78}~,
\end{equation*}
where the $\theta_i$ are nonzero for $\xi$ to be everywhere
spacelike.  Notice that the anti-de~Sitter component is never zero,
although its norm vanishes.  It integrates to the following action on
$\RR^{2,3}$:
\begin{equation*}
  \begin{pmatrix}
    x^1\\x^2\\x^3\\x^4\\x^5
  \end{pmatrix}
  \mapsto
  \begin{pmatrix}
    x^1 + t (x^3 - x^2)\\
    x^2 + t (x^1 - x^4)\\
    x^3 + t (x^1 - x^4)\\
    x^4 + t (x^3 - x^2)\\
    x^5
  \end{pmatrix}
\end{equation*}
which are straight lines ruling the quadric in $\RR^{2,3}$ which is
locally isometric to $\AdS_4$.  The combined Killing vector $\xi$
therefore integrates to an action of $\Gamma \cong \RR$.  It is easy
to see that it has trivial stabilisers, since it does on $\AdS_4$.
Therefore the resulting background $(\AdS_4 \times S^7)/\Gamma$ is
smooth and, since $\Gamma \cong \RR$, also spin.

The action of $\xi$ on the representation carried by the Killing
spinors breaks up into a sum of commuting nilpotent and semisimple
pieces: $N + S$.  The $\Gamma$-invariant Killing spinors are precisely
those which are annihilated by both $N$ and $S$.

The nilpotent piece
\begin{equation*}
  N=\gamma_{13}-\gamma_{34} + \gamma_{12}-\gamma_{24}
\end{equation*}
in turn breaks up as a product of two anticommuting nilpotent
operators $N= N_1 N_2$, where
\begin{equation*}
  N_1 = \gamma_2 + \gamma_3 \qquad\text{and}\qquad N_2 = \gamma_1 +
  \gamma_4~.
\end{equation*}
$N$ acts on $\Delta^{2,3}$ and its kernel is the subspace generated by
the kernels of $N_1$ and $N_2$.  It is three-dimensional.

The semisimple piece is given by
\begin{equation*}
  S = \theta_1 \Gamma_{12} + \theta_2 \Gamma_{34} + \theta_3
  \Gamma_{56} + \theta_4 \Gamma_{78}
\end{equation*}
and its action on $\Delta^{8,0}_-$ is determined by the weights of
that representation: $\pm \theta_1 \pm \theta_2 \pm \theta_3 \pm
\theta_4$ with uncorrelated signs whose product is negative and with
multiplicity one.  The existence of zero weights define hyperplanes in
the space of the $\theta_i$.  Taking into account the ordering of the
$\theta_i$ and the fact $\theta_{1,2,3}$ are positive, we find that
there are three possible hyperplanes:
\begin{itemize}
\item $\theta_1 - \theta_2 + \theta_3 + \theta_4 = 0$;
\item $\theta_1 + \theta_2 - \theta_3 + \theta_4 = 0$; and
\item $\theta_1 - \theta_2 - \theta_3 - \theta_4 = 0$.
\end{itemize}
Whenever the $\theta_i$ belong to one (and only one) of these
hyperplanes, there is a two-dimensional subspace of $\Delta^{8,0}_-$
which is annihilated by $S$.  The resulting quotient therefore has six
supersymmetries, and hence preserves a fraction $\nu = \frac3{16}$ of the
supersymmetry of the eleven-dimensional vacuum.  When the $\theta_i$
belong to the intersection of the first and third hyperplanes, which
happens when $\theta_1 = \theta_2$ and $\theta_3 = - \theta_4$, the
kernel of $S$ is four-dimensional.  The resulting quotient has $12$
supersymmetries, and hence preserves a fraction $\nu = \frac38$ of the
supersymmetry of the eleven-dimensional vacuum.  Finally when
$\theta_i$ lies in the intersection of all three hyperplanes, which
happens when $\theta_1=\theta_2=\theta_3= -\theta_4$, the kernel of
$S$ is eight-dimensional and the resulting quotient has $18$
supersymmetries, corresponding to a fraction $\nu = \frac9{16}$.

\subsubsection{$\xi = \xi_A^{(11)} + \xi_S$}

We let $\beta_1 = \beta_2 = \beta$ in case (11) to arrive at the
Killing vector
\begin{equation*}
  \xi = \beta( \be_{13} + \be_{24}) + \theta_1 R_{12} +
  \theta_2 R_{34} + \theta_3 R_{56} + \theta_4 R_{78}~,
\end{equation*}
where now the $\theta_i$ are allowed to be zero provided that $\beta$
is not.  Since we already discussed the case $\beta=0$, we will now
concentrate on $\beta\neq 0$, whence we can rescale $\xi$ such that
$\beta=1$.  The anti-de~Sitter component integrates to the following
action on $\RR^{2,3}$:
\begin{equation*}
  \begin{pmatrix}
    x^1\\x^2\\x^3\\x^4\\x^5
  \end{pmatrix}
  \mapsto
  \begin{pmatrix}
    x^1 \cosh t  - x^3 \sinh t \\
    x^2 \cosh t - x^4 \sinh t \\
    x^3 \cosh t  - x^1 \sinh t \\
    x^4 \cosh t  - x^2 \sinh t \\
    x^5
  \end{pmatrix}
\end{equation*}
which is clearly a free action of $\Gamma \cong \RR$ on the quadric
and hence already on the simply-connected $\AdS_4$ for all values of
$\theta_i$.  The resulting background $(\AdS_4 \times S^7)/\Gamma$ is
therefore smooth and, since $\Gamma \cong \RR$, also spin.

The action of $\xi$ on the Killing spinors breaks up into two
commuting semisimple pieces $S_1 + S_2$, where
\begin{equation*}
  S_1 = \gamma_{13} + \gamma_{24}
\end{equation*}
and
\begin{equation*}
  S_2 =  \theta_1 \Gamma_{12} + \theta_2 \Gamma_{34} + \theta_3
  \Gamma_{56} + \theta_4 \Gamma_{78}
\end{equation*}
is just as before, except that now the $\theta_i$ are allowed to
vanish.  If we complexify the Killing spinors we may diagonalise $S_1$
and $S_2$: with $S_1$ having real eigenvalues and $S_2$ imaginary
eigenvalues.  Therefore a spinor is annihilated by $S_1+S_2$ if and
only if it is annihilated by $S_1$ and $S_2$ separately.

The kernel of $S_1$ on $\Delta^{2,3}$ is clearly two-dimensional,
since it corresponds to those spinors $\varepsilon$ obeying
$\gamma_{1234}\, \varepsilon= \varepsilon$.

The analysis of $S_2$ is similar to that in the previous section
except for the possibility that one or more of the $\theta_i$ may
vanish.  We have the same cases we had before, but where the fractions
of supersymmetry are now two thirds of the previous fractions, namely
$\nu=\frac18$, $\nu=\frac14$ and $\nu=\frac38$.  In addition we have
the case in which all $\theta_i$ vanish, which corresponds to a
half-BPS quotient.

\subsubsection{Summary}

The smooth spacelike supersymmetric quotients of $\AdS_4 \times S^7$
are summarised in Table~\ref{tab:AdS4S7}, where we have also indicated
the fraction $\nu$ of the eleven-dimensional supersymmetry preserved
by the quotient.

\begin{table}[h!]
  \centering
  \setlength{\extrarowheight}{3pt}
  \renewcommand{\arraystretch}{1.3}
  \begin{tabular}{|>{$}l<{$}|>{$}l<{$}|>{$}l<{$}|>{$}c<{$}|}\hline
    & \multicolumn{2}{c|}{Conditions for} & \\
    \multicolumn{1}{|c|}{Killing vector} & 
    \multicolumn{1}{c|}{Causality/Weyl} &
    \multicolumn{1}{c|}{Supersymmetry} & 
    \multicolumn{1}{c|}{$\nu$}\\
      \hline\hline
      \xi_S = \theta_1 R_{12} + \theta_2 R_{34} & & & \\
      \hfill + \theta_3 R_{56} + \theta_4 R_{78} & \theta_1 \geq
      \theta_2 \geq \theta_3 \geq |\theta_4| > 0 & 
      \text{$\theta_{1,2,3} = 1$, $\theta_4=-1$} & \frac34\\ \hline
      \varphi \be_{34} + \xi_S & \varphi \geq 0 & \text{$\varphi=2$,
        $\theta_i=1$} & \half\\
      & \theta_1 \geq \theta_2 \geq \theta_3 \geq |\theta_4| > 0 &
      \text{$\varphi=4$, $\theta_{1,2,3}=1$, $\theta_4=-1$} &
      \frac18\\ \hline
      -\be_{12} - \be_{13} & \theta_1 \geq \theta_2 \geq \theta_3 \geq
      |\theta_4| > 0 & \theta_1 \mp \theta_2 \pm \theta_3 = - \theta_4
      & \frac3{16}\\
      \hfill + \be_{24} + \be_{34} + \xi_S & & \theta_1 - \theta_2 -
      \theta_3 = \theta_4 & \frac3{16}\\
      & & \text{$\theta_1=\theta_2$, $\theta_3 = - \theta_4$} &
      \frac38 \\
      & & \theta_1 = \theta_2 = \theta_3 = - \theta_4 & \frac9{16}\\
      \hline
      \be_{13} + \be_{24} + \xi_S & \theta_1 \geq \theta_2 \geq
      \theta_3 \geq |\theta_4| \geq 0 & \theta_1 \mp \theta_2 \pm
      \theta_3 = - \theta_4 & \frac18 \\
      & & \theta_1 - \theta_2 - \theta_3 = \theta_4 & \frac18\\
      & & \text{$\theta_1=\theta_2$, $\theta_3 = - \theta_4$} & \frac14
      \\
      & & \theta_1 = \theta_2 = \theta_3 = - \theta_4 & \frac38\\
      & & \theta_i = 0 & \half\\ \hline
    \end{tabular}
  \vspace{8pt}
  \caption{Smooth spacelike supersymmetric quotients of $\AdS_4 \times
    S^7$}
  \label{tab:AdS4S7}
\end{table}

\subsection{Supersymmetric quotients of $\AdS_5 \times S^5$}
\label{sec:ads5s5}

In the $\AdS_5 \times S^5$ vacuum of ten-dimensional type IIB
supergravity, both spaces have equal radii of curvature $R$.  Letting
$s>0$ be the scalar curvature of the sphere, the selfdual $5$-form is
given by
\begin{equation*}
  F = \sqrt{\tfrac1{80} s} \left( \dvol_A + \dvol_S \right)~,
\end{equation*}
where $\dvol_A$ and $\dvol_S$ are the volume forms of $\AdS_5$ and
$S^5$, respectively.  (The bizarre-looking factor of $80$ is a
consequence of our chosen value for the (constant) dilaton.)  This
vacuum solution has thirty-two supersymmetries.  In this section we
will study the smooth supersymmetric quotients.

The situation here is very close to the one in the previous sections.
Since the sphere is odd-dimensional, it also admits Killing vectors
which are bounded below by a positive number and therefore we will be
able to admit Killing vectors in $\AdS_5$ which are not necessarily
everywhere spacelike, provided their norms are bounded below.  From
the results in Section~\ref{sec:so24}, the only Killing vectors
$\xi_A$ on $\AdS_5$ which satisfy this condition are those labelled
(4), (5), (6), (11) for $\beta_1 = \beta_2$, (16), (17), (20) for
$|\varphi_2| \geq \varphi_1 > 0$, (21), (23), (24) for $\varphi_2 \geq
\varphi_3 \geq |\varphi_1|>0$, and (25) for $\beta_1 = \beta_2$.  We
now analyse each case in more detail.

Throughout this section the spherical component $\xi_S$ of the Killing
vector is given by
\begin{equation*}
  \xi_S  = \theta_1 R_{12} + \theta_2 R_{34} + \theta_3 R_{56}~,
\end{equation*}
whose norm obeys $|\xi_S|^2 \geq \min_i\theta_i^2 R^2$, where $R$ is
the radius of curvature of both $S^5$ and $\AdS_5$.  The action of the
Weyl group allows us to order them as $\theta_1 \geq \theta_2 \geq
|\theta_3|$.

\subsubsection{$\xi=\xi_S$}

This case is very similar to the analogous reduction of $\AdS_4 \times
S^7$, so we will be brief.  The action is locally free provided all
$\theta_i$ are nonzero.  To analyse the stabilisers it is
convenient to think of $S^5$ as embedded in $\CC^3$ with standard
coordinates $z_i$, $i=1,2,3$, as the quadric
\begin{equation*}
  |z_1|^2 + |z_2|^2 + |z_3|^2 = R^2~.
\end{equation*}
The action of $\xi$ integrates to
\begin{equation*}
  (z_1,z_2,z_3) \mapsto (e^{i\theta_1 t} z_1, e^{i\theta_2 t} z_2,
  e^{i\theta_3 t} z_3)~,
\end{equation*}
where we demand periodicity for the quotient to be Hausdorff.  Let $T$
be the period, so that there are integers $p_i$ such that $\theta_i T
= 2\pi p_i$ and since $T$ is the smallest such positive integer, the
$p_i$ are coprime.  Using the freedom to rescale the Killing vector,
we can take
\begin{equation*}
  \xi = p_1 R_{12} + p_2 R_{34} + p_3 R_{56}
  \quad\text{with $\gcd(p_1,p_2,p_3)=1$},
\end{equation*}
which integrates to a circle action
\begin{equation*}
  (z_1,z_2,z_3) \mapsto (e^{ip_1 t} z_1, e^{ip_2 t} z_2,
  e^{ip_3 t} z_3)~
\end{equation*}
with period $2\pi$.  Demanding that the stabilisers be trivial one
again obtains that the $p_i$ have to be $\pm 1$ giving $2^4 = 8$
possibilities which, taking into account the action of the Weyl group,
are reduced to two: $p_1=p_2= \pm p_3=1$.  The resulting quotient is
in either case $\CP^2$ (although with opposite orientations) which, as
recalled in Section~\ref{sec:spin}, is not spin, whence these reductions
do not give rise to any geometrically realised supersymmetry.  The
same conclusion can be reached by computing the weights of $\xi_S$ on
the representation
\begin{equation*}
  \left( \Delta^{2,4} \otimes \Delta^{6,0} \right) \oplus
  \left( \overline\Delta^{2,4} \otimes \overline\Delta^{6,0} \right)
\end{equation*} 
of (complexified) Killing spinors, where the bar denotes complex
conjugation.  From now on we will adopt the notation $\Delta^{2,4}_+$
and $\Delta^{6,0}_+$ for $\Delta^{2,4}$ and $\Delta^{6,0}$,
respectively, and $\Delta^{2,4}_-$ and $\Delta^{6,0}_-$ for the
complex conjugate representations.  The weights are $\pm\frac32$ and
$\pm\frac12$ with multiplicities $4$ and $12$, respectively.  This
calculation serves to illustrate the general discussion of
Section~\ref{sec:susy} about the fact that not only is $\CP^2$ not
spin, but indeed all spinors in the quotient spacetime are charged
under the circle subgroup associated with the compact dimension.

We now turn our attention to the eleven possible families of
reductions where the anti-de~Sitter component of the Killing vector is
present.

\subsubsection{$\xi=\xi^{(I)}_A + \xi_S$, for $I=4,16$}

Both of these cases can be treated simultaneously since (4) is a
specialisation of (16) where one of the angles of rotation in $\AdS_5$
is put to zero.  The analysis is very similar to the one above: it
shows that for $j=1,2$, $\varphi_j T = 2\pi n_j$ for some integers
$n_j$ which are coprime to the $p_i$, whence after rescaling the
Killing vector it becomes
\begin{equation*}
  \xi = n_1 \be_{34} +  n_2 \be_{56} + p_1 R_{12} + p_2 R_{34} + p_3
  R_{56}~.
\end{equation*}
As before, the triviality of the stabilisers demands that $p_i=\pm 1$,
and hence rescaling $\xi$ and up to conjugation by the Weyl group one
has
\begin{equation*}
  \xi^\pm = n_1 \be_{34} +  n_2 \be_{56} + R_{12} + R_{34} \pm R_{56}
  \quad\text{with $n_1 \geq n_2 =0,1,2,\dots$}
\end{equation*}
The case where $n_1 = n_2 = 0$ was treated above, so we will assume
that they cannot both be zero.  Let us once again employ the
convention that $\gamma_i$ denote the $\gamma$-matrices for $\Cl(2,4)$
and $\Gamma_a$ denote those of $\Cl(6)$.  Then the element in
$\fso(2,4) \oplus \fso(6) \subset \Cl(2,10)$ corresponding to $\xi$ is
given by
\begin{equation*}
  X = n_1 \gamma_{34} + n_2 \gamma_{56} + \Gamma_{12} + \Gamma_{34}
  \pm \Gamma_{56}~.
\end{equation*}
Exponentiating this element we get
\begin{multline*}
  \widehat g(t) = \exp\frac{tX}2 = \left(\id \cos\frac{n_1t}2 +
    \gamma_{34} \sin \frac{n_1t}2\right) \left(\id \cos\frac{n_2t}2 +
    \gamma_{56} \sin \frac{n_2t}2\right)\\
  \times \left(\id \cos\frac{t}2 + \Gamma_{12} \sin\frac{t}2\right)
  \left(\id \cos\frac{t}2 + \Gamma_{34} \sin\frac{t}2\right)
  \left(\id \cos\frac{t}2 \pm \Gamma_{56} \sin\frac{t}2\right)
\end{multline*}
and this can only possibly be $-\id$ at $t=2\pi$, where we get
\begin{equation*}
  \widehat g(2\pi) = (-1)^{n_1+n_2+1} \id~;
\end{equation*}
whence only for $n_1 + n_2$ odd do we have a spin structure in the
quotient.

In order to analyse the amount of supersymmetry preserved by this set
of quotients, we shall the action of $X$ on the Killing spinors.  The
weights of $X$ in that representation are given by
\begin{equation*}
  \pm \frac{n_1}2 \pm \frac{n_2}2 \pm \frac32 \qquad \text{and} \qquad
  \pm \frac{n_1}2 \pm \frac{n_2}2 \pm \frac12
\end{equation*}
with uncorrelated signs and with multiplicities $1$ and $3$,
respectively.

It is now straightforward to find out for which values of $n_1 \geq
n_2 =0,1,2,\dots$ are there zero weights and how many.  One finds the
following values for $(n_1,n_2)$:
\begin{itemize}
\item $(1,0)$ with $12$ zero weights;
\item $(2,1)$ with $8$;
\item $(n+1,n>1)$ with $6$;
\item $(3,0)$ with $4$; and
\item $(n+3,n>0)$ with $2$.
\end{itemize}
The resulting quotients have fractions $\nu=\frac38$, $\nu=\frac14$,
$\nu=\frac3{16}$, $\nu=\frac18$ and $\nu=\frac1{16}$, respectively.

\subsubsection{$\xi=\xi^{(I)}_A + \xi_S$, for $I=5,17$}

These cases can again be treated simultaneously because (5) is the
specialisation of (17) in which the angle of the spatial rotation  is
put to zero.  Just as in the case of $\AdS_4 \times S^7$ this case
does no lead to any smooth reductions due to some points having
nontrivial stabiliser: there are non-periodic orbits, whence $\xi$
integrates to an $\RR$-action, but there exist periodic orbits at
whose points the stabiliser is not trivial.

\subsubsection{$\xi=\xi^{(I)}_A + \xi_S$ for $I=6,21,20$}

These cases are all specialisations of (20) where one or both of the
rotation angles in the $\AdS$ space are put to zero.

The Killing vector is
\begin{equation*}
  \xi = -\be_{12} - \be_{13} + \be_{24} + \be_{34} +
  \varphi_1(\be_{12} + \be_{34}) + \varphi_2 \be_{56}
  + \theta_1 R_{12} + \theta_2 R_{34} + \theta_3 R_{56}~,
\end{equation*}
with $\varphi_2 \geq |\varphi_1| > 0$ and $\theta_i > |\varphi_1|$
for all $i=1,2,3$.  The anti-de~Sitter component breaks up into two
\emph{commuting} vector fields: $-\be_{12} - \be_{13} + \be_{24} +
\be_{34}$ and $\varphi_1(\be_{12} + \be_{34}) + \varphi_2 \be_{56}$;
whence the combined vector field integrates to the following
$\RR$-action on $\RR^{2,4}$:
\begin{equation*}
  \begin{pmatrix}
    x^1\\x^2\\x^3\\x^4\\x^5\\x^6
  \end{pmatrix}
  \mapsto
  \begin{pmatrix}
    x^1 \cos\varphi t  + x^2 \sin\varphi t +
    t\left(\left(x^1-x^4\right)\,\sin\varphi t
      + \left(-x^2 + x^3\right)\,\cos\varphi t\right) \\
    x^2 \cos\varphi t  - x^1 \sin\varphi t +
    t\left(\left(x^1-x^4\right)\,\cos\varphi t +
      \left(x^2-x^3\right)\,\sin\varphi t\right) \\
    x^3 \cos\varphi t  - x^4 \sin\varphi t +
    t\left(\left(x^1-x^4\right)\,\cos\varphi t +
      \left(x^2-x^3\right)\,\sin\varphi t\right) \\
    x^4 \cos\varphi t  + x^3 \sin\varphi t +
    t\left(\left(x^1-x^4\right)\,\sin\varphi t +
      \left(-x^2 + x^3\right)\,\cos\varphi t\right) \\
    x^5\cos \varphi_2 t - x^6 \sin \varphi_2 t\\
    x^6\cos \varphi_2 t + x^5 \sin \varphi_2 t
  \end{pmatrix}~.
\end{equation*}
It is clear that there are no fixed points and that the stabilisers
are trivial in the quadric and hence also in the universal cover
$\AdS_5$, whence the quotient is smooth and, since $\Gamma\cong \RR$,
also spin.  Notice that in case (6), when both $\varphi_1=\varphi_2=0$
the resulting orbits are straight lines ruling the quadric.

The action of $\xi$ on the (complexified) Killing spinors breaks up
into a nilpotent piece
\begin{equation*}
  N = - \gamma_{12} - \gamma_{13} + \gamma_{24} + \gamma_{34}
\end{equation*}
and a semisimple piece
\begin{equation*}
  S = \varphi_1 (\gamma_{12} + \gamma_{34} ) + \varphi_2 \gamma_{56} +
  \theta_1 \Gamma_{12} + \theta_2 \Gamma_{34} + \theta_3 \Gamma_{56}~,
\end{equation*}
whence the invariant Killing spinors lie in the kernel of both $N$ and
$S$.  The analysis of the kernel of $N$ is very similar to the one
carried out in Section~\ref{sec:N=N1N2}.  Indeed, $N$ is essentially
what we called $N^+$ there.  Let us decompose
\begin{equation*}
  \Delta^{2,4}_\pm = \left( \Delta^{2,2}_+ \otimes \Delta^{0,2}_\pm
  \right) \oplus  \left( \Delta^{2,2}_- \otimes \Delta^{0,2}_\mp
  \right)
\end{equation*}
and observe that as in Section~\ref{sec:N=N1N2}, $N$ acts trivially in
$\Delta^{2,2}_-$ and has a one-dimensional kernel in $\Delta^{2,2}_+$
on which the term $\varphi_1 (\gamma_{12} + \gamma_{34})$ acts
trivially.  It is then easy to compute the weights of $S$ on the
kernel of $N$.  They are given by the weights of
\begin{equation*}
 \varphi_2 \gamma_{56} + \theta_1 \Gamma_{12} + \theta_2 \Gamma_{34} +
 \theta_3 \Gamma_{56}
\end{equation*}
acting on
\begin{equation*}
 \left( \Delta^{0,2}_+ \otimes \Delta^{6,0}_+ \right) \oplus 
 \left( \Delta^{0,2}_- \otimes \Delta^{6,0}_- \right)
\end{equation*}
and the weights of $S$ acting on
\begin{equation*}
 \left( \Delta^{2,2}_- \otimes \Delta^{0,2}_- \otimes \Delta^{6,0}_+
 \right) \oplus 
 \left(  \Delta^{2,2}_- \otimes \Delta^{0,2}_+ \otimes \Delta^{6,0}_-
 \right)~.
\end{equation*}
The calculation is routine and we obtain the following $24$ weights
\begin{align*}
  & \pm\half \varphi_2 \pm\half \theta_1 \pm \half \theta_2 \pm\half
  \theta_3\\
   \varphi_1 & \pm\half \varphi_2 \pm\half \theta_1 \pm \half
  \theta_2 \pm\half \theta_3\\
  -\varphi_1 & \pm\half \varphi_2 \pm\half \theta_1 \pm \half
  \theta_2 \pm\half \theta_3
\end{align*}
where the product of the signs is always negative for a total of $8$
weights per line.  These $24$ weights define $12$ hyperplanes.  For
parameters lying in one and only one such hyperplane there are two
invariant Killing spinors.  Three of the hyperplanes contain no points
due to the conditions on the parameters coming from causality and the
ordering due to the action of the Weyl group.  The other nine
hyperplanes are defined by the conditions:
\begin{equation}
  \label{eq:634h1}
  \begin{aligned}
    \varphi_2 \pm (\theta_1 - \theta_2) + \theta_3 &= 0\\
    \varphi_2 - \theta_1 - \theta_2 - \theta_3 &= 0\\
    \varphi_2 \pm 2 \varphi_1 - \theta_1 - \theta_2 + \theta_3 &= 0\\
    \varphi_2 \pm 2 \varphi_1 \pm (\theta_1 - \theta_2) - \theta_3 &=
    0~,
  \end{aligned}
\end{equation}
where the signs in the last line are uncorrelated.  Each of these
reductions preserve a fraction $\nu= \frac1{16}$ of the
ten-dimensional supersymmetry.

The analysis of the possible intersections of these hyperplanes has
been done with the help of \textsc{Mathematica}.  The following
conditions, where each line is one condition, guarantee that the
parameters lie in the intersection of precisely two hyperplanes:
\begin{equation}
  \label{eq:634h2}
  \begin{aligned}
    \theta_1 &= \theta_2  & \qquad \theta_3 &= - \varphi_2\\
    \theta_1 &= \theta_2  & \qquad \theta_3 &= \pm 2\varphi_1 + \varphi_2\\
    \theta_1 &= \varphi_2  & \qquad \theta_2 &= - \theta_3\\
    \theta_1 \pm \varphi_1 &= \theta_2 \pm \theta_3 & \qquad 
     \varphi_2 &= - \varphi_1\\
    \theta_1 \pm \varphi_1 &= \theta_2 \mp \theta_3 & \qquad 
     \varphi_2 &= \varphi_1\\
    \theta_1 + \varphi_1 &= - \theta_2 - \theta_3 & \qquad 
     \varphi_2 &= \pm \varphi_1\\
    \theta_1 &= \theta_2 \pm 2\varphi_1 & \qquad 
    \theta_3 &= \varphi_2\\
    \theta_1 &= \theta_3 \pm 2\varphi_1 & \qquad
    \theta_2 &= \varphi_2\\
    \theta_1 &= \varphi_2 \pm 2\varphi_1 & \qquad
    \theta_2 &= \theta_3\\
    \theta_2 &= \theta_3 \pm 2\varphi_1 & \qquad
    \theta_1 &= \varphi_2~.
  \end{aligned}
\end{equation}
Each of these reductions preserve $\tfrac18$ of the ten-dimensional
supersymmetry.

Similarly, parameters satisfying the following conditions lie in
precisely three hyperplanes:
\begin{equation}
  \label{eq:634h3}
  \begin{aligned}
    \theta_1 &= \theta_2 = \pm \theta_3 = \varphi_2\\
    \theta_1 &= \theta_2 = \theta_3 = \pm 2 \varphi_1 + \varphi_2\\
    \theta_1 &= \theta_2 = \theta_3 \pm 2 \varphi_1 = \varphi_2\\
    \theta_1 \pm 2 \varphi_1 &= \theta_2 = \theta_3 = \varphi_2~.
  \end{aligned}
\end{equation}
Each of these reductions preserves $\frac3{16}$ of the supersymmetry.
The causality constraints on the parameters, which do not allow them
to vanish, forbid intersections of four or more hyperplanes.

\subsubsection{$\xi=\xi^{(I)}_A + \xi_S$ for $I=11,25$}

These two cases are treated simultaneously because (11) is the
specialisation of (25) where $\varphi =0$.  The Killing vector is
\begin{equation*}
  \xi = \beta( \be_{13} + \be_{24}) + \varphi \be_{56} + \theta_1
  R_{12} + \theta_2 R_{34} + \theta_3 R_{56}~,
\end{equation*}
where $\varphi \geq 0$ and where $\theta_1 \geq \theta_2 \geq
|\theta_3|$ are now allowed to be zero provided that $\beta$ is not.
The case with $\beta = 0$ was treated before, so we will assume
$\beta\neq 0$ and hence rescale $\xi$ such that $\beta=1$.  The
anti-de~Sitter component integrates to the following action on
$\RR^{2,4}$:
\begin{equation*}
  \begin{pmatrix}
    x^1\\x^2\\x^3\\x^4\\x^5\\x^6
  \end{pmatrix}
  \mapsto
  \begin{pmatrix}
    x^1 \cosh t  - x^3 \sinh t \\
    x^2 \cosh t  - x^4 \sinh t \\
    x^3 \cosh t  - x^1 \sinh t \\
    x^4 \cosh t  - x^2 \sinh t \\
    x^5\cos \varphi t - x^6 \sin \varphi t\\
    x^6\cos \varphi t + x^5 \sin \varphi t
  \end{pmatrix}
\end{equation*}
which is clearly a free action of $\Gamma \cong \RR$ on the quadric
and hence on $\AdS_5 \times S^5$ for all values of $\theta_i$.  The
resulting quotient is therefore smooth and, since $\Gamma \cong \RR$,
also spin.

To analyse the supersymmetry, notice that the action of $\xi$ on the
(complexified) Killing spinors breaks up into two commuting semisimple
pieces:
\begin{equation*}
  S_1 = \gamma_{13} + \gamma_{24}~,
\end{equation*}
having real eigenvalues, and
\begin{equation*}
  S_2 = \varphi \gamma_{56} + \theta_1 \Gamma_{12} + \theta_2
  \Gamma_{34} + \theta_3 \Gamma_{56}~,
\end{equation*}
having imaginary eigenvalues.  Therefore a Killing spinor is
invariant under $\xi$ if and only if it is simultaneously annihilated
by $S_1$ and $S_2$.  The kernel of $S_1$ consists of those spinors
satisfying $\gamma_{1234} \varepsilon = \varepsilon$, or equivalently
$\gamma_{56} \varepsilon = \pm i \varepsilon$ in $\Delta^{2,4}_\pm$.
The weights of $S_2$ on the kernel of $S_1$ are easy to compute, and
one obtains
\begin{equation*}
  \pm \half \varphi \pm \half \theta_1 \pm \half \theta_2 \pm \half
  \theta_3
\end{equation*}
with uncorrelated signs whose product is negative.  Their vanishing
determines four hyperplanes:
\begin{equation}
  \label{eq:635h1}
  \begin{aligned}
    \varphi + \theta_1 + \theta_2 - \theta_3 &= 0\\
    \varphi - \theta_1 - \theta_2 - \theta_3 &= 0\\
    \varphi + \theta_1 - \theta_2 + \theta_3 &= 0\\
    \varphi - \theta_1 + \theta_2 + \theta_3 &= 0~.
  \end{aligned}
\end{equation}
Whenever the parameters are such that they lie on precisely one of
these hyperplanes, the reduction preserves four supersymmetries.
These reductions preserve $\frac18$ of the supersymmetry.  This
happens whenever the second, third or fourth equations are satisfied,
since the first equation implies that $\varphi = \theta_1 =
\theta_2 = \theta_3 = 0$ due to the constraints on these parameters,
and these values satisfy all four equations.

Whenever the following conditions are satisfied, the parameters belong
to precisely two such hyperplanes and the reduction preserves
$\frac14$ of the supersymmetry:
\begin{equation}
  \label{eq:635h2}
  \begin{aligned}
    \varphi &= \theta_1  \neq \theta_2 = - \theta_3\\
    \theta_1 &= \theta_2 \neq  \varphi = - \theta_3~.
  \end{aligned}
\end{equation}
Similarly, whenever the condition
\begin{equation*}
  \theta_1 = \theta_2 = \varphi = - \theta_3 \neq 0
\end{equation*}
are met, the parameters belong to precisely three such hyperplanes and
the reduction preserves $\frac38$ of the supersymmetry.  Finally, if
all the parameters vanish, we obtain a half-BPS reduction.

\subsubsection{$\xi=\xi^{(23)}_A + \xi_S$}
\label{sec:doublenull}

The Killing vector in this case is
\begin{equation*}
  \xi = \be_{13} - \be_{34} + \be_{25} - \be_{56} + \theta_1 R_{12} +
  \theta_2 R_{34} + \theta_3 R_{56}~.
\end{equation*}
The anti-de~Sitter component integrates to the following $\RR$-action
on $\RR^{2,4}$:
\begin{equation*}
  \begin{pmatrix}
    x^1\\x^2\\x^3\\x^4\\x^5\\x^6
  \end{pmatrix}
  \mapsto
  \begin{pmatrix}
    x^1 - t x^3 + \half t^2 (x^1-x^4)\\
    x^2 - t x^5 + \half t^2 (x^2-x^6)\\
    x^3 + t (x^4 - x^1)\\
    x^4 - t x^3 + \half t^2 (x^1 - x^4) \\
    x^5 + t (x^6 - x^2)\\
    x^6 - t x^5 + \half t^2 (x^2 - x^6)
  \end{pmatrix}~,
\end{equation*}
which is clearly free of fixed points on $\AdS_5$ and has trivial
stabilisers.  The resulting quotient $(\AdS_5 \times S^5)/\Gamma$ is
therefore smooth and, since $\Gamma\cong \RR$, it is also spin.

To analyse the supersymmetry of this class of quotients, notice that
the action of $\xi$ on the (complexified) Killing spinors splits into
two commuting pieces: one nilpotent,
\begin{equation*}
  N = \gamma_{13} - \gamma_{34} + \gamma_{25} - \gamma_{56}~,
\end{equation*}
and one semisimple
\begin{equation*}
  S = \theta_1\Gamma_{12} + \theta_2 \Gamma_{34} + \theta_3
  \Gamma_{56}~.
\end{equation*}
The nilpotent operator is the spinorial representation of a double
null rotation and we will now show that it preserves one-half of the
AdS supersymmetry.  Let us rewrite $N$ as
\begin{equation*}
  N = (\gamma_1 + \gamma_4) \gamma_3 + (\gamma_2 + \gamma_6)
  \gamma_5~,
\end{equation*}
which suggests the following decomposition of the spinorial
representations $\Delta^{2,4}_\pm$:
\begin{equation*}
  \Delta^{2,4}_\pm \cong V_\pm \oplus (\gamma_1 + \gamma_4) V_\mp
  \oplus (\gamma_2 + \gamma_6) V_\mp \oplus (\gamma_1 +
  \gamma_4)(\gamma_2 + \gamma_6) V_\pm~,
\end{equation*}
where $V_\pm \subset \Delta^{2,4}_\pm$ is the subspace
\begin{equation*}
  V_\pm = \ker (\gamma_1 - \gamma_4) \cap \ker (\gamma_2 -
  \gamma_6)~.
\end{equation*}
This decomposition is nothing else but the decomposition of the
Clifford modules $\Delta^{2,4}_\pm$ as fermionic Fock spaces
associated to the annihilation and creation operators $\gamma_1 \pm
\gamma_4$ and $\gamma_2 \pm \gamma_5$, with $V_\pm$ the
one-dimensional subspace spanned by the Clifford vacua.  Accordingly,
every $\varepsilon \in \Delta^{2,4}_\pm$ has a unique decomposition of
the form
\begin{equation*}
  \varepsilon = \varepsilon_1 + (\gamma_1 + \gamma_4) \varepsilon_2 +
  (\gamma_2 + \gamma_6) \varepsilon_3 + (\gamma_1 + \gamma_4)(\gamma_2
  + \gamma_6) \varepsilon_4~,
\end{equation*}
where $\varepsilon_1,\varepsilon_4 \in V_\pm$ and
$\varepsilon_2,\varepsilon_3 \in V_\mp$.  It is now clear that
\begin{equation*}
  N\varepsilon = 0  \iff \varepsilon_1 = 0 \quad\text{and}\quad
  \varepsilon_3 = \gamma_{35} \varepsilon_2~.
\end{equation*}
Notice that if $\varepsilon\in V_\pm$ then $\gamma_{35}\varepsilon =
\pm i \varepsilon$, whence we can write the most general spinor
$\varepsilon \in \Delta^{2,4}_\pm$ in the kernel of $N$ as
\begin{equation*}
 \varepsilon = \left((\gamma_1 + \gamma_4) \mp i (\gamma_2 +
   \gamma_6)\right) \varepsilon_2 + (\gamma_1 + \gamma_4)(\gamma_2 +
 \gamma_6) \varepsilon_4~,
\end{equation*}
whence $\ker N \subset \Delta^{2,4}_\pm$ is two-dimensional.

The calculation of weights of the semisimple piece $S$ on $\ker N$ is
routine, and we obtain the following eight weights,
\begin{equation*}
  \pm \half \theta_1   \pm \half \theta_2  \pm \half \theta_3~,
\end{equation*}
with uncorrelated signs and each with multiplicity $2$.  Supersymmetry
requires the existence of zero weights.  These conditions determine
four hyperplanes
\begin{equation*}
  \theta_1 \pm \theta_2 \pm \theta_3 = 0~,
\end{equation*}
and if the parameters lie in precisely one of these hyperplanes, the
resulting quotient has four supersymmetries.  Two of the hyperplanes
have no points due to the conditions on the parameters.  The remaining
two hyperplanes are
\begin{equation*}
  \theta_1  - \theta_2 \pm \theta_3 = 0~,
\end{equation*}
and the resulting quotients preserve a fraction $\nu = \tfrac18$ of
the supersymmetry.  The intersection of these two hyperplanes consists
of parameters satisfying
\begin{equation*}
  \theta_1 = \theta_2 \qquad\text{and}\qquad \theta_3 = 0~.
\end{equation*}
Such reductions preserve $\frac14$ of the supersymmetry.  Finally, if
$\theta_i=0$ then the reduction preserves one half of the
supersymmetry.  The discrete quotient associated to this last case was
discussed in \cite{JoanNullAdS}\footnote{It was wrongly argued in
\cite{JoanNullAdS} that the supersymmetry of the double null rotation
quotient was $\nu=\frac{1}{4}$.}

\subsubsection{$\xi=\xi^{(24)}_A + \xi_S$}

In this case we are adding a timelike rotation to the case
$\xi^{(16)}_A$ treated above, but in the simply-connected $\AdS_5$
these orbits are not periodic.  Therefore this integrates to an action
of $\Gamma \cong \RR$ and as there are no points on which
$\xi^{(24)}_A=0$, the stabilisers are all trivial.  The most general
spacelike smooth reduction of this type corresponds to the Killing
vector
\begin{equation*}
  \xi = \be_{12} + \varphi_1 \be_{34} + \varphi_2 \be_{56} + \theta_1
  R_{12} + \theta_2 R_{34} + \theta_3 R_{56}~,
\end{equation*}
where we have relabelled the $\varphi_i$ and rescaled the vector and
where $\theta_1 \geq \theta_2 \geq \theta_3 > 1$ and $\varphi_1 \geq
\varphi_2 \geq 1$.  Since $\Gamma \cong \RR$, the quotient is spin. 

To analyse the supersymmetry of this class of quotients, we notice
that the action of $\xi$ on the (complexified) Killing spinors is
semisimple with weights:
\begin{equation*}
  \pm \half \varphi_1 \pm \half \varphi_2 \pm \half \varphi_3
  \pm \half \theta_1 \pm \half \theta_2 \pm \half \theta_3~,
\end{equation*}
where the signs are uncorrelated except that their product is
positive and where we have reintroduced $\varphi_3$, which can be set
to $1$ as was done before if desired.  Supersymmetry requires the
vanishing of one or more of these weights.  These conditions define
$16$ hyperplanes, of which $6$ have no points for our choice of
ordering of the parameters.  If the parameters lie in precisely one of
the remaining $10$ hyperplanes, the corresponding quotient has two
supersymmetries.  This happens whenever one of the following $10$
equations is satisfied:
\begin{equation}
  \label{eq:637h1}
  \begin{aligned}
    \varphi_1 \pm (\varphi_2 - \varphi_3) - \theta_1 \pm (\theta_2 +
    \theta_3) &= 0 \\
    \varphi_1 \pm (\varphi_2 + \varphi_3) + \theta_1 - \theta_2 - 
    \theta_3 &= 0 \\
    \varphi_1 \pm (\varphi_2 + \varphi_3) - \theta_1 \pm (\theta_2 - 
    \theta_3) &= 0~.
  \end{aligned}
\end{equation}
Such reductions preserve a fraction $\nu=\frac1{16}$ of the
supersymmetry.

The analysis of the possible intersections of these hyperplanes has
again been done with the help of \textsc{Mathematica}.  The following
$16$ conditions guarantee that the parameters lie in the intersection
of precisely two hyperplanes:
\begin{equation}
  \label{eq:637h2}
  \begin{aligned}
    \varphi_2 &= \varphi_3 \qquad & \theta_1 &= \varphi_1 \pm
    (\theta_2 + \theta_3) \\
    \theta_1 &= \varphi_1 \qquad & \theta_2 &= \varphi_2 \pm (\theta_3
    + \varphi_3) \\
    \theta_2 &= \varphi_2 \qquad & \theta_1 &= \varphi_1 \pm (\theta_3
    + \varphi_3) \\
    \theta_3 &= \varphi_2 \qquad & \theta_1 &= \varphi_1 \pm (\theta_2
    + \varphi_3) \\
    \theta_1 &= \theta_2 \qquad & \theta_3 &= \varphi_1 \pm (\varphi_2
    + \varphi_3) \\
    \theta_2 &= \theta_3 \qquad & \theta_1 &= \varphi_1 \pm (\varphi_2
    + \varphi_3) \\
    \varphi_1 &= \varphi_2 \qquad & \theta_1 &= \theta_2 + \theta_3 +
    \varphi_3 \\
    \theta_1 &= \varphi_2 \qquad & \varphi_1 &= \theta_2 + \theta_3 +
    \varphi_3 \\
    \theta_2 &= \varphi_1 \qquad & \theta_1 &= \theta_3 + \varphi_2 +
    \varphi_3 \\
    \theta_3 &= \varphi_1 \qquad & \theta_1 &= \theta_2 + \varphi_2 +
    \varphi_3~.
  \end{aligned}
\end{equation}
Each of these reductions preserve $\tfrac18$ of the ten-dimensional
supersymmetry.

Similarly, parameters satisfying the following $13$ conditions lie in
precisely three hyperplanes:
\begin{equation}
  \label{eq:637h3}
  \begin{aligned}
  \varphi_1 = \varphi_2 = \varphi_3 &= \theta_1 - \theta_2 - \theta_3
  \\
  \theta_1 = \theta_2 = \theta_3 &= \varphi_1 \pm (\varphi_2 +
  \varphi_3) \\
  \theta_1 = \varphi_1 = \varphi_2 &= \theta_2 + \theta_3 + \varphi_3
  \\
  \theta_3 = \varphi_1 = \varphi_2 &= \theta_1 - \theta_2 - \varphi_3
  \\
  \theta_2 = \varphi_1 = \varphi_2 &= \theta_1 - \theta_3 - \varphi_3
  \\
  \theta_1 = \theta_2 = \varphi_2 &= \varphi_1 - \theta_3 - \varphi_3
  \\
  \theta_2 = \theta_3 = \varphi_2 &= \pm (\theta_1 - \varphi_1) -
  \varphi_3 \\
  \theta_1 = \theta_2 = \varphi_1 &= \theta_3 \pm (\varphi_2 +
  \varphi_3) \\
  \theta_2 = \theta_3 = \varphi_1 &= \theta_1 \pm (\varphi_2 +
  \varphi_3)~.
  \end{aligned}
\end{equation}
Each of these reductions preserves $\frac3{16}$ of the supersymmetry.

Finally there are reductions preserving $\frac14$ of the supersymmetry
corresponding to parameters lying in precisely four of the
hyperplanes, which happens whenever either one of the two conditions
hold:
\begin{equation}
  \label{eq:637h4}
  \begin{aligned}
    \theta_1 & = 2\varphi_2 + \varphi_3 \qquad & \theta_2 &= \theta_3 =
    \varphi_1 = \varphi_2 \\
    \varphi_1 & = 2 \varphi_2 + \varphi_3 \qquad & \theta_1 &=
    \theta_2 = \theta_3 = \varphi_2~.
  \end{aligned}
\end{equation}
The causality constraints on the parameters, which do not allow them
to vanish, forbid intersections of five or more hyperplanes.

\subsubsection{Summary}

The smooth spacelike supersymmetric quotients of $\AdS_5 \times S^5$
are summarised in Table~\ref{tab:AdS5S5}, where we have also indicated
the fraction $\nu$ of the IIB supersymmetry preserved by the quotient.

\begin{table}[h!]
  \centering
  \setlength{\extrarowheight}{3pt}
  \renewcommand{\arraystretch}{1.3}
  \begin{tabular}{|>{$}l<{$}|>{$}l<{$}|>{$}l<{$}|>{$}c<{$}|}\hline
    & \multicolumn{2}{c|}{Conditions for} & \\
    \multicolumn{1}{|c|}{Killing vector} & 
    \multicolumn{1}{c|}{Causality/Weyl} &
    \multicolumn{1}{c|}{Supersymmetry} & 
    \multicolumn{1}{c|}{$\nu$}\\
    \hline\hline
    \varphi_1 \be_{34} + \varphi_2 \be_{56} & \varphi_1 \geq \varphi_2
    \geq 0 & \text{$\varphi_1 = 1$, $\varphi_2 = 0$} & \frac38 \\
    \hfill + R_{12} + R_{34} \pm R_{56} & & \text{$\varphi_1 = 2$,
      $\varphi_2 = 1$} & \frac14 \\
    & & \text{$\varphi_1 = 1+n$, $\varphi_2 = n>1$} & \frac3{16}\\
    & & \text{$\varphi_1 = 3$, $\varphi_2 = 0$} & \frac18 \\
    & & \text{$\varphi_1 = 3+n$, $\varphi_2 = n>0$} & \frac1{16}\\
    \hline
    -\be_{12} - \be_{13} + \be_{24} + \be_{34} & \varphi_2 \geq
    |\varphi_1| > 0 & \multicolumn{1}{c|}{\textbf{\eqref{eq:634h1}}} &
    \frac1{16}\\
    \hfill + \varphi_1(\be_{12} + \be_{34}) + \varphi_2 \be_{56}
    \hfill & \theta_1 \geq \theta_2 \geq |\theta_3| > |\varphi_1| &
    \multicolumn{1}{c|}{\textbf{\eqref{eq:634h2}}} & \frac18\\
    \hfill + \theta_1 R_{12} + \theta_2 R_{34} + \theta_3 R_{56} & 
    & \multicolumn{1}{c|}{\textbf{\eqref{eq:634h3}}} &
    \frac3{16}\\ \hline
    \be_{13} + \be_{24} + \varphi \be_{56} & \theta_1 \geq \theta_2
    \geq |\theta_3| & \multicolumn{1}{c|}{\textbf{\eqref{eq:635h1}}} &
    \frac18\\
    \hfill + \theta_1 R_{12} + \theta_2 R_{34} + \theta_3 R_{56} &
    \varphi \geq 0 & \multicolumn{1}{c|}{\textbf{\eqref{eq:635h2}}} &
    \frac14\\
    & & \varphi = \theta_1 = \theta_2 = - \theta_3 \neq 0 & \frac38\\
    & & \theta_i = \varphi = 0 & \half\\ \hline
    \be_{13} - \be_{34} + \be_{25} - \be_{56} & \theta_1 \geq \theta_2
    \geq |\theta_3| & \theta_1  - \theta_2 \pm \theta_3 = 0 & \frac18\\
    \hfill +  \theta_1 R_{12} + \theta_2 R_{34} + \theta_3 R_{56} & &
    \text{$\theta_1 = \theta_2$, $\theta_3 = 0$} & \frac14 \\
    & & \theta_i = 0 & \frac12\\ \hline
    \be_{12} + \varphi_1 \be_{34} + \varphi_2 \be_{56} &
    \varphi_1 \geq \varphi_2 \geq 1 &
    \multicolumn{1}{c|}{\textbf{\eqref{eq:637h1}} with $\varphi_3=1$}
    & \frac1{16}\\
    \hfill + \theta_1 R_{12} + \theta_2 R_{34} + \theta_3 R_{56} &
    \theta_1 \geq \theta_2 \geq \theta_3 > 1 &
    \multicolumn{1}{c|}{\textbf{\eqref{eq:637h2}} with $\varphi_3=1$}
    & \frac18\\
    & & \multicolumn{1}{c|}{\textbf{\eqref{eq:637h3}} with
      $\varphi_3=1$} & \frac3{16}\\
    & & \multicolumn{1}{c|}{\textbf{\eqref{eq:637h4}} with
      $\varphi_3=1$} & \frac14\\ \hline
  \end{tabular}
  \vspace{8pt}
  \caption{Smooth spacelike supersymmetric quotients of $\AdS_5 \times
    S^5$}
  \label{tab:AdS5S5}
\end{table}

\subsection{Supersymmetric quotients of $\AdS_7 \times S^4$}
\label{sec:ads7s4}

The $\AdS_7 \times S^4$ vacuum of eleven-dimensional supergravity is
such that $S^4$ has radius of curvature $R_S = R$ and $\AdS_7$ has
radius of curvature $R_A = 2R$.  Letting $s>0$ denote the scalar
curvature of the sphere, the $4$-form is given by
\begin{equation*}
  F = \sqrt{\tfrac34 s} \dvol_S ~,
\end{equation*}
where $\dvol_S$ is the volume form of $S^4$.  This vacuum solution has
thirty-two supersymmetries.  In this section we will study the smooth
supersymmetric quotients.

In this case the sphere is even-dimensional, whence every (Killing)
vector has zeroes.  This means that in contrast to the cases treated
previously, we require the component $\xi_A$ of the Killing vector
tangent to $\AdS_7$ to be everywhere spacelike.  The results in
Section~\ref{sec:so26} tell us that they are the ones labelled by
(11), (25) and (39) all three with $\beta_1 = \beta_2$, (23) and (37),
which we proceed to investigate.

Throughout this section the spherical component $\xi_S$ of the Killing
vector is given by
\begin{equation*}
  \xi_S = \theta_1 R_{12} + \theta_2 R_{34}~,
\end{equation*}
where we can choose $\theta_1 \geq \theta_2 \geq 0$.  This vector
field clearly vanishes at the points $(0,0,0,0,\pm R)$.

\subsubsection{$\xi=\xi^{(I)}_A + \xi_S$, for $I=11,25,39$}

These three cases can be treated simultaneously, since they are all
specialisations of (39) where one or both of the rotation angles in
$\AdS_7$ vanish.  The Killing vector is given by
\begin{equation*}
  \xi = \beta( \be_{13} + \be_{24}) + \varphi_1 \be_{56} + \varphi_2
  \be_{78} + \theta_1 R_{12} + \theta_2 R_{34}~,
\end{equation*}
where $\beta>0$, whence we can set it equal to $1$ by rescaling
$\xi$.  The action on $\RR^{2,6}$ integrates to
\begin{equation*}
  \begin{pmatrix}
    x^1\\x^2\\x^3\\x^4\\x^5\\x^6\\x^7\\x^8
  \end{pmatrix}
  \mapsto
  \begin{pmatrix}
    x^1 \cosh t  - x^3 \sinh t \\
    x^2 \cosh t  - x^4 \sinh t \\
    x^3 \cosh t  - x^1 \sinh t \\
    x^4 \cosh t  - x^2 \sinh t \\
    x^5\cos \varphi_1 t - x^6 \sin \varphi_1 t\\
    x^6\cos \varphi_1 t + x^5 \sin \varphi_1 t\\
    x^7\cos \varphi_2 t - x^8 \sin \varphi_2 t\\
    x^8\cos \varphi_2 t + x^7 \sin \varphi_2 t
  \end{pmatrix}
\end{equation*}
which is clearly a free action of $\Gamma \cong \RR$ on the quadric
and hence on $\AdS_7 \times S^4$ for all values of $\theta_i$ and
$\varphi_i$.  The resulting background $(\AdS_7 \times S^4)/\Gamma$ is
therefore smooth and, since $\Gamma \cong \RR$, also spin.

Let us discuss the supersymmetry preserved by these quotients.
As in previous sections, the action of $\xi$ on Killing spinors given
by equation \eqref{eq:susyeq} splits into two commuting semisimple
pieces:
\begin{equation*}
  S_1 = \gamma_{13} + \gamma_{24}
\end{equation*}
and
\begin{equation*}
  S_2 = \varphi_{1}\,\gamma_{56} + \varphi_2\,\gamma_{78} +
  \theta_1\,\Gamma_{12} + \theta_2\,\Gamma_{34}~,
\end{equation*}
where $\gamma_{ij}$ are the gamma matrices of $\Cl(2,6)$ and
$\Gamma_{ij}$ those of $\Cl(5,0)$.

The kernel of $S_1$ consists of spinors $\varepsilon$ such that
\begin{equation}
  \gamma_{1234}\,\varepsilon = \varepsilon~,
\end{equation}
which on $\Delta^{2,6}_-$ implies the relation
\begin{equation*}
  \gamma_{5678}\,\varepsilon = - \varepsilon~.
\end{equation*}
The weights of $S_2$ on $\ker S_1 \otimes \Delta^{5,0} \subset
\Delta^{2,6}_- \otimes \Delta^{5,0}$ are easily computed to be
\begin{equation*}
  \pm \half (\varphi_1 + \varphi_2) \pm \half \theta_1 \pm \half
  \theta_2
\end{equation*}
with uncorrelated signs and each with multiplicity $2$.  The
zero-weight conditions give four linear equations on the parameters.
If one (and only one) of these equations is satisfied, the
corresponding quotient preserves four supersymmetries.  There are cases
in which one or more equations can be satisfied simultaneously and
there is enhancement of supersymmetry.  This is easily analysed
yielding the following supersymmetric quotients:
\begin{itemize}
\item $\varphi_1 + \varphi_2 = \theta_1 \pm \theta_2$, with $4$
  supersymmetries;
\item $\varphi_1 + \varphi_2 =0$ and $\theta_1 = \theta_2\neq 0$, with
  $8$ supersymmetries; and
\item $\varphi_1 + \varphi_2 =0$ and $\theta_1 = \theta_2 =0$, with
  $16$ supersymmetries.
\end{itemize}
Note that the last supersymmetric quotient is another example of the
phenomenon first encountered in quotients of $\AdS_3\times S^3$ in
which, supersymmetry wise, one can deform the action of the quotient
in a non-trivial way for free, that is, without decreasing the amount
of supersymmetry preserved by the quotient.

\subsubsection{$\xi=\xi^{(I)}_A + \xi_S$, for $I=23,37$}

Both of these cases can be treated simultaneously since (23) is the
specialisation of (37) where the rotation angle in $\AdS_7$ is set to
zero.  The Killing vector in this case is
\begin{equation*}
  \xi = \be_{13} - \be_{34} + \be_{25} - \be_{56} + \varphi \be_{78} +
  \theta_1 R_{12} +   \theta_2 R_{34}~.
\end{equation*}
The anti-de~Sitter component integrates to the following $\RR$-action
on $\RR^{2,6}$:
\begin{equation*}
  \begin{pmatrix}
    x^1\\x^2\\x^3\\x^4\\x^5\\x^6\\x^7\\x^8
  \end{pmatrix}
  \mapsto
  \begin{pmatrix}
    x^1 - t x^3 + \half t^2 (x^1-x^4)\\
    x^2 - t x^5 + \half t^2 (x^2-x^6)\\
    x^3 + t (x^4 - x^1)\\
    x^4 - t x^3 + \half t^2 (x^1 - x^4) \\
    x^5 + t (x^6 - x^2)\\
    x^6 - t x^5 + \half t^2 (x^2 - x^6) \\
    x^7 \cos \varphi t - x^8 \sin\varphi t\\
    x^8 \cos \varphi t + x^7 \sin\varphi t\\
  \end{pmatrix}~,
\end{equation*}
which is clearly free of fixed points on $\AdS_7$ and has trivial
stabilisers.  The resulting quotient is smooth and also spin.

The supersymmetry analysis departs from the observation that the
action of $\xi$ on the (complexified) Killing spinors $\Delta^{2,6}_-
\otimes \Delta^{5,0}$ splits into a nilpotent piece
\begin{equation*}
  N = \gamma_{13} - \gamma_{34} + \gamma_{25} - \gamma_{56}
\end{equation*}
and a semisimple piece
\begin{equation*}
  S = \varphi \gamma_{78} +  \theta_1 \Gamma_{12} +   \theta_2
  \Gamma_{34}~.
\end{equation*}
The invariant Killing spinors are therefore in the intersection of the
kernels of both $N$ and $S$.  The nilpotent operator $N$ corresponds
to a double null rotation and its kernel was analysed already in
Section~\ref{sec:doublenull}, where we found that it is
half-dimensional, independent of chirality.  In other words, under the
split
\begin{equation*}
  \Delta^{2,6}_- = \left( \Delta^{2,4}_+ \otimes \Delta^{0,2}_+
  \right) \oplus \left( \Delta^{2,4}_- \otimes \Delta^{0,2}_-
  \right)
\end{equation*}
the kernels of $N$ restricted to $\Delta^{2,4}_\pm$ are
two-dimensional and $S$ acts trivially on them.  The weights of $S$ on
$\ker N$ are therefore those of $S$ restricted to $\Delta^{0,2}
\otimes \Delta^{5,0}$ but with multiplicity $2$.  These weights are
\begin{equation*}
  \pm \half \varphi \pm \half \theta_1 \pm \half \theta_2
\end{equation*}
with uncorrelated signs.  Their vanishing defines four hyperplanes in
the parameter space, each contributing four supersymmetries to the
supersymmetry of the quotient, with possible enhancement when the
parameters lie in more than one such hyperplane.  The analysis is
again routine and yields the following supersymmetric quotients:
\begin{itemize}
\item $\varphi = \theta_1 \pm \theta_2$, with four supersymmetries;
\item $\varphi = 0$ and $\theta_1 = \theta_2$, with eight
  supersymmetries;
\item $\varphi = \theta_1$ and $\theta_2 = 0$, with eight
  supersymmetries; and
\item $\varphi = \theta_i = 0$, with sixteen supersymmetries.
\end{itemize}

\subsubsection{Summary}

The smooth spacelike supersymmetric quotients of $\AdS_7 \times S^4$
are summarised in Table~\ref{tab:AdS7S4}, where we have also indicated
the fraction $\nu$ of the eleven-dimensional supersymmetry preserved
by the quotient.

\begin{table}[h!]
  \centering
  \setlength{\extrarowheight}{3pt}
  \renewcommand{\arraystretch}{1.3}
  \begin{tabular}{|>{$}l<{$}|>{$}l<{$}|>{$}l<{$}|>{$}c<{$}|}\hline
    & \multicolumn{2}{c|}{Conditions for} & \\
    \multicolumn{1}{|c|}{Killing vector} & 
    \multicolumn{1}{c|}{Causality/Weyl} &
    \multicolumn{1}{c|}{Supersymmetry} & 
    \multicolumn{1}{c|}{$\nu$}\\
    \hline\hline
    \be_{13} + \be_{24} + \varphi_1 \be_{56} + \varphi_2 \be_{78} &
    \varphi_1 \geq |\varphi_2| \geq 0 & \varphi_1 + \varphi_2 =
    \theta_1 \pm \theta_2 & \frac18\\
     \hfill + \theta_1 R_{12} + \theta_2 R_{34} &
     \theta_1 \geq \theta_2 \geq 0 & \text{$\varphi_1 = - \varphi_2$,
       $\theta_1 = \theta_2 \neq 0$} & \frac14\\
    & & \text{$\varphi_1 = - \varphi_2$, $\theta_1 = \theta_2 = 0$} &
    \half\\ \hline
    \be_{13} - \be_{34} + \be_{25} - \be_{56} & \theta_1 \geq \theta_2
    \geq 0 & \varphi = \theta_1 \pm \theta_2 & \frac18\\
    \hfill + \varphi \be_{78} + \theta_1 R_{12} + \theta_2 R_{34} &
    \varphi \geq 0 & \text{$\varphi=0$, $\theta_1 = \theta_2$} &
    \frac14 \\
    & & \text{$\varphi=\theta_1$, $\theta_2=0$} & \frac14 \\
    & & \varphi=\theta_i=0 & \half\\
    \hline
  \end{tabular}
  \vspace{8pt}
  \caption{Smooth spacelike supersymmetric quotients of $\AdS_7 \times
    S^4$}
  \label{tab:AdS7S4}
\end{table}

\section{Supersymmetry of singular quotients}
\label{sec:singular}

Although in the main body of this work we have focused on the
supersymmetry preserved by smooth and everywhere spacelike quotients
of Freund--Rubin backgrounds of the form $\AdS_{p+1}\times S^q$, the
technology we have developed applies equally to quotients with
singularities: either of causal or differential/topological type.  It
is beyond the scope of this work to study in detail the singular
quotients of such Freund--Rubin backgrounds.  Instead we provide in
this section an indication of the conditions under which supersymmetry
is preserved in a singular quotient.

A preliminary observation is that it \emph{does} make sense to talk
about spin structure and supersymmetry for a quotient $M/\Gamma$ of a
spin manifold $M$ by a group $\Gamma$ acting via
orientation-preserving isometries, even if the quotient is singular,
at least for the geometries of interest.  Indeed, it follows from our
discussion in Section~\ref{sec:spin} on the existence of spin
structures on the quotient of $M=\AdS_{p+1} \times S^q$ by $\Gamma$,
that the action of $\Gamma$ on $P_{\SO}(M)$ is free even when the
action on $M$ is not.  This means that we can \emph{define} the frame
bundle $P_{\SO}(M/\Gamma)$ of a singular quotient simply as
$P_{\SO}(M)/\Gamma$.  This definition agrees with the standard
definition when $M/\Gamma$ is regular.  In the same way, provided that
the action of $\Gamma$ lifts to the spin bundle $P_{\Spin}(M)$, this
action will be free, and we can again define a spin structure on
$M/\Gamma$ simply by $P_{\Spin}(M)/\Gamma$, which again agrees with
the standard definition if the quotient is regular.

As explained in Section~\ref{sec:spin}, if $\Gamma \cong \RR$, the
quotient will always admit a spin structure, whereas if $\Gamma \cong
S^1$, there is a possible obstruction which can be tested by examining
the weights of $\Gamma$ on the (complexified) Killing spinors.  These
weights can either be integral or half-integral.  If the weights are
half-integral there is no spin structure on $M/\Gamma$, but only a
$\Spin^c$ structure, whereas if the weights are integral, there is a
spin structure and moreover the (geometrically realised) supersymmetry
corresponds to the zero weights.

In this section we will examine the weights of $\Gamma$, or
equivalently, of the Lie algebra element generating $\Gamma$, using
the classification of adjoint orbits of $\fso(2,p) \oplus \fso(q+1)$
in terms of direct sums of elementary blocks.  In
Section~\ref{sec:eblksing} we consider the elementary blocks
themselves and in Section~\ref{sec:dblksing} we consider their direct
sums.  We shall focus on the existence of zero weights; that is, on
the conditions for the preservation of supersymmetry.

\subsection{Elementary blocks}
\label{sec:eblksing}

In this section we will examine the action of each elementary block
$B^{m,n} \in \fso(m,n)$ acting on the spinor representation
$\Delta^{m,n}$, or in its complexification.  We recall that every
element $\be_{ij} \in \fso(m,n)$ acts on $\Delta^{m,n}$ as $\half
\gamma_{ij}$.

\subsubsection{$B^{(0,2)}(\varphi)$}

This block corresponds to a spatial rotation.  It is represented by
$\half\varphi \gamma_{34}$, a semisimple operator with eigenvalue
$\frac{i\varphi}2$ on $\Delta^{0,2} \cong \CC$ and eigenvalue
$-\frac{i\varphi}2$ on the conjugate representation
$\overline\Delta^{0,2}$.

\subsubsection{$B^{(1,1)}(\beta)$}

This block corresponds to a boost.  It is representated by a
semisimple operator $\half\beta\gamma_{13}$, whose eigenvalues are
real and equal to $\pm \frac{\beta}2$ on $\Delta^{1,1}_\pm \cong
\RR$.

\subsubsection{$B^{(2,0)}(\varphi)$}

This block is a timelike rotation.  It is representated in
by $\half\varphi\gamma_{12}$, a semisimple operator with imaginary
eigenvalues $\frac{i\varphi}2$ on $\Delta^{2,0}\cong \CC$ and 
$-\frac{i\varphi}2$ on $\overline\Delta^{2,0}$.

Quotienting any $\AdS_{p+1} \times S^q$ by the action of any one of
the above three elementary blocks clearly breaks all supersymmetry.

\subsubsection{$B^{(1,2)}$}

This block is a null rotation.  It is represented in $\Delta^{1,2}
\cong \RR^2$ by a nilpotent operator $\half(\gamma_{13} -
\gamma_{34})$, which has a one-dimensional kernel.

\subsubsection{$B^{(2,1)}$}

This block is also null rotation, but `timelike'.  It is represented
in $\Delta^{2,1} \cong \RR^2$ by a nilpotent operator
$\half(\gamma_{12}-\gamma_{23})$, whose kernel, as in the previous
case, is one-dimensional.

In summary the quotients of any $\AdS_{p+1} \times S^q$ associated to
these two blocks preserve half the supersymmetry.

\subsubsection{$B^{(2,2)}_\pm$}

This block is a `rotation' in a totally null $2$-plane. It is
represented in $\Delta^{2,2}\cong \RR^2 \oplus \RR^2$ as a nilpotent
operator $N^\pm = \half(\mp \gamma_{12} - \gamma_{13} \pm \gamma_{24}
+ \gamma_{34})$ which is itself the product of two nilpotent operators
$N^\pm = \half N_2^\pm N_1 = \half (\pm \gamma_2 + \gamma_3)(\gamma_1
+ \gamma_4)$.  As we saw in Section~\ref{sec:nilprod}, the kernel is
the subspace generated by the kernels of $N_1$ and $N_2^\pm$, which is
three-dimensional.  More precisely, since
\begin{equation*}
  N^\pm (\id \mp \gamma_{1234}) = 0~,
\end{equation*}
$N^\pm$ acts trivially in $\Delta^{2,2}_\mp \cong \RR^2$ and hence has
a one-dimensional kernel in $\Delta^{2,2}_\pm$.

The supersymmetry preserved by such a quotient depends on whether the
Killing spinors belong to $\Delta^{2,2}_\pm$ or to $\Delta^{2,2}$.
The quotient preserves all the supersymmetry in $\Delta^{2,2}_\mp$ and
half the supersymmetry in $\Delta^{2,2}_\pm$ and hence three fourths
of the one in $\Delta^{2,2}$.

\subsubsection{$B^{(2,2)}_\pm(\beta)$}

This block is a `deformation' of the previous one by a selfdual (or
anti-selfdual) boost.  It is represented in $\Delta^{2,2}$ by $N^\pm +
S^\pm$, where $N^\pm$ is as above and $S^\pm = \half\beta (\gamma_{14}
\mp \gamma_{23})$ is semisimple.  It is easy to see that $\ker S^\pm =
\Delta^{2,2}_\pm$ and that $S^\pm$ has eigenvalues $\beta$ and
$-\beta$ on $\Delta^{2,2}_\mp$.  Since $N^\pm S^\pm = S^\pm N^\pm =
0$, it follows that the kernel of $N^\pm + S^\pm$ is a one-dimensional
subspace of $\Delta^{2,2}_\pm$.

The quotient preserves half the supersymmetry in $\Delta^{2,2}_\pm$ and
none of the supersymmetry in $\Delta^{2,2}_\mp$, whence it preserves
one fourth of the one in $\Delta^{2,2}$.

\subsubsection{$B^{(2,2)}_\pm(\varphi)$}

This block is a different deformation of $B^{(2,2)}_\pm$, this time by
a selfdual (or anti-selfdual) rotation.  It is represented in
$\Delta^{2,2}$ by $N^\pm + S^\pm$, where $N^\pm$ is as above and
$S^\pm = \half\varphi (\pm \gamma_{12} + \gamma_{34})$ is semisimple.
It is not hard to see that $N^\pm S^\pm = S^\pm N^\pm = 0$ and that
$\ker S^\pm = \Delta^{2,2}_\pm$. It follows that the kernel of $N^\pm
+ S^\pm$ is a one dimensional subspace of $\Delta^{2,2}_\pm$.

The quotient preserves half the supersymmetry in $\Delta^{2,2}_\pm$ and
none of the supersymmetry in $\Delta^{2,2}_\mp$ and hence one fourth of
the one in $\Delta^{2,2}$.

\subsubsection{$B^{(2,2)}_\pm(\beta,\varphi)$}

This block consists of a selfdual (resp.~anti-selfdual) rotation and a
commuting anti-selfdual (resp.~selfdual) boost.  It is represented in
$\Delta^{2,2}$ by the sum $S^\pm_1 + S^\pm_2$ of two commuting
semisimple operators, where $S^\pm_1 = \half\varphi(\pm\gamma_{12} -
\gamma_{34})$ and $S^\pm_2 = \half\beta(\gamma_{14} \mp \gamma_{23})$.
On the complexification of $\Delta^{2,2}$, $S^\pm_1$ has pure
imaginary eigenvalues whereas $S^\pm_2$ has real eigenvalues, hence
the kernel of their sum is the intersection of their kernels.  Since
the selfduality properties of $S^\pm_1$ and $S^\pm_2$ are opposite,
they leave invariant different subspaces: $\ker S^\pm_1 =
\Delta^{2,2}_\mp$ whereas $\ker S^\pm_2 = \Delta^{2,2}_\pm$.  In other
words, all supersymmetry is broken in any quotient of $\AdS_{p+1}
\times S^q$ by the action of this block; although as we will see in
the next section, there are decomposable blocks containing
$B^{(2,2)}_\pm(\beta,\varphi)$ in which supersymmetry is partially
restored.

\subsubsection{$B^{(2,3)}$}

The block $B^{2,3}$ is represented in $\Delta^{2,3} \cong \RR^4$ by
the nilpotent operator
\begin{equation*}
  N = \half \left( \gamma_{12} + \gamma_{13} + \gamma_{15} -
    \gamma_{24} - \gamma_{34} - \gamma_{45} \right)
\end{equation*}
which can be shown to satisfy $N^4 =0$.  We can rewrite $N$ as
\begin{equation*}
  N =\half (\gamma_1 + \gamma_4) (\gamma_2 + \gamma_3) + \half
  (\gamma_1 - \gamma_4) \gamma_5~,
\end{equation*}
which suggests the following decomposition
\begin{equation*}
  \Delta^{2,3} = V \oplus (\gamma_2 + \gamma_3) V \oplus (\gamma_1 +
  \gamma_4) V \oplus (\gamma_2 + \gamma_3)(\gamma_1 + \gamma_4) V~,
\end{equation*}
analogous to the one in Section~\ref{sec:doublenull}, where now $V =
\ker(\gamma_1-\gamma_4) \cap \ker(\gamma_2 - \gamma_3)$.  Let
$\varepsilon \in \Delta^{2,3}$ be decomposed accordingly as
\begin{equation*}
  \varepsilon = \varepsilon_1 + (\gamma_2 + \gamma_3) \varepsilon_2 + 
  (\gamma_1 + \gamma_4) \varepsilon_3 + (\gamma_2 + \gamma_3)
  (\gamma_1 + \gamma_4) \varepsilon_4~,
\end{equation*}
where $\varepsilon_i \in V$, whence
\begin{equation*}
  N \varepsilon = 2 \gamma_5 \varepsilon_3 + 2
  (\gamma_2+\gamma_3)\gamma_5 \varepsilon_4 + \half (\gamma_1 +
  \gamma_4)(\gamma_2 + \gamma_3) \varepsilon_1~,
\end{equation*}
where $\gamma_5 \varepsilon_i \in V$.  Therefore we see that
$N\varepsilon = 0$ if and only if $\varepsilon_1 = \varepsilon_3 =
\varepsilon_4 = 0$.  In other words, $\varepsilon  = (\gamma_2 +
\gamma_3) \varepsilon_2$, and $\ker N$ is one-dimensional.

Any quotient of $\AdS_{p+1} \times S^q$, for $p\geq 3$, by the action
of this block, preserves one fourth of the supersymmetry.

\subsubsection{$B^{(2,4)}_\pm(\varphi)$}

This block is a deformation of the double null rotation.  It is
represented in $\Delta^{2,4} \cong \CC^4$ by a sum of a nilpotent
operator
\begin{equation*}
  N^\pm = \half (\gamma_1 - \gamma_3)\gamma_5 + (\pm \gamma_2 -
  \gamma_4) \gamma_6~,
\end{equation*}
which is itself the sum of two commuting nilpotent operators, and a
commuting semisimple operator
\begin{equation*}
  S^\pm = \half \varphi (\mp \gamma_{12} + \gamma_{34} +
  \gamma_{56})~.
\end{equation*}
The kernel of $N^\pm + S^\pm$ is thus the intersection of the kernels
of $N^\pm$ and $S^\pm$.  However it is easy to see that $S^\pm$ has no
zero weights in $\Delta^{2,4}$.  Therefore this quotient breaks all
supersymmetry; although as we will see in the next section, there are
decomposable blocks containing $B^{(2,4)}_\pm(\varphi)$ in which
supersymmetry is again partially restored.

The supersymmetry fractions preserved by quotients corresponding to
elementary blocks is summarised in Table~\ref{tab:susyblocks}.  The
notation is such that $\nu$ is the fraction of the supersymmetry
preserved by a block $B^{m,n}$ on $\Delta^{m,n}$, whereas $\nu_\pm$ is
the fraction relative to $\Delta^{m,n}_\pm$ whenever this refinement
is necessary.

\begin{table}[h!]
  \centering
  \setlength{\extrarowheight}{3pt}
  \renewcommand{\arraystretch}{1.3}
    \begin{tabular}[t]{|>{$}l<{$}|>{$}l<{$}|}\hline
      \multicolumn{1}{|c|}{Block} & \nu \\
      \hline\hline
      B^{(0,2)}(\varphi) & 0 \\
      B^{(1,1)}(\beta) &  0 \\ 
      B^{(2,0)}(\varphi) &  0 \\
      B^{(1,2)} & \half \\
      B^{(2,1)} & \half \\
      B^{(2,3)} & \frac14 \\
      B^{(2,4)}_+ (\varphi) & 0 \\
      B^{(2,4)}_- (\varphi) & 0 \\[3pt]
      \hline
    \end{tabular}\hfil
    \begin{tabular}[t]{|>{$}l<{$}|>{$}l<{$}|}\hline
      \multicolumn{1}{|c|}{Block} & \nu\quad(\nu_+,\nu_-) \\
      \hline\hline
      B^{(2,2)}_+ &  \frac34 \quad (\half,1)\\
      B^{(2,2)}_- &  \frac34 \quad (1,\half)\\
      B^{(2,2)}_+(\beta) &  \frac14 \quad (\half,0)\\
      B^{(2,2)}_-(\beta) &  \frac14 \quad (0,\half)\\
      B^{(2,2)}_+(\varphi) & \frac14 \quad (\half,0) \\
      B^{(2,2)}_-(\varphi) & \frac14 \quad (0,\half) \\
      B^{(2,2)}_+(\beta,\varphi) & 0 \\
      B^{(2,2)}_-(\beta,\varphi) & 0 \\[3pt]
      \hline
    \end{tabular}
  \vspace{8pt}
  \caption{The supersymmetry preserved by the elementary blocks.}
  \label{tab:susyblocks}
\end{table}

\subsection{Decomposable blocks}
\label{sec:dblksing}

As discussed in detail in the paper, generic quotients of $\AdS_{p+1}
\times S^q$ are not generated simply by elementary blocks, but by sums
of elementary blocks acting on orthogonal subspaces giving rise to an
element of $\fso(2,p) \oplus \fso(q+1)$.  The precise blocks which can
occur depend crucially on the parameters $p$ and $q$, of course, as
we have seen in Section~\ref{sec:so2p}.

To each decomposable block there is associated an operator $B$ acting
on the relevant spinorial representation of $\fso(2,p) \oplus
\fso(q+1)$, in whose kernel (if any) are contained the supersymmetries
preserved in the corresponding quotient.  As we have seen in
Section~\ref{sec:squots} and indeed in Section~\ref{sec:eblksing}, $B$
decomposes into a sum of commuting operators which can be of three
types: nilpotent ($N$), semisimple with real eigenvalues ($S_R$) and
semisimple with pure imaginary eigenvalues ($S_I$).  In all, there are
fourteen different types of decomposable blocks, which we arrange
according to the form of $B$.  Four cases arise: $N + S_R + S_I$, $N +
S_I$, $S_R + S_I$ and $S_I$, which will be discussed in turn after the
following general remark.

The supersymmetries preserved by the quotient corresponding to $B$ are
given by the kernel of $B$ acting on the relevant spinorial
representation.  It is a standard fact in linear algebra that the
kernel of $B = N + S_R + S_I$ is the intersection of the kernels of
$N$, $S_R$ and $S_I$.  The kernel of $S_I$ can be determined by a
detailed analysis of the weights of $S_I$ in the spinorial
representation.  This has been done for the smooth spacelike quotients
in Section~\ref{sec:squots} and we will not perform this analysis for
the singular quotients in what follows.  We will simply point out when
preservation of supersymmetry is possible provided that the angles in
the expression for $S_I$ are suitably fine-tuned.

The results are summarised in Tables~\ref{tab:N+SR+SI},
\ref{tab:N+SI}, \ref{tab:SR+SI} and \ref{tab:SI}, in which the
notation $\bigoplus_i B^{(0,2)}(\varphi_i)$ denotes a block which can
contain sub-blocks acting on the sphere.  The second column in the
tables contain the corresponding adjoint orbits of $\fso(2,6)$
following the enumeration in Section~\ref{sec:so26}.  Some of these
adjoint orbits also exist in smaller $\fso(2,p<6)$.

\subsubsection{$B = N + S_R + S_I$}
\label{sec:N+SR+SI}

In Table~\ref{tab:N+SR+SI} we list the possible decomposable blocks
giving rise to a spinor operator of the form $N + S_R + S_I$.

\begin{table}[h!]
  \centering
  \setlength{\extrarowheight}{3pt}
  \renewcommand{\arraystretch}{1.3}
    \begin{tabular}{|>{$}l<{$}|>{$}l<{$}|}\hline
      \multicolumn{1}{|c|}{Block} & \multicolumn{1}{c|}{Cases} \\
      \hline\hline
      B^{(2,2)}_\pm(\beta) \oplus \bigoplus_i B^{(0,2)}(\varphi_i) &
      7, 19, 34\\
      B^{(1,1)}(\beta) \oplus B^{(1,2)} \oplus \bigoplus_i
      B^{(0,2)}(\varphi_i) & 15, 29\\
      \hline
    \end{tabular}
  \vspace{8pt}
  \caption{Decomposable blocks of the form $N+S_R+S_I$}
  \label{tab:N+SR+SI}
\end{table}

The quotient by $B^{(2,2)}_\pm(\beta) \oplus \bigoplus_i
B^{(0,2)}(\varphi_i)$ preserves supersymmetry if and only if $\bigoplus_i
B^{(0,2)}(\varphi_i)$ has zero weights in the spinorial
representation, whereas the boost in $B^{(1,1)}(\beta) \oplus
B^{(1,2)} \oplus \bigoplus_i B^{(0,2)}(\varphi_i)$ breaks all the
supersymmetry.

\subsubsection{$B = N + S_I$}
\label{sec:N+SI}

In Table~\ref{tab:N+SI} we list the possible decomposable blocks
giving rise to a spinor operator of the form $N + S_I$.

\begin{table}[h!]
  \centering
  \setlength{\extrarowheight}{3pt}
  \renewcommand{\arraystretch}{1.3}
    \begin{tabular}{|>{$}l<{$}|>{$}l<{$}|}\hline
      \multicolumn{1}{|c|}{Block} & \multicolumn{1}{c|}{Cases}\\
      \hline\hline
      B^{(1,2)} \oplus \bigoplus_i B^{(0,2)}(\varphi_i) & 5, 17, 31\\
      B^{(1,2)} \oplus B^{(1,2)} \oplus \bigoplus_i
      B^{(0,2)}(\varphi_i) & 23, 37\\
      B^{(2,1)} \oplus \bigoplus_i B^{(0,2)}(\varphi_i) & 3, 14, 28\\
      B^{(2,2)}_\pm \oplus \bigoplus_i B^{(0,2)}(\varphi_i) & 6, 21,
      33\\
      B^{(2,2)}_\pm(\varphi) \oplus \bigoplus_i B^{(0,2)}(\varphi_i) &
      8, 20, 35\\
      B^{(2,3)} \oplus \bigoplus_i B^{(0,2)}(\varphi_i) & 13, 27\\
      B^{(2,4)}_\pm(\varphi) \oplus \bigoplus_i B^{(0,2)}(\varphi_i) &
      18, 32\\
      \hline
    \end{tabular}
  \vspace{8pt}
  \caption{Decomposable blocks of the form $N+S_I$}
  \label{tab:N+SI}
\end{table}

The blocks $B^{(1,2)} \oplus \bigoplus_i B^{(0,2)}(\varphi_i)$,
$B^{(1,2)} \oplus B^{(1,2)} \oplus \bigoplus_i B^{(0,2)}(\varphi_i)$
and $B^{(2,1)} \oplus \bigoplus_i B^{(0,2)}(\varphi_i)$ preserve one
half of the supersymmetry preserved by $\bigoplus_i
B^{(0,2)}(\varphi_i)$, if any.

The blocks $B^{(2,2)}_\pm \oplus \bigoplus_i B^{(0,2)}(\varphi_i)$ and
$B^{(2,2)}_\pm(\varphi)\oplus \bigoplus_i B^{(0,2)}(\varphi_i)$
preserve supersymmetry if and only if $\bigoplus_i
B^{(0,2)}(\varphi_i)$ does.  The precise fraction depends on the
particular spinorial representations to which the Killing spinors
belong.

The block $B^{(2,3)} \oplus \bigoplus_i B^{(0,2)}(\varphi_i)$
preserves one fourth of the supersymmetry preserved by $\bigoplus_i
B^{(0,2)}(\varphi_i)$, if any.

Finally, the block $B^{(2,4)}_\pm(\varphi) \oplus \bigoplus_i
B^{(0,2)}(\varphi_i)$ deserves special attention.  Although the
elementary block $B^{(2,4)}_\pm(\varphi)$ breaks all the
supersymmetry, it is possible to restore supersymmetry by adding
$\bigoplus_i B^{(0,2)}(\varphi_i)$ and fine-tuning the parameters
$\varphi$ and $\varphi_i$ in such a way that the corresponding
semisimple operator $S_I(\varphi,\varphi_i)$ have zero weights in the
relevant spinor representation.

\subsubsection{$B = S_R + S_I$}
\label{sec:SR+SI}

In Table~\ref{tab:SR+SI} we list the possible decomposable blocks
giving rise to a spinor operator of the form $S_R + S_I$.

\begin{table}[h!]
  \centering
  \setlength{\extrarowheight}{3pt}
  \renewcommand{\arraystretch}{1.3}
    \begin{tabular}{|>{$}l<{$}|>{$}l<{$}|}\hline
      \multicolumn{1}{|c|}{Block} & \multicolumn{1}{c|}{Cases}\\
      \hline\hline
      B^{(1,1)}(\beta) \oplus \bigoplus_i B^{(0,2)}(\varphi_i) & 2,
      12, 26\\
      B^{(1,1)}(\beta_1) \oplus B^{(1,1)}(\beta_2) \oplus \bigoplus_i
      B^{(0,2)}(\varphi_i) & 11, 25, 39\\
      B^{(2,2)}_\pm(\beta,\varphi) \oplus
      \bigoplus_iB^{(0,2)}(\varphi_i) & 9, 22, 36\\
      \hline
    \end{tabular}
  \vspace{8pt}
  \caption{Decomposable blocks of the form $S_R+S_I$}
  \label{tab:SR+SI}
\end{table}

The boost in $B^{(1,1)}(\beta) \oplus \bigoplus_i
B^{(0,2)}(\varphi_i)$ breaks all supersymmetry and so do the boosts in
$B^{(1,1)}(\beta_1) \oplus B^{(1,1)}(\beta_2) \oplus \bigoplus_i
B^{(0,2)}(\varphi_i)$, unless $\beta_1 = \beta_2$, in which case this
block preserves one half of the supersymmetry preserved by
$\bigoplus_i B^{(0,2)}(\varphi_i)$, if any.

Finally, there is the possibility of restoration of supersymmetry in
the block $B^{(2,2)}_\pm(\beta,\varphi) \oplus
\bigoplus_iB^{(0,2)}(\varphi_i)$, provided that the angles $\varphi,
\varphi_i$ belong to the loci where $S_I(\varphi,\varphi_i)$ has zero
weights.  In this case, the block preserves one half of the
supersymmetry preserved by $S_I$.

\subsubsection{$B = S_I$}
\label{sec:SI}

In Table~\ref{tab:SI} we list the possible decomposable blocks
giving rise to a spinor operator of the form $S_I$.

\begin{table}[h!]
  \centering
  \setlength{\extrarowheight}{3pt}
  \renewcommand{\arraystretch}{1.3}
    \begin{tabular}{|>{$}l<{$}|>{$}l<{$}|}\hline
      \multicolumn{1}{|c|}{Block} & \multicolumn{1}{c|}{Cases}\\
      \hline\hline
      \bigoplus_i B^{(0,2)}(\varphi_i) & 4, 16, 30\\
      B^{(2,0)}(\varphi) \oplus \bigoplus_i B^{(0,2)}(\varphi_i) & 1,
      10, 24, 38\\
      \hline
    \end{tabular}
  \vspace{8pt}
  \caption{Decomposable blocks of the form $S_I$}
  \label{tab:SI}
\end{table}

In both of these cases, the supersymmetry preserved by the quotient is
dictated by the zero weights of the corresponding spinor operator.

It should be remarked that in the case of $\bigoplus_i
B^{(0,2)}(\varphi_i)$, the quotient will not be Hausdorff unless the
orbits of the Killing vector are periodic, which requires the angles
$\varphi_i$ to be rationally related.

\section*{Acknowledgments}

We would like to acknowledge useful conversations with Charles
Frances, Chris Hull, Takashi Kimura, Felipe Leitner, Andrei Moroianu
and Elmer Rees.  We are also grateful to Owen Madden and Simon Ross
for making the results of their paper \cite{MaddenRoss} known to us
prior to publication.

This work and \cite{FOMRS} conclude a project conceived while the
authors were participating in the programme \emph{Mathematical Aspects
  of String Theory} which took place in the Fall of 2001 at the Erwin
Schrödinger Institute in Vienna, and it is again our pleasure to thank
them for support and for providing such a stimulating environment in
which to do research.  JMF's participation in that programme was made
possible in part by a travel grant from the UK PPARC.  JMF would like
to acknowledge the support and hospitality of the Weizmann Institute
and CERN which he visited during the time that it took to complete and
write up this work.  In particular he would like to thank Micha
Berkooz for the invitation to visit the Weizmann Institute.  JMF is a
member of EDGE, Research Training Network HPRN-CT-2000-00101,
supported by The European Human Potential Programme, and his research
is partially supported by the UK EPSRC grant GR/R62694/01.  JS would
like to thank the School of Mathematics of the University of Edinburgh
and the Aspen Center for Physics for kind hospitality during the
initial and intermediate stages, respectively, of this work.  During
the time it took to complete this work, JS was supported by a Marie
Curie Fellowship of the European Community programme `Improving the
Human Research Potential and the Socio-Economic Knowledge Base' under
the contract number HPMF-CT-2000-00480, by grants from the United
States--Israel Binational Science Foundation (BSF), the European RTN
network HPRN-CT-2000-00122, Minerva, the Phil Zacharia fellowship, and
by the United States Department of Energy under the grant number
DE-FG02-95ER40893 and the National Science Foundation under Grant
No.~PHY99-07949.

\bibliographystyle{utphys}
\bibliography{AdS,ESYM,Sugra,Geometry,CaliGeo,Algebra}

\end{document}